\documentclass[twocolumn]{aastex62}
\usepackage{amsmath}
\usepackage{csquotes}
\usepackage{bm}
\usepackage{savesym}

\savesymbol{tablenum} % credit: https://tex.stackexchange.com/a/269074
\usepackage{siunitx}
\restoresymbol{SI}{tablenum}

\usepackage{grfext}
\graphicspath{{./}{figures/}}  % NOTE: the trailing '/' matters

\newcommand{\R}[1]{\mathrm{#1}}
\newcommand{\D}[1]{\R{d} #1}
\newcommand{\diff}[2]{\frac{\D{#1}}{\D{#2}}}
\newcommand{\pdiff}[2]{\frac{\partial #1}{\partial #2}}
\newcommand{\lcdm}{$\Lambda$CDM}
\newcommand{\Halpha}{\text{H$\alpha$}}
\newcommand{\Hi}{H\,\textsc{i}}

\newcommand{\klos}{\text{$k_{\parallel}$}}
\newcommand{\kperp}{\text{$k_{\bot}$}}
\newcommand{\fov}{\text{Fo\!V}}

\DeclareMathOperator{\erf}{erf}

\DeclareSIUnit\arcsec{arcsec}
\DeclareSIUnit\arcmin{arcmin}
\DeclareSIUnit\cMpc{cMpc}  % comoving Mpc
\DeclareSIUnit\cGpc{cGpc}  % comoving Gpc
\DeclareSIUnit\deg{deg}
\DeclareSIUnit\erg{erg}
\DeclareSIUnit\gauss{G}
\DeclareSIUnit\hour{hr}  % overwrite hour from 'h' to 'hr'
\DeclareSIUnit\hubble{\text{$h$}}
\DeclareSIUnit\jansky{Jy}
\DeclareSIUnit\lightyear{ly}
\DeclareSIUnit\parsec{pc}
\DeclareSIUnit\rayleigh{Rayleigh}
\DeclareSIUnit\solarmass{\text{M$_{\odot}$}}
\DeclareSIUnit\Msun{\solarmass}
\DeclareSIUnit\year{yr}
\DeclareSIUnit\keV{\kilo\electronvolt}
\DeclareSIUnit\kpc{\kilo\parsec}
\DeclareSIUnit\mJy{\milli\jansky}
\DeclareSIUnit\mK{\milli\kelvin}
\DeclareSIUnit\uG{\micro\gauss}
\DeclareSIUnit\Gyr{\giga\year}
\DeclareSIUnit\MHz{\mega\hertz}
\DeclareSIUnit\Mpc{\mega\parsec}
\DeclareSIUnit\Gpc{\giga\parsec}

\def\equationautorefname~#1\null{Equation~(#1)\null}

\newcounter{sssseccount}
\newcommand{\sssseclabel}{\alph{sssseccount}}
\newcommand{\ssssec}[1]{%
  \vspace{1ex}%
  \stepcounter{sssseccount}%
  \noindent\emph{\sssseclabel. #1}%
}

\received{2018 August 14}
\revised{2019 March 26}
\accepted{2019 May 13}
\submitjournal{ApJ}

\shorttitle{EoR Foreground Contribution of Radio Halos}
\shortauthors{Li~\textit{et~al.}}

\begin{document}

\title{\bfseries%
  Contribution of Radio Halos to the Foreground for SKA EoR Experiments
}

\correspondingauthor{Haiguang Xu}
\email{hgxu@sjtu.edu.cn}

\author[0000-0002-7527-380X]{Weitian Li}
\email{liweitianux@sjtu.edu.cn}
\affiliation{School of Physics and Astronomy,
  Shanghai Jiao Tong University,
  800 Dongchuan Road, Shanghai 200240, China}

\author{Haiguang Xu}
\affiliation{School of Physics and Astronomy,
  Shanghai Jiao Tong University,
  800 Dongchuan Road, Shanghai 200240, China}
\affiliation{Tsung-Dao Lee Institute,
  Shanghai Jiao Tong University,
  800 Dongchuan Road, Shanghai 200240, China}
\affiliation{IFSA Collaborative Innovation Center,
  Shanghai Jiao Tong University,
  800 Dongchuan Road, Shanghai 200240, China}

\author[0000-0003-1263-9453]{Zhixian Ma}
\affiliation{Department of Electronic Engineering,
  Shanghai Jiao Tong University,
  800 Dongchuan Road, Shanghai 200240, China}
\author{Dan Hu}
\affiliation{School of Physics and Astronomy,
  Shanghai Jiao Tong University,
  800 Dongchuan Road, Shanghai 200240, China}
\author{Zhenghao Zhu}
\affiliation{School of Physics and Astronomy,
  Shanghai Jiao Tong University,
  800 Dongchuan Road, Shanghai 200240, China}
\author{Chenxi Shan}
\affiliation{School of Physics and Astronomy,
  Shanghai Jiao Tong University,
  800 Dongchuan Road, Shanghai 200240, China}
\author{Jingying Wang}
\affiliation{Department of Physics and Astronomy,
  University of the Western Cape,
  Cape Town 7535, South Africa}
\author[0000-0001-9765-6521]{Junhua Gu}
\affiliation{National Astronomical Observatories,
  Chinese Academy of Sciences,
  20A Datun Road, Beijing 100012, China}
\author{Dongchao Zheng}
\affiliation{School of Physics and Astronomy,
  Shanghai Jiao Tong University,
  800 Dongchuan Road, Shanghai 200240, China}
\author{Xiaoli Lian}
\affiliation{School of Physics and Astronomy,
  Shanghai Jiao Tong University,
  800 Dongchuan Road, Shanghai 200240, China}
\author{Qian Zheng}
\affiliation{Shanghai Astronomical Observatory,
  Chinese Academy of Sciences,
  80 Nandan Road, Shanghai 200030, China}
\author{Yu Wang}
\affiliation{School of Mathematics, Physics and Statistics,
  Shanghai University of Engineering Science,
  333 Longteng Road, Shanghai 201620, China}
\author{Jie Zhu}
\affiliation{Department of Electronic Engineering,
  Shanghai Jiao Tong University,
  800 Dongchuan Road, Shanghai 200240, China}
\author{Xiang-Ping Wu}
\affiliation{National Astronomical Observatories,
  Chinese Academy of Sciences,
  20A Datun Road, Beijing 100012, China}

%%%%%%%%%%%%%%%%%%%%%%%%%%%%%%%%%%%%%%%%%%%%%%%%%%%%%%%%%%%%%%%%%%%%%%%%%

%% The abstract should summarize concisely the content and conclusions
%% of the article.  The abstract should be a single paragraph of not more
%% than 250 words.

\begin{abstract}
\noindent
The overwhelming foreground contamination is one of the primary impediments
to probing the Epoch of Reionization (EoR) through measuring the redshifted 21~cm signal.
Among various foreground components, radio halos are less studied and
their impacts on the EoR observations are still poorly understood.
In this work, we employ the Press--Schechter formalism, merger-induced
turbulent reacceleration model, and the latest SKA1-Low layout
configuration to simulate the SKA \enquote{observed} images of radio halos.
We calculate the one-dimensional power spectra from simulated images and
find that radio halos can be about \numlist{e4; e3; e2.5} times more luminous
than the EoR signal on scales of $\SI{0.1}{\per\Mpc} < k < \SI{2}{\per\Mpc}$
in the \numrange{120}{128}, \numrange{154}{162}, and \numrange{192}{200}
\si{\MHz} bands, respectively.
By examining the two-dimensional power spectra inside properly defined
EoR windows, we find that the power leaked by radio halos can still be
significant, as the power ratios of radio halos to the EoR signal on scales
of $\SI{0.5}{\per\Mpc} \lesssim k \lesssim \SI{1}{\per\Mpc}$ can be up to
about \numrange{230}{800}\%, \numrange{18}{95}\%, and \numrange{7}{40}\%
in the three bands when the 68\% uncertainties caused by
the variation of the number density of bright radio halos are considered.
Furthermore, we find that radio halos located inside the far side lobes
of the station beam can also impose strong contamination within the EoR
window.
In conclusion, we argue that radio halos are severe foreground sources
and need serious treatments in future EoR experiments.
\end{abstract}

%% Authors should select subject keywords from the given list.  No more
%% than 6 keywords should be given, and they should be listed in
%% alphabetical order.
%% http://journals.aas.org/authors/keywords2013.html

%########################################################################
\section{Introduction}
\label{sec:intro}
%########################################################################

The Epoch of Reionization (EoR; $z \sim \numrange{6}{15}$) refers to
a period of our universe preceded by the Cosmic Dawn
($z \sim \numrange{15}{30}$) and the Dark Ages ($z \sim \numrange{30}{200}$)
and is expected to last from about 300 million to about 1 billion years
after the big bang (see \citealt{koopmans2015rev} and references therein).
During the EoR, the reionization of neutral hydrogen (\Hi), which was
caused primarily by the ultraviolet and soft X-ray photons emitted from
the first-generation celestial objects, efficiently surpassed the cooling
of the gas \citep{dayal2018}.
As a result, the majority of baryonic matter was again in a highly ionized
state.
Comparing with the observations of distant quasars and cosmic microwave
background (CMB), which have provided some important but loose constraints
on the reionization process (see \citealt{fan2006rev} for a review),
the 21~cm line emission of \Hi{} that is redshifted to frequencies below
\SI{200}{\MHz} is regarded as the decisive probe to directly explore the EoR
\citep[see][for reviews]{furlanetto2006rev,zaroubi2013rev,furlanetto2016rev}.

In order to probe the EoR, a number of radio interferometers working
at the low-frequency radio bands (\SIrange{\sim 50}{200}{\MHz}) have been
designed to target the redshifted 21~cm signal, among which there are
the Square Kilometre Array (SKA; \citealt{mellema2013rev,koopmans2015rev}),
the Hydrogen Epoch of Reionization Array (HERA; \citealt{deboer2017}),
and their pathfinders, such as
the LOw Frequency ARray (LOFAR; \citealt{vanHaarlem2013}),
the Murchison Widefield Array (MWA; \citealt{bowman2013,tingay2013}),
the Precision Array for Probing the Epoch of Reionization
(PAPER; \citealt{parsons2010}),
and the 21 CentiMeter Array (21CMA; \citealt{zheng2016}).
The challenges met in these experiments, however, are immense
due to a variety of complicated instrumental effects,
ionospheric distortions, radio frequency interference, and the
strong celestial foreground contamination that overwhelms the
redshifted 21~cm signal by about four to five orders of magnitude
\citep[see][for a review]{morales2010rev}.

Among various contaminating foreground components, the Galactic diffuse
radiation (including both the synchrotron and free-free emissions)
and extragalactic point sources are the most prominent and contribute
the majority of the foreground contamination \citep[e.g.,][]{%
  shaver1999,diMatteo2004,gleser2008,liu2012,murray2017,spinelli2018}.
At about \SI{150}{\MHz}, it is estimated that they may account for
about 71\% and 27\% of the total foreground
contamination, respectively \citep{shaver1999}.
Most of the remaining foreground contamination arises from the emission
from the extragalactic diffuse sources that include the large-scale
filaments embedded in cosmic webs \citep[e.g.,][]{vazza2015},
the intergalactic medium located at cluster outskirts
\citep[e.g.,][]{keshet2004rev},
and the intracluster medium (ICM) of galaxy clusters (radio halos,
relics, and mini-halos; e.g., \citealt{feretti2012rev}).
There is only limited observational evidence, especially in the
low-frequency regime, of these diffuse sources.
Among them, radio halos have gained relatively more observational
constraints and theoretical understanding, which enables us to
effectively evaluate their contamination on the EoR observations.

First discovered in the Coma cluster \citep{large1959}, radio halos
have been observed in about 80 merging galaxy clusters, exhibiting
relatively regular morphologies and about \si{Mpc} spatial extensions.
It should be noted that the angular sizes of radio halos appear to be
several to tens of arcminutes, which coincide with those of the
ionizing bubbles during the EoR.
This, complemented by the potentially large number (several to tens of
thousands in the whole sky) of radio halos to be revealed by the
forthcoming low-frequency radio telescopes \citep[e.g.,][]{cassano2015},
indicates that radio halos might be important contaminating foreground
sources \citep[e.g.,][]{diMatteo2004,gleser2008}.
As of today, however, only very few works have been dedicated to this
topic and are all based on relatively straightforward modeling methods,
such as using the \SI{1.4}{\GHz} radio flux function or radio--X-ray
scaling relations that are barely constrained by the very limited
observations \citep[e.g.,][]{gleser2008,jelic2008}.
In this work, we focus on the radio halos and employ a semi-analytic
approach to derive their low-frequency emission maps with the SKA1-Low's
instrumental effects incorporated.
By making use of the power spectra and the EoR window, the contamination
of radio halos on the EoR observations is quantitatively evaluated for
both foreground removal and avoidance methods, which are the two major
categories of methods that proposed to tackle the foreground
contamination (see \citealt{chapman2016} and references therein).

This paper is prepared as follows.
In \autoref{sec:sky-simu}, we simulate the low-frequency radio sky, where a
semi-analytic simulation of radio halos is developed by employing
the Press--Schechter formalism and turbulent reacceleration model.
In \autoref{sec:obs-simu}, we adopt the latest SKA1-Low layout
configuration to incorporate the practical instrumental effects into
the simulated sky maps.
We briefly introduce the power spectra and the EoR window in
\autoref{sec:ps-eorw}
and then quantitatively evaluate the contamination caused by radio halos
on the EoR measurements in \autoref{sec:results}.
In \autoref{sec:discussions}, we discuss how the EoR detection is
affected by radio halos due to the instrumental frequency artifacts
and the far side lobes of the station beam.
Finally, we summarize our work in \autoref{sec:summary}.
Throughout this work we adopt a flat \lcdm{} cosmology with
$H_0 = \SI{100}{\hubble} = \SI{71}{\km\per\second\per\Mpc}$,
$\Omega_m = 0.27$, $\Omega_{\Lambda} = 1 - \Omega_m = 0.73$,
$\Omega_b = 0.046$, $n_s = 0.96$, and $\sigma_8 = 0.81$.
The quoted uncertainties are at 68\% confidence level unless
otherwise stated.

%########################################################################
\section{Simulation of Radio Sky}
\label{sec:sky-simu}
%########################################################################

Based on our previous works \citep{wang2010,wang2013}, we have developed
the \texttt{FG21sim}\footnote{%
  FG21sim: \url{https://github.com/liweitianux/fg21sim}}
software to simulate the low-frequency
radio sky by taking into account the contributions of our Galaxy,
extragalactic point sources, and radio halos in galaxy clusters.
We choose three representative frequency bands, namely
\numrange{120}{128}, \numrange{154}{162}, and \numrange{192}{200}
\si{\MHz}, and perform simulations for a sky patch of size
\SI[product-units=repeat]{10 x 10}{\degree}.
The \SI{8}{\MHz} bandwidth is chosen to limit the effect of
cosmological evolution of the EoR signal when calculating power
spectra \citep[e.g.][]{wyithe2004,thyagarajan2013}.
The simulated sky maps are pixelized into \num{1800 x 1800} with a pixel
size of \SI{20}{\arcsecond}.

%========================================================================
\subsection{Radio Halos in Galaxy Clusters}
\label{sec:cluster-halos}
%========================================================================

As a significant improvement over our past works \citep{wang2010,wang2013},
we model the radio halos in galaxy clusters by employing the
Press--Schechter formalism and turbulent reacceleration model,
which was pioneered by \citet{brunetti2001} and \citet{petrosian2001}
and further developed in many works
\citep[e.g.,][]{fujita2003,brunetti2004,cassano2005,brunetti2007,brunetti2011}
to explain the observed properties and formation of radio halos
(see \citealt{brunetti2014rev} for a recent review).
In the framework of reacceleration model, relativistic electrons in the
ICM are reaccelerated by the turbulence generated in merger events via the
second-order Fermi process, and lose energies due to mechanisms including
synchrotron radiation, inverse Compton scattering off the CMB photons, and
Coulomb collisions with the thermal ICM.
For a galaxy cluster, we first simulate its merging history according to
the extended Press--Schechter theory and then derive the temporal evolution
of the relativistic electron spectrum by applying the reacceleration
model.
Finally, the radio halo associated with the galaxy cluster is identified
and its synchrotron radiation is determined.

%------------------------------------------------------------------------
\subsubsection{Mass Function}
\label{sec:mass-function}
%------------------------------------------------------------------------

The Press--Schechter formalism was originally advanced as one of the standard
methods to predict the mass function of galaxy clusters and their evolution
in the universe \citep{press1974} and has been extended to combine with
the cold dark matter (CDM) models \citep[e.g.,][]{bond1991,lacey1993}.
In this formalism, the number of galaxy clusters per unit of comoving volume
at redshift $z$ in the mass range $[M, M + \R{d}M]$ is
\begin{multline}
  \label{eq:ps-mass-func}
  n(M, z) \,\D{M} = \sqrt{\frac{2}{\pi}} \frac{\langle{\rho}\rangle}{M}
  \frac{\delta_c(z)}{\sigma^2(M)} \left| \diff{\sigma(M)}{M} \right| \\
  \times \exp\!\left[ -\frac{\delta_c^2(z)}{2\sigma^2(M)} \right] \D{M},
\end{multline}
where $M$ is the virial mass of galaxy clusters,
$\langle {\rho} \rangle$ is the current mean density of the universe,
$\delta_c(z)$ is the critical linear overdensity for a region to collapse
at redshift $z$ [see \autoref{eq:delta-crit}],
and $\sigma(M)$ is the current root-mean-square (rms) density
fluctuations within a sphere of mean mass $M$.

Considering the CDM model and the mass range covered by galaxy clusters,
it is reasonable to adopt the following power-law distribution for the
density perturbations \citep{randall2002,sarazin2002}
\begin{equation}
  \label{eq:sigma-mass}
  \sigma(M) = \sigma_8 \left( \frac{M}{M_8} \right)^{-\alpha},
\end{equation}
where $\sigma_8$ is the current rms density fluctuation on
a scale of \SI{8}{\per\hubble\Mpc},
$M_8 = (4\pi/3)(8 \,\si{\per\hubble\Mpc})^3 \langle{\rho}\rangle$
is the mass contained in a sphere of radius \SI{8}{\per\hubble\Mpc},
and the exponent $\alpha = (n+3)/6$ with $n = -7/5$ \citep{randall2002}
is related to the fluctuation pattern whose power spectrum varies
with wave number $k$ as $k^n$.

\begin{figure}
  \centering
  \includegraphics[width=\columnwidth]{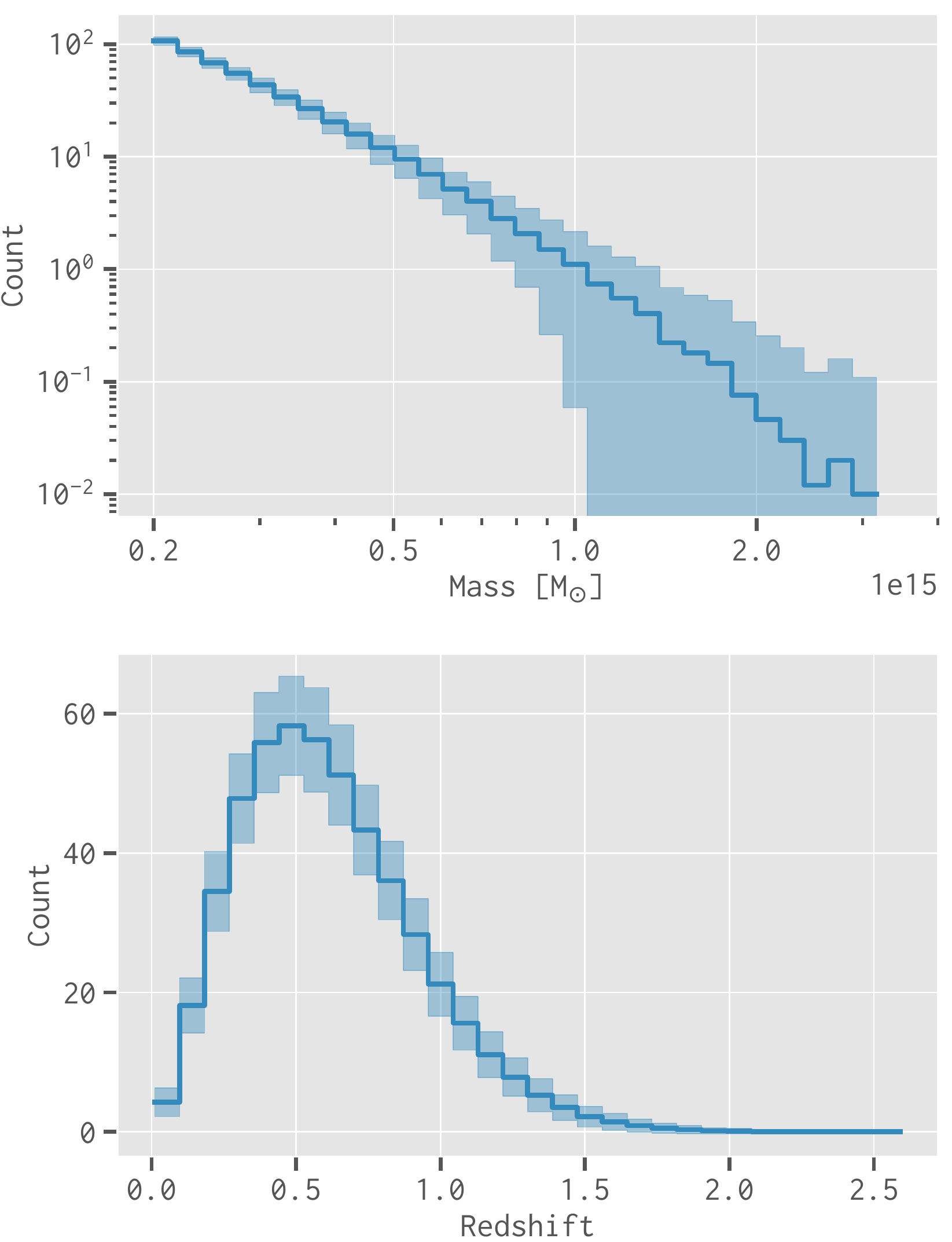}
  \caption{\label{fig:m-z-dist}%
    The mass (upper panel) and redshift (lower panel) histograms of the
    simulated galaxy clusters in a
    \SI[product-units=repeat]{10 x 10}{\degree} sky patch.
    The solid lines and shaded regions represent the means and
    68\% uncertainties derived from 500 simulation runs,
    respectively.
  }
\end{figure}

With a minimum galaxy cluster mass of
$M_{\R{min}} = \SI{2e14}{\solarmass}$
and a maximum redshift cut at $z_{\R{max}} = 4$,
we apply \autoref{eq:ps-mass-func} and
calculate that the total number of galaxy clusters in a
\SI[product-units=repeat]{10 x 10}{\degree} sky patch is 504.
Then the galaxy cluster sample is built by randomly drawing mass and
redshift pairs $(M_{\R{sim}}, z_{\R{sim}})$ from the
mass and redshift distributions as shown in \autoref{fig:m-z-dist},
which are determined by the Press--Schechter mass function
[\autoref{eq:ps-mass-func}].

%------------------------------------------------------------------------
\subsubsection{Merging History}
\label{sec:merging-history}
%------------------------------------------------------------------------

The extended Press--Schechter theory outlined in \citet{lacey1993} provides
a way to describe the growth history of galaxy clusters in terms of the
merger tree.
In order to build the merger tree for a galaxy cluster, we start with
its \enquote{current} mass $M_{\R{sim}}$ and redshift $z_{\R{sim}}$ obtained
in \autoref{sec:mass-function}, and trace its growth history back in time
by running Monte Carlo simulations to randomly determine the mass change
$\Delta M$ at each step, which may be regarded either as a merger event
(if $\Delta M > \Delta M_c$) or as an accretion event
(if $\Delta M \leq \Delta M_c$).
Since radio halos are usually associated with major mergers, we choose
$\Delta M_c = \SI{e13}{\solarmass}$ \citep[e.g.,][]{cassano2005}.

We assume that, during each growth step, the cluster mass increases
from $M_1$ at time $t_1$ to $M_2$ at a later time $t_2$ ($> t_1$).
Given $M_2$ and $t_2$, the conditional probability of the cluster had
a progenitor of mass in the range $[M_1, M_1 + \D{M_1}]$ at an earlier
time $t_1$ can be expressed as
\begin{multline}
  \label{eq:eps-condprob}
  \R{Pr}(M_1, t_1 | M_2, t_2) \,\D{M_1} =
    \frac{1}{\sqrt{2\pi}} \frac{M_2}{M_1}
    \frac{\delta_{c1} - \delta_{c2}}{(\sigma_1^2 - \sigma_2^2)^{3/2}} \\
    \times \left| \diff{\sigma_1^2}{M_1} \right|
    \exp \!\left[ -\frac{(\delta_{c1} - \delta_{c2})^2}
      {2(\sigma_1^2 - \sigma_2^2)} \right] \D{M_1},
\end{multline}
where
$\delta_{ci} \equiv \delta_c(t_i)$, $\sigma_i \equiv \sigma(M_i)$, and
$i = 1, 2$ are used to denote parameters defined at time $t_1$ and $t_2$,
respectively \citep{lacey1993,randall2002}.
By further introducing $S \equiv \sigma^2(M)$ and
$\omega \equiv \delta_c(t)$, this equation reduces to
\begin{equation}
  \label{eq:eps-condprob-simp}
  \R{Pr}(\Delta S, \Delta \omega) \,\D{\Delta S} = \frac{1}{\sqrt{2\pi}}
  \frac{\Delta\omega}{(\Delta S)^{3/2}}
  \exp \!\left[ -\frac{(\Delta\omega)^2}{2 \Delta S} \right] \D{\Delta S}.
\end{equation}

In order to resolve mergers with a mass change $\Delta M_c \ll M$
during the backward tracing of a galaxy cluster, a time step $\Delta t$
(i.e., $\Delta\omega$) that satisfies
\begin{equation}
  \label{sec:dw-step}
  \Delta\omega \lesssim \Delta\omega_{\R{max}} = \left[
    S \left| \diff{\ln \sigma^2}{\ln M} \right|
    \left( \frac{\Delta M_c}{M} \right) \right]^{1/2}
\end{equation}
is required \citep{lacey1993}, and we adopt an adaptive step of
$\Delta\omega = \Delta\omega_{\R{max}} / 2$ \citep{randall2002}.
At a certain step when $\Delta\omega$ is given, the mass change
$\Delta S$ can be randomly drawn from the following cumulative
probability distribution of subcluster masses
\begin{align}
  \label{sec:cdf-sub-masses}
  \R{Pr}(<\!\Delta S, \Delta\omega)
    & = \int_0^{\Delta S} \R{Pr}(\Delta S', \Delta\omega) \,\D{\Delta S'} \\
    & = 1 - \erf \!\left( \frac{\Delta \omega}{\sqrt{2 \Delta S}} \right),
\end{align}
where
$\erf(x) = (2/\!\sqrt{\pi}) \int_0^x \R{e}^{-t^2} \,\D{t}$
is the error function.
Then, the cluster's progenitor mass $M_1$ is obtained as
$S_1 = S_2 + \Delta S$.

\begin{figure}
  \centering
  \includegraphics[width=\columnwidth]{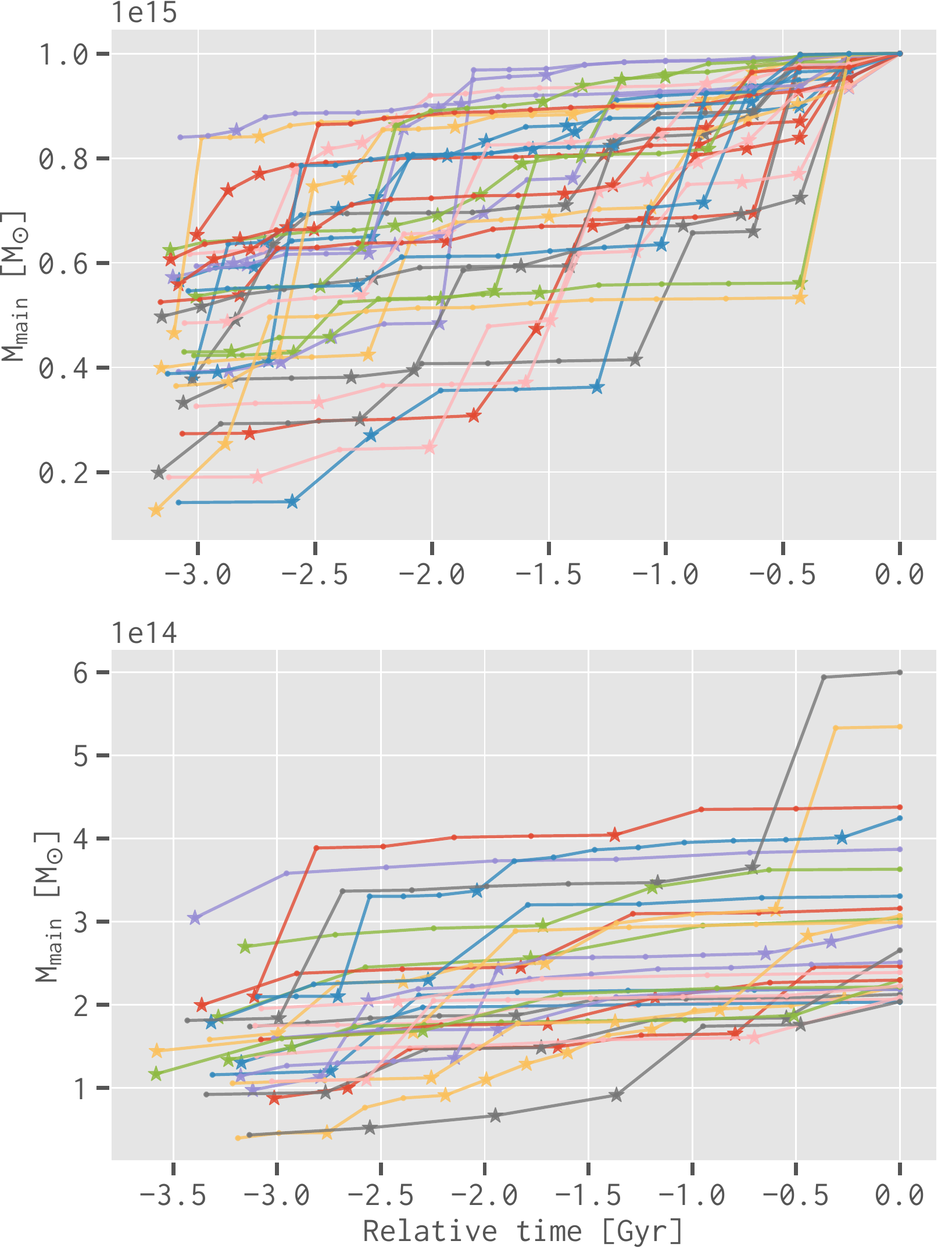}
  \caption{\label{fig:merging-history}%
    \textbf{(Upper)} Merger trees for one galaxy cluster of mass
    \SI{e15}{\solarmass} obtained by repeating the random build process
    for 30 times.
    \textbf{(Lower)} Example merger trees for 30 galaxy clusters randomly
    drawn from the sample constructed in \autoref{sec:mass-function}.
    Asterisks mark merger events and dots represent accretion events.
  }
\end{figure}

Given that observable radio halos are regarded to be associated
with recent (in the observer's frame) major mergers
and have typical lifetimes of $\tau_{\R{halo}} \lesssim \SI{1}{\Gyr}$
at \SI{1.4}{\GHz} \citep[e.g.,][]{brunetti2009,cassano2016},
we trace the merging history of each galaxy cluster for
$t_{\R{back}} = \SI{3}{\Gyr}$ from its \enquote{current}
age $t_{\R{sim}}$ (corresponding to $z_{\R{sim}}$).
For each built merger tree, we extract the information of all
the mergers associated with the main cluster to carry out the
subsequent simulation of radio halos.
As shown in the upper panel of \autoref{fig:merging-history},
we took one galaxy cluster of mass \SI{e15}{\solarmass} as an example and
repeated the random merger tree build process for 30 times.
We also randomly drew 30 galaxy clusters from the sample constructed
in \autoref{sec:mass-function} and built one merger tree instance for each
galaxy cluster, as shown in the lower panel.

%------------------------------------------------------------------------
\subsubsection{Evolution Modeling}
\label{sec:halo-evo}
%------------------------------------------------------------------------

According to the reacceleration model, there exists a population of
primary (or fossil) high-energy electrons, which permeate the ICM and
are thought to be injected by multiple processes, such as active
galactic nucleus (AGN) activities and star formations
(see \citealt{blasi2007rev} for a review).
When a cluster experiences a major merger, the turbulence is generated
throughout the ICM and can accelerate the primary electrons to be highly
relativistic, resulting in the observed radio halo.
On the other hand, relativistic electrons in the ICM lose energy via
mechanisms that include synchrotron radiation, inverse Compton scattering
off the CMB photons, and Coulomb collisions \citep{sarazin1999}.
For a population of electrons with isotropic energy distribution, the
temporal evolution of the number density distribution $n(\gamma, t)$
is governed by the following Fokker--Planck diffusion--advection equation
\citep{eilek1991,schlickeiser2002}
\begin{multline}
  \label{eq:fokkerplanck}
  \pdiff{n(\gamma,t)}{t} = \pdiff{}{\gamma} \left[ n(\gamma,t) \left(
      \left| \diff{\gamma}{t} \right| -
      \frac{2}{\gamma} D_{\gamma\gamma}(\gamma, t) \right) \right] \\
    + \pdiff{}{\gamma} \left[ D_{\gamma\gamma} \pdiff{n(\gamma,t)}{\gamma}
    \right] + Q_e(\gamma,t),
\end{multline}
where $\gamma$ is the Lorentz factor of electrons,
$D_{\gamma\gamma}(\gamma, t)$ is the diffusion coefficient describing
the interactions between the turbulence and electrons,
$|\R{d}\gamma / \R{d}t|$ is the energy-loss rate,
and $Q_e(\gamma, t)$ describes the electron injection.

%........................................................................
\setcounter{sssseccount}{0}
\ssssec{Thermal Properties of the ICM}

The number density of thermal electrons $n_{\R{th}}$ in the ICM can be
calculated as
\begin{equation}
  \label{eq:n-th}
  n_{\R{th}} \simeq
    \frac{3 f_{\R{gas}} M_{\R{vir}}}{4\pi \mu m_u \,r^3_{\R{vir}}},
\end{equation}
where
$\mu \simeq 0.6$ is the mean molecular weight \citep[e.g.,][]{ettori2013},
$m_u$ is the atomic mass unit,
$M_{\R{vir}}$ is the cluster's virial mass,
$r_{\R{vir}}$ is the virial radius [see \autoref{eq:radius-virial}],
and
$f_{\R{gas}} \simeq \Omega_b/\Omega_m$ is the assumed gas mass fraction.
Then, the corresponding ICM thermal energy density $\epsilon_{\R{th}}$
is given by
\begin{equation}
  \label{eq:e-th}
  \epsilon_{\R{th}} = \frac{3}{2} \,n_{\R{th}} k_B T_{\R{cl}}.
\end{equation}
The ICM mean temperature $T_{\R{cl}}$ is approximately given by
\begin{equation}
  \label{eq:t-icm}
  T_{\R{cl}} \simeq T_{\R{vir}} + \frac{3}{2} \,T_{\R{out}}
\end{equation}
\citep{cavaliere1998},
where
$T_{\R{vir}} = \mu m_u G M_{\R{vir}} / (2 \,r_{\R{vir}})$ is the virial
temperature and $T_{\R{out}} \simeq \SI{0.5}{\keV}$ is the temperature
of the gas flowing into the cluster from its outskirts \citep{fujita2003}.

%........................................................................
\ssssec{Electron Injection}

As primary electrons are continuously injected into the ICM via multiple
processes, it is reasonable to assume an average injection rate and a
power-law spectrum for the electron injection process
\citep[e.g.,][]{cassano2005,donnert2014}, i.e.,
\begin{equation}
  \label{eq:electron-inj}
  Q_e(\gamma, t) \simeq Q_e(\gamma) = K_e \,\gamma^{-s},
\end{equation}
where the spectral index $s$ is adopted to be 2.5 \citep{cassano2005}.
Moreover, the energy density of the injected electrons can be assumed to
account for a fraction ($\eta_e$) of the ICM thermal energy density
\citep{cassano2005}, i.e.,
\begin{equation}
  \tau_{\R{cl}} \int_{\gamma_{\R{min}}}^{\gamma_{\R{max}}}
  Q_e(\gamma') \gamma'\epsilon_e \,\D{\gamma'}
  = \eta_e \,\epsilon_{\R{th}},
\end{equation}
where $\tau_{\R{cl}} \simeq t_{\R{sim}}$ is the cluster's age at its
\enquote{current} redshift $z_{\R{sim}}$,
and $\epsilon_e = m_e c^2$ is the electron's rest energy.
Given $\gamma_{\R{min}} \ll \gamma_{\R{\max}}$, the injection rate $K_e$
is derived to be
\begin{equation}
  \label{eq:injrate}
  K_e \simeq \frac{(s-2)\,\eta_e\,\epsilon_{\R{th}}}{\epsilon_e\,\tau_{\R{cl}}}
    \gamma_{\R{min}}^{s-2}.
\end{equation}

%........................................................................
\ssssec{Stripping Radius}

When a subcluster merges into the main cluster, the gas at its outer
regions is stripped due to the ram pressure \citep{gunn1972}.
The stripping radius $r_s$ of the subcluster, outside which the stripping
is efficient, can be obtained from the equipartition between the ram
pressure and the hydrostatic pressure \citep{cassano2005}, i.e.,
\begin{equation}
  \label{eq:rs-eqp}
  \bar{\rho}_m v_{\R{imp}}^2 = \frac{\rho_s(r_s)}{\mu m_u} k_B T_{\R{cl,s}},
\end{equation}
where
$\bar{\rho}_m = \mu m_u n_{\R{th,m}}$ is the mean gas density of the main
cluster,
$v_{\R{imp}}$ is the impact velocity of the two merging clusters,
and $\rho_s(r)$ and $T_{\R{cl,s}}$ are the gas density profile and
temperature of the subcluster, respectively.

Starting from a sufficiently large distance with zero velocity,
the impact velocity $v_{\R{imp}}$ of two merging clusters with
masses $M_{\R{vir,m}}$ and $M_{\R{vir,s}}$ is given by
\begin{equation}
  \label{eq:v-imp}
  v_{\R{imp}} \simeq \left[
    \frac{2G (M_{\R{vir,m}} + M_{\R{vir,s}})}{r_{\R{vir,m}}}
    \left( 1 - \frac{1}{\eta_v} \right)\right]^{1/2},
\end{equation}
where $\eta_v \simeq 4 \,(1 + M_{\R{vir,s}}/M_{\R{vir,m}})^{1/3}$
\citep{sarazin2002,cassano2005}.

The gas density profile $\rho_s(r)$ can be well approximated with
a standard $\beta$-model \citep{cavaliere1976}:
\begin{equation}
  \label{eq:beta-model}
  \rho_s(r) = \rho_s(0) \left[1 + (r / r_{\R{c,s}})^2 \right]^{-3\beta/2},
\end{equation}
where
$r_{\R{c,s}}$ and $\beta$ are the core radius and slope parameter,
respectively, and we adopt $r_{\R{c,s}} = 0.1 \,r_{\R{vir,s}}$
\citep[e.g.,][]{sanderson2003} and
$\beta = 2/3$ \citep[e.g.,][]{jones1984}.
The central gas density $\rho_s(0)$ can then be determined by the total gas
mass ($M_{\R{gas,s}} = f_{\R{gas}} M_{\R{vir,s}}$).

%........................................................................
\ssssec{Turbulent Acceleration}

The details of interactions between the turbulence and both thermal and
relativistic particles are complicated and still poorly understood.
Among several particle acceleration mechanisms that can be potentially
triggered by the turbulence, the most important one is the transit time
damping process, i.e., the turbulence dissipates its energy and
accelerates particles by interacting with the relativistic particles
(e.g., cosmic rays) in the ICM
\citep[and references therein]{brunetti2007,brunetti2011}.
The associated diffusion coefficient is derived to be
\citep{miniati2015,pinzke2017}
\begin{equation}
  \label{eq:dpp}
  D_{\gamma\gamma} = 2 \gamma^2 \zeta \,k_L
    \frac{\langle (\delta v_t)^2 \rangle^2}{\chi_{\R{cr}} \, c_s^3},
\end{equation}
where
$\zeta$ is an efficiency factor characterizing the ICM plasma instabilities
(e.g., due to spatial or temporal intermittency),
$\chi_{\R{cr}} = \epsilon_{\R{cr}} / \epsilon_{\R{th}}$ is the relative
energy density of cosmic rays with respect to the thermal ICM,
$k_L \simeq 2\pi / r_{\R{turb}}$ is the turbulence injection scale
with $r_{\R{turb}}$ being the radius of the turbulence region,
$\langle (\delta v_t)^2 \rangle$ is the turbulence velocity dispersion,
and $c_s$ is the sound speed in the ICM
\begin{equation}
  \label{eq:sound-speed}
  c_s = \sqrt{\gamma_{\R{gas}} k_B T_{\R{cl}} / (\mu m_u)}
\end{equation}
with $\gamma_{\R{gas}} = 5/3$ being the adiabatic index
of ideal monatomic gas.

In addition to mergers, mechanisms such as outflows from AGNs and galactic
winds can introduce turbulence in the ICM, which is found to account for
$\lesssim 5\%$ of the thermal energy in the central regions
of relaxed clusters \citep[e.g.,][]{vazza2011}.
Therefore, the base velocity dispersion $\langle (\delta v_0)^2 \rangle$
of the turbulence in the absence of mergers is
\begin{equation}
  \label{eq:v-turb-base}
  \langle (\delta v_0)^2 \rangle
    = 3 \chi_{\R{turb}} \frac{k_B T_{\R{cl,m}}}{\mu m_u} ,
\end{equation}
where
$\chi_{\R{turb}}$ is the ratio of energy density between the base
turbulence and the thermal ICM.
A merger will contribute a significant part of its energy to the turbulence
and greatly increase the turbulence velocity dispersion
$\langle (\delta v_t)^2 \rangle$, which leads to
\begin{equation}
  \label{eq:energy-turb}
  E_{\R{turb}} =
    \frac{1}{2} M_{\R{turb}} \langle (\delta v_t)^2 \rangle =
    \frac{1}{2} M_{\R{turb}} \langle (\delta v_0)^2 \rangle + \eta_t E_m ,
\end{equation}
where
$E_m$ is the energy injected by the subcluster during the merger,
$\eta_t$ is the fraction of injected energy ($E_m$) transferred into
turbulent waves,
and $M_{\R{turb}}$ is the gas mass enclosed in the turbulence region
of radius $r_{\R{turb}}$, i.e.,
\begin{equation}
  \label{eq:mass-turb}
  M_{\R{turb}} = \int_0^{r_{\R{turb}}} \! \rho(r) 4\pi r^2 \,\D{r},
\end{equation}
where $\rho(r)$ is the gas density profile of the merged cluster
characterized by the $\beta$-model [see \autoref{eq:beta-model}].
The injected energy $E_m$ is approximated as the work done by the infalling
subcluster, i.e.,
\begin{equation}
  \label{eq:energy-inj}
  E_m \simeq \bar{\rho}_m v_{\R{imp}}^2 V_{\R{turb}},
\end{equation}
with $V_{\R{turb}} \simeq \pi r_s^2 \,r_{\R{vir,m}}$ being the swept volume
\citep{fujita2003,cassano2005}.
Therefore, the turbulence velocity dispersion during a merger is obtained
as
\begin{equation}
  \label{eq:v-turb}
  \langle (\delta v_t)^2 \rangle
    = \langle (\delta v_0)^2 \rangle
    + 2 \pi\,\eta_t\, \bar{\rho}_m r_{\R{vir,m}}
      \,\frac{r_s^2 v_{\R{imp}}^2}{M_{\R{turb}}} .
\end{equation}

One remaining parameter is the turbulence region radius $r_{\R{turb}}$,
which is estimated to be
\begin{equation}
  \label{eq:radius-turb}
  r_{\R{turb}} = r_s + r_{\R{c,m}} ,
\end{equation}
where
$r_{\R{c,m}} = 0.1 \,r_{\R{vir,m}}$ is the core radius of the main cluster,
and $r_s$ is the stripping radius of the subcluster
[see \autoref{eq:rs-eqp}]
with a value of $\sim \numrange{1}{2} \,r_{\R{c,m}}$
for major mergers ($M_{\R{vir,m}} / M_{\R{vir,s}} \lesssim 3$)
and $< r_{\R{c,m}}$ for minor mergers
($M_{\R{vir,m}} / M_{\R{vir,s}} \sim \numrange{3}{10}$).
This assumption is well consistent with previous simulation studies, which
show that mergers introduce turbulence in regions of radius about
$\numrange{0.1}{0.3} \,r_{\R{vir,m}}$
\citep[e.g.,][]{vazza2011,vazza2012,miniati2015ss}.
We note that minor mergers can also generate
a relatively large turbulence region of radius about $r_{\R{c,m}}$
due to the core gas sloshing induced by the infalling subcluster
\citep{vazza2012}.
However, the generated turbulence by a minor merger is rather weak because
the injected energy $E_m$ is much less than a major one
[see \autoref{eq:energy-inj}].

%........................................................................
\ssssec{Energy Losses}

Among the mechanisms through which relativistic electrons
in the ICM can lose energy, we take into account the following three
major mechanisms in this work \citep{sarazin1999}.
The first one is the inverse Compton scattering off the CMB photons,
the energy-loss rate of which is
\begin{equation}
  \label{eq:eloss-ic}
  \left( \diff{\gamma}{t} \right)_{\R{IC}} =
    \num{-4.32e-4} \,\gamma^2 (1+z)^4
    \quad [\si{\per\Gyr}].
\end{equation}

Secondly, with the \si{\uG}-level magnetic field permeating the ICM
\citep[e.g.,][]{govoni2004,ryu2008}, relativistic electrons will
produce synchrotron radiation and lose energy at a rate of
\begin{equation}
  \label{eq:eloss-syn}
  \left( \diff{\gamma}{t} \right)_{\R{syn}} =
    \num{-4.10e-5} \,\gamma^2 \left( \frac{B}{\SI{1}{\uG}} \right)^2
    \quad [\si{\per\Gyr}],
\end{equation}
where $B$ is the magnetic field strength.
We assume that the magnetic field is uniform and its energy density reaches
equipartition with that of cosmic rays, i.e.,
$\epsilon_B = B^2/(8\pi) \simeq \epsilon_{\R{cr}} = \chi_{\R{cr}}\,\epsilon_{\R{th}}$
\citep{beck2005}.

The last mechanism considered is that relativistic electrons interact
with the thermal electrons via Coulomb collisions, the energy-loss rate
of which is
\begin{multline}
  \label{eq:eloss-coul}
  \left( \diff{\gamma}{t} \right)_{\R{Coul}} =
    \num{-3.79e4} \left( \frac{n_{\R{th}}}{\SI{1}{\per\cm\cubed}} \right)
    \\ \times
    \left[ 1 + \frac{1}{75} \ln \left(
        \gamma \,\frac{\SI{1}{\per\cm\cubed}}{n_{\R{th}}} \right) \right]
    \quad [\si{\per\Gyr}].
\end{multline}

The inverse Compton scattering and synchrotron radiation dominate
the energy losses at the high-energy regime ($\gamma \gtrsim 1000$),
while Coulomb collisions are the main energy-loss mechanism for electrons
with lower energies ($\gamma \lesssim 100$).
Therefore, electrons with intermediate energies (e.g., $\gamma \sim 300$)
have a long lifetime (\SI{\sim 3}{\Gyr}) and can accumulate in the ICM
as the cluster grows \citep{sarazin1999}.

%------------------------------------------------------------------------
\subsubsection{Numerical Implementation}
\label{sec:numerical}
%------------------------------------------------------------------------

\begin{figure*}
  \centering
  \includegraphics[width=0.8\textwidth]{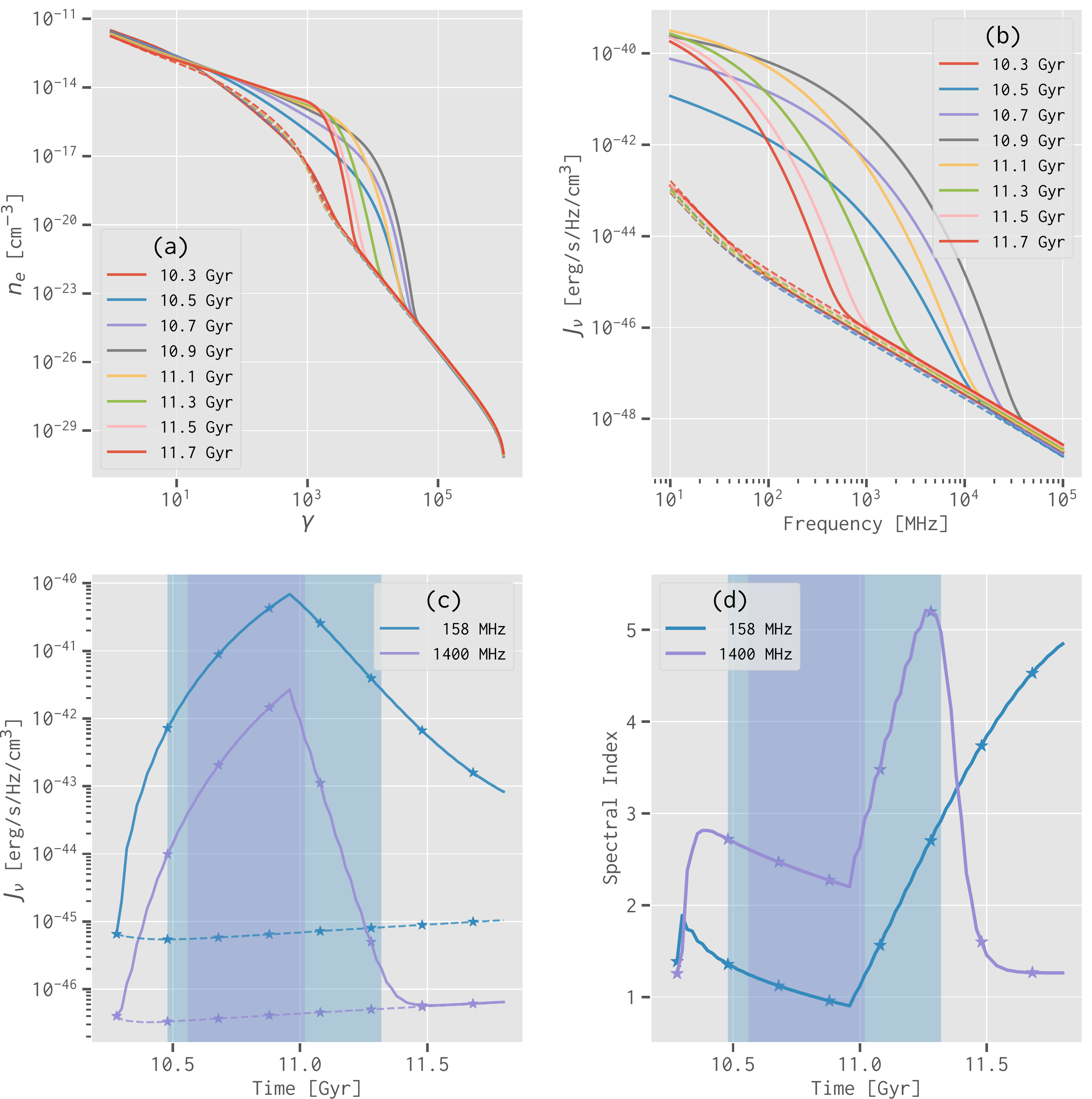}
  \caption{\label{fig:spec-evo}%
    The temporal evolution of the electron and synchrotron
    emission spectra for an example cluster with one major merger,
    which begins at redshift $z = 0.3$ (i.e., $t \simeq \SI{10.3}{\Gyr}$)
    and is tracked until $z = 0.15$ (i.e., $t \simeq \SI{11.8}{\Gyr}$).
    \textbf{(a)} The relativistic electron spectra (solid lines) and the
    corresponding reference electron spectra (dashed lines; see
    \autoref{sec:halo-size}).
    \textbf{(b)} The synchrotron emission spectra (solid lines) and the
    corresponding reference synchrotron spectra (dashed lines).
    \textbf{(c)} The variation of \SI{158}{\MHz} (solid blue line) and
    \SI{1400}{\MHz} (solid purple line) synchrotron emissivity as well as
    the corresponding reference emissivity (dashed lines) with time.
    \textbf{(d)} The temporal variation of spectral indices at
    \SI{158}{\MHz} (blue line) and \SI{1400}{\MHz} (purple line).
    Shaded regions show the periods during which the radio halo exists
    (see \autoref{sec:halo-size}).
    Asterisks mark the time points corresponding to the spectra presented
    in panels (a) and (b).
  }
\end{figure*}

In order to solve the Fokker--Planck equation [\autoref{eq:fokkerplanck}],
we apply an efficient numerical method proposed by \citet{chang1970}
and adopt the no-flux boundary condition \citep{park1996}.
To avoid the unphysical pile up of electrons around the lower boundary
caused by the boundary condition,
we define a buffer region below $\gamma_{\R{buf}}$, within which
the spectral data are replaced by extrapolating the data above
$\gamma_{\R{buf}}$ as a power-law spectrum \citep{donnert2014}.
We adopt a logarithmic grid with 256 cells for $\gamma \in [1, 10^6]$,
and let the buffer region span 10 cells.

By making use of the same Fokker--Planck equation but with the
merger-induced turbulent acceleration turned off [i.e., $E_m \equiv 0$ and
$\langle (\delta v_t)^2 \rangle \equiv \langle (\delta v_0)^2 \rangle$
in \autoref{eq:energy-turb}],
the initial electron spectrum $n_e(\gamma, t_0)$ is derived by evolving the
accumulated electron spectrum $\tilde{n}_e(\gamma) = Q_e(\gamma) \,\tau_0$
for \SI{1}{\Gyr} \citep[e.g.,][]{brunetti2007}, where $\tau_0$ is the
cluster's age at the beginning of the earliest merger.

Although the whole process of a single merger can last for about
\SIrange{2}{3}{\Gyr} \citep[e.g.,][]{tormen2004,cassano2016}, the period
during which the turbulence is intense enough to effectively accelerate
electrons is relatively short.
An appropriate estimation of the turbulent acceleration period is
$\tau_{\R{turb}} \simeq 2 \,r_{\R{turb}} / v_{\R{imp}}$ \citep{miniati2015}.

A galaxy cluster may experience multiple mergers in the past
$t_{\R{back}} = \SI{3}{\Gyr}$.
For each merger event
$\left( M^{(i)}_{\R{vir,m}}, M^{(i)}_{\R{vir,s}}, t^{(i)}_{\R{begin}} \right)$,
where $t^{(i)}_{\R{begin}}$ denotes the beginning time of this merger,
it can induce effective turbulent acceleration (i.e., being active) during
the period $\left[ t^{(i)}_{\R{begin}}, t^{(i)}_{\R{end}} \right]$ with
$t^{(i)}_{\R{end}} = t^{(i)}_{\R{begin}}+\tau^{(i)}_{\R{turb}}$ being the
time when this merger becomes inactive.
At other times (i.e., no active merger), only the base turbulence
contributes to the acceleration of electrons, which, however,
is insufficient to balance the energy loss due to synchrotron radiation
and inverse Compton scattering.

Over the history of a galaxy cluster that has multiple mergers, the
turbulence region has a different radius during different mergers.
To take this variation into account, we identify the radius of the largest
turbulence region ($R_{\R{turb}}$) and properly diffuse the electron
spectrum to the sphere of radius $R_{\R{turb}}$ for mergers with a smaller
turbulence region.
Specifically, for a merger that is active during 
$\left[ t^{(i)}_{\R{begin}}, t^{(i)}_{\R{end}} \right]$ and has a
turbulence region of radius $r^{(i)}_{\R{turb}}$, the accelerated part of
the electron spectrum during this merger [i.e., the difference between
spectra at $t^{(i)}_{\R{end}}$ and at $t^{(i)}_{\R{begin}}$] is rescaled
by a volume ratio given by
\begin{equation}
  \label{eq:ratio-v}
  R_{\R{vol}} = \left[ r^{(i)}_{\R{turb}} \Big/ R_{\R{turb}} \right]^3 .
\end{equation}

Once the desired electron spectrum $n_e(\gamma, t)$ is obtained, the
synchrotron emissivity at a frequency $\nu$ is given by
\begin{equation}
  \label{sec:jnu-sync}
  % unit: [erg/s/Hz/cm^3]
  J(\nu) = \frac{\sqrt{3} \, e^3 B}{m_e c^2}
    \!\int_{\gamma_{\R{min}}}^{\gamma_{\R{max}}} \!\!\!\int_0^{\pi/2}\!
    n_e(\gamma, t) F(\nu/\nu_c) \sin^2 \!\theta \,\D{\theta} \,\D{\gamma}
\end{equation}
\citep{rybicki1979},
where
$c$ is the speed of light,
$e$ is the elementary charge,
$\theta$ is the pitch angle of electrons with respect to the magnetic
field, $\nu_c = (3/2) \,\gamma^2 \nu_L \sin\theta$ is the electron's
critical frequency with $\nu_L = e B / (2\pi m_e c)$ being the Larmor
frequency, and $F(\cdot)$ is the synchrotron kernel:
\begin{equation}
  \label{eq:sync-kernel}
  F(x) = x \int_x^{\infty} K_{5/3}(y) \,\D{y} ,
\end{equation}
where $K_{5/3}(\cdot)$ is the modified Bessel function of order 5/3.

In \autoref{fig:spec-evo}, we present the temporal evolution of the
relativistic electron and synchrotron emission spectra for an example
galaxy cluster with one major merger.
The cluster has a mass of \SI{e15}{\solarmass} and merges with a
subcluster of mass \SI{6e14}{\solarmass} at redshift $z = 0.3$
(i.e., $t \simeq \SI{10.3}{\Gyr}$), from which we solve the Fokker--Planck
equation to track the electron and synchrotron emission spectra until
redshift $z = 0.15$ (i.e., $t \simeq \SI{11.8}{\Gyr}$).
As demonstrated in this figure, the merger-induced turbulence is active
for a period of $\tau_{\R{turb}} \simeq \SI{0.67}{\Gyr}$ and efficiently
accelerates electrons to extremely high energies
($\gamma \gtrsim \num{e4}$), which gives rise to the radio halo even at
high frequencies ($> \si{\GHz}$).
However, once the turbulence becomes inactive ($t > \SI{10.9}{\Gyr}$), the
high-energy electrons quickly lose energy and the radio halo fades out
shortly, especially at high frequencies.

%------------------------------------------------------------------------
\subsubsection{Halo Identification and Size}
\label{sec:halo-size}
%------------------------------------------------------------------------

Radio halos cannot form or will rapidly disappear if there is no active
turbulent acceleration.
In order to determine whether or not there exists a radio halo at frequency
$\nu$, we employ the following two criteria:
(1) the synchrotron emissivity $J(\nu)$ of the final electron spectrum
$n_e(\gamma, t_{\R{sim}})$ is at least \num{1000} times larger than the
emissivity $J'(\nu)$ of the reference electron spectrum
$n'_e(\gamma, t_{\R{sim}})$, which is obtained by solving the identical
Fokker--Planck equation but without merger-induced turbulent acceleration,
similar to the way of deriving the initial electron spectrum
(\autoref{sec:numerical});
(2) the spectral index\footnote{%
  We adopt a power-law spectrum of form $J(\nu) \propto \nu^{-\alpha}$.}
at frequency $\nu$ satisfies $\alpha_{\nu} \le 3$.
For the example as shown in \autoref{fig:spec-evo}(c,d), a radio halo
is identified from about 10.6 to 11.0 \si{\Gyr} at \SI{1.4}{\GHz} and
from about 10.5 to 11.3 \si{\Gyr} at \SI{158}{\MHz}.
The spectral indices at \SI{1.4}{\GHz} and \SI{158}{\MHz} reach about 2.1
and 1.0, respectively.
These results demonstrate that radio halos have longer lifetimes at low
frequencies and thus we expect to observe more radio halos in
low-frequency radio bands.
We note that the \SI{1.4}{\GHz} spectral index ($\alpha_{1400} \sim 2.1$)
is slightly larger than the general result from observations,
because we calculate the spectral index around a specific frequency while
observed spectral indices are generally obtained from two separated
frequencies (e.g., 0.3 and 1.4 GHz; \citealt{feretti2012rev}).

Previous studies \citep[e.g.,][]{cassano2007,basu2012}
have shown that the radius of radio halos ($r_{\R{halo}}$)
increases nonlinearly with the cluster's virial radius ($r_{\R{vir}}$),
which may be caused by the distributions of relativistic electrons and
magnetic fields \citep[e.g.,][]{dolag2002}.
Therefore, we assume the following scaling relation for $r_{\R{halo}}$:
\begin{equation}
  \label{eq:r-halo}
  r_{\R{halo}} = f_r R_{\R{turb}}
    \left( \frac{r_{\R{vir}}}{r_{\R{vir,*}}} \right)^b ,
\end{equation}
where
$R_{\R{turb}}$ is the radius of the largest turbulence region as also used
in \autoref{eq:ratio-v},
$r_{\R{vir,*}}$ is the virial radius of a reference cluster of mass
\SI{e15}{\solarmass},
and $f_r$ and $b$ are the scaling normalization and slope, respectively.
After comparing with the observed scaling relation of
$r_{\R{halo}} \propto r_{\R{vir}}^{2.63 \pm 0.50}$ \citep{cassano2007},
we obtain $f_r = 0.7$ and $b = 1.8$.

Then, the power of a radio halo at frequency $\nu$ is
\begin{equation}
  \label{eq:halo-power}
  P(\nu) = \frac{4\pi}{3} r_{\R{halo}}^3 J(\nu),
\end{equation}
and the flux density at the same frequency is
\begin{equation}
  \label{eq:halo-flux}
  S(\nu) = \frac{(1+z_{\R{sim}}) P(\nu(1+z_{\R{sim}}))}
    {4\pi D_{\!L}^2(z_{\R{sim}})} ,
\end{equation}
where $D_{\!L}(z_{\R{sim}})$ is the luminosity distance to the halo,
and the factor $(1 + z_{\R{sim}})$ accounts for the $K$ correction
\citep[e.g.,][]{hogg1999}.

%------------------------------------------------------------------------
\subsubsection{Model Parameters and Results}
\label{sec:halo-results}
%------------------------------------------------------------------------

Our model has the following parameters:
(1) $\eta_e$, the ratio of the energy density of injected electrons to
the thermal energy density;
(2) $\eta_t$, the fraction of the merger energy transferred into the
turbulence;
(3) $\chi_{\R{cr}}$, the relative energy density of cosmic rays to
the thermal component;
(4) $\chi_{\R{turb}}$, the relative energy density of the base turbulence;
(5) $\zeta$, the efficiency of the ICM plasma instabilities.
Since currently no reasonable constraints on these parameters can be
obtained from either observational or theoretical studies,
it is necessary to tune them to make the model predictions (e.g., the halo
flux function, the scaling relation between the halo power and the hosting
cluster mass) consistent with observations.

\begin{figure}
  \centering
  \includegraphics[width=\columnwidth]{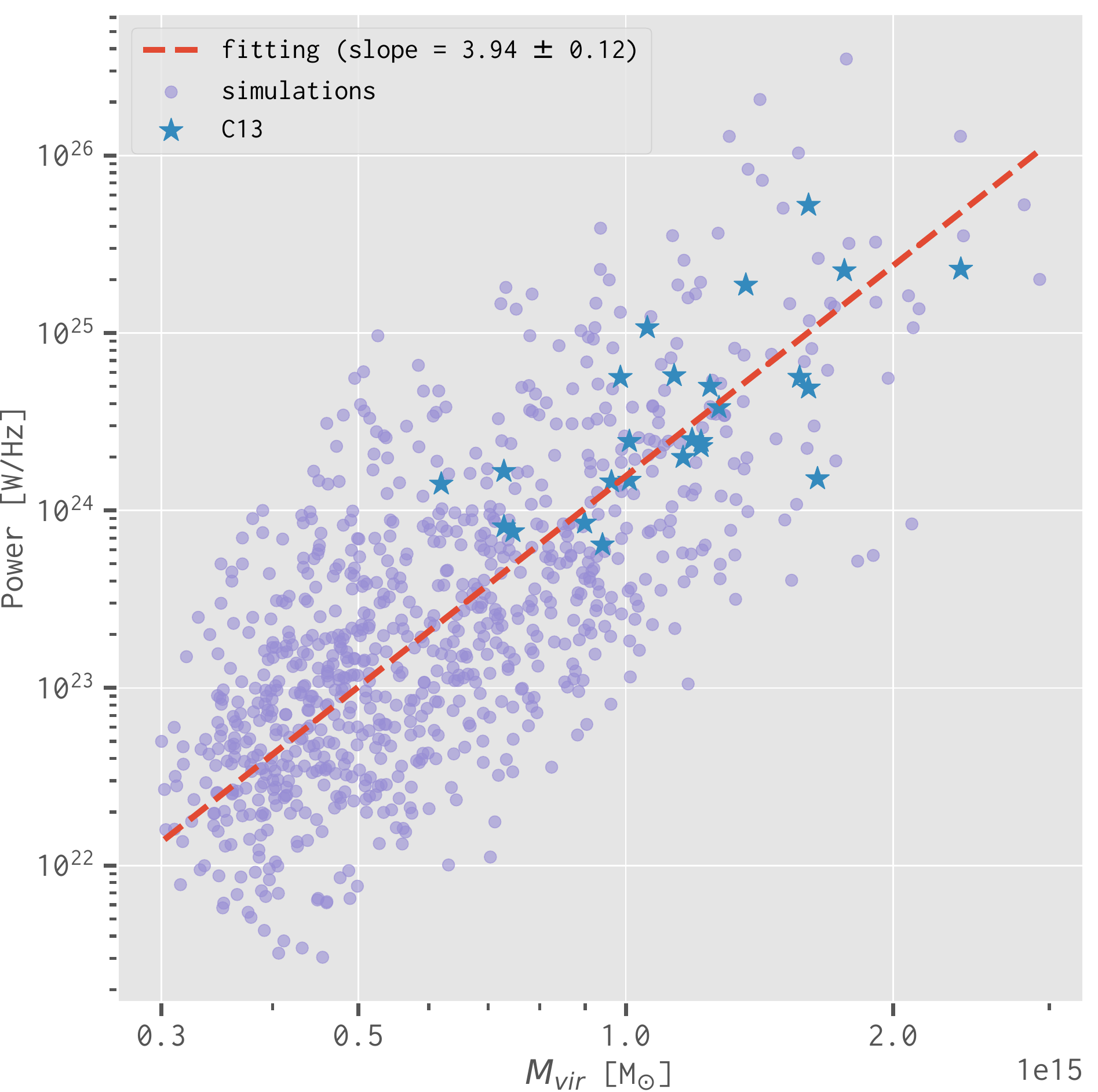}
  \caption{\label{fig:halo-power}%
    Simulated scaling relation between the radio halo power at
    \SI{1.4}{\GHz} ($P_{1400}$) and the cluster mass ($M_{\R{vir}}$).
    Blue asterisks mark the observation data from \citet{cassano2013}.
    Purple dots represent the results of 500 simulation runs
    and the dashed red line shows the fitted relation of
    $P_{1400} \propto M_{\R{vir}}^{3.94 \pm 0.12}$.
  }
\end{figure}

\begin{figure}
  \centering
  \includegraphics[width=\columnwidth]{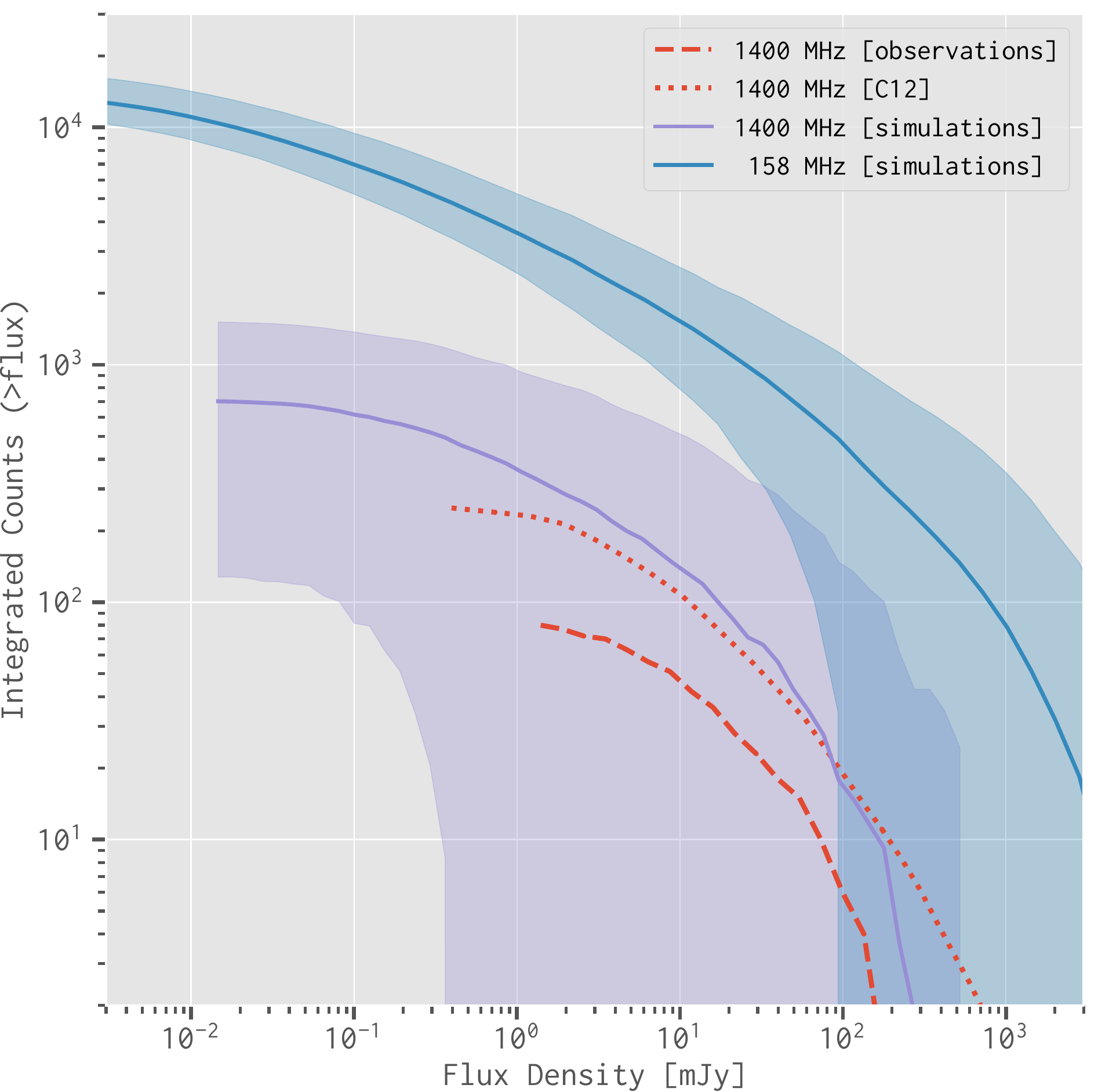}
  \caption{\label{fig:halos-simucomp}%
    The \SI{1.4}{\GHz} all-sky integrated flux function comparison
    between the observed (dashed red line) and simulated
    (solid purple line) radio halos.
    The dotted red line shows the \SI{1.4}{\GHz} flux function
    predicted by \citet{cassano2012}.
    The solid blue line represents the \SI{158}{\MHz} flux function for
    the simulated halos as a comparison.
    Shaded regions mark the 68\% uncertainties of the
    simulated radio halos estimated from the 500 simulation runs.
  }
\end{figure}

We perform two comparisons between our simulations and observations.
The first comparison involves the observed scaling relation between the
radio halo power at \SI{1.4}{\GHz} ($P_{1400}$) and the cluster mass
($M_{\R{vir}}$).
We make use of the observation data presented by \citet{cassano2013},
who reported a scaling relation of
$P_{1400} \propto M_{500}^{3.70 \pm 0.56}$.
We convert their mass $M_{500}$ to virial mass by assuming an NFW density
profile \citep{navarro1997} and employing the mass--concentration relation
derived by \citet{duffy2008}.
Second, we compare the \SI{1.4}{\GHz} all-sky integrated flux function
between the simulated radio halos and observations.
To this end, we have collected all currently observed radio halos
(\autoref{tab:halos-observed} in \autoref{sec:halos-collection};
71 identified halos and 9 candidates; as of 2018 January).
Considering that current observations are far from complete,
especially at the low-flux end, our strategy is to require that the
flux function of simulated radio halos agrees with the
observed one at the high-flux end.

We have explored various parameter configurations and,
for each configuration, we have repeated the simulation for 500 times
in order to take into account the distribution variations of bright
radio halos across the sky.
By comparing the simulation results with the observed
$P_{1400}$--$M_{\R{vir}}$ scaling relation and the \SI{1.4}{\GHz} flux
function, we finally choose a set of model parameters with
$\eta_e = 0.01\%$,
$\eta_t = 15\%$,
$\chi_{\R{cr}} = 1.5\%$,
$\chi_{\R{turb}} = 1.5\%$,
and $\zeta = 0.1$.
As shown in \autoref{fig:halo-power}, the radio halos simulated by our
model with the tuned parameters show a scaling relation of $P_{1400}
\propto M_{\R{vir}}^{3.94 \pm 0.12}$, which is consistent with the
findings of \citet{cassano2013} on both the slope and normalization.
In \autoref{fig:halos-simucomp}, we present the \SI{1.4}{\GHz} flux
functions for the simulated radio halos and the observed halos,
which agree with each other at the high-flux end.
The \SI{1.4}{\GHz} flux function given by our tuned model also matches
the prediction of \citet{cassano2012}.

\begin{figure}
  \centering
  \includegraphics[width=\columnwidth]{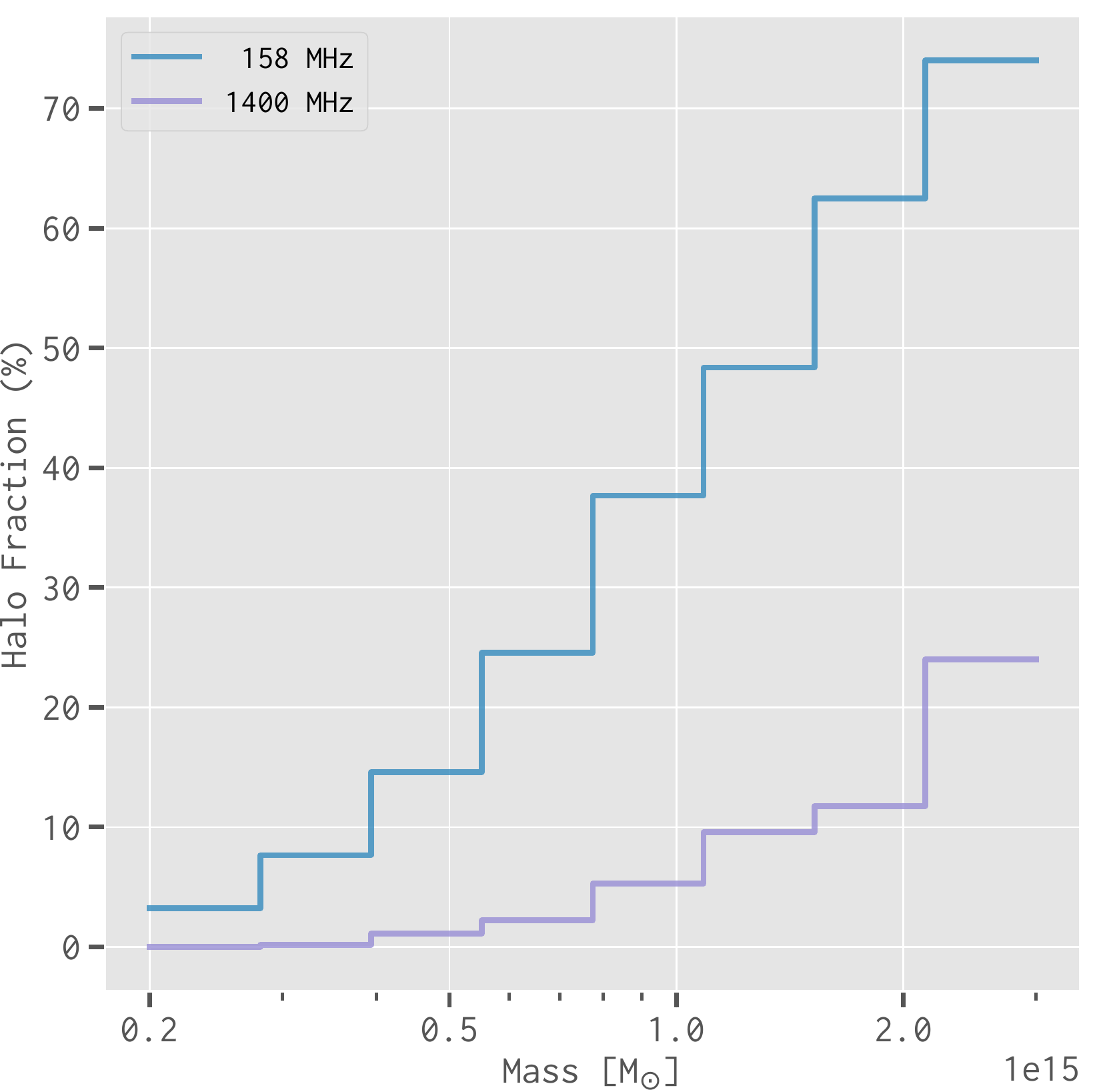}
  \caption{\label{fig:halo-fraction}%
    The fraction of clusters with radio halos as a function of the cluster
    mass.
    The blue and purple lines represent the fraction of halos identified at
    \SI{158}{\MHz} and \SI{1.4}{\GHz}, respectively.
  }
\end{figure}

Furthermore, we display the fraction of clusters with radio halos as a
function of the cluster mass in \autoref{fig:halo-fraction}.
It clearly shows that more massive clusters tend to have higher
probabilities of hosting radio halos.
Meanwhile, we expect to observe many more radio halos at low
frequencies (e.g., $\sim \SIrange{100}{200}{\MHz}$), which could be
generated by less intense mergers and have longer lifetimes (see also
\autoref{sec:numerical} and \autoref{fig:spec-evo}).

%------------------------------------------------------------------------
\subsubsection{Sky Map Generation}
\label{sec:skymaps}
%------------------------------------------------------------------------

To generate images for the simulated radio halos, we adopt an
exponential profile for the azimuthally averaged brightness distribution
\citep{murgia2009}:
\begin{equation}
  \label{eq:halo-profile}
  I_{\nu}(\theta) = I_{\nu,0} \exp(-3 \,\theta / \theta_{\R{halo}}),
\end{equation}
where $\theta = r / D_{\!A}(z_{\R{sim}})$ is the angular radius from the
halo center with $D_{\!A}(z_{\R{sim}})$ being the angular diameter
distance to the halo
and $I_{\nu,0} = 9 S(\nu) / (2\pi \theta^2_{\R{halo}})$ is the central
brightness.

\begin{figure}
  \centering
  \includegraphics[width=\columnwidth]{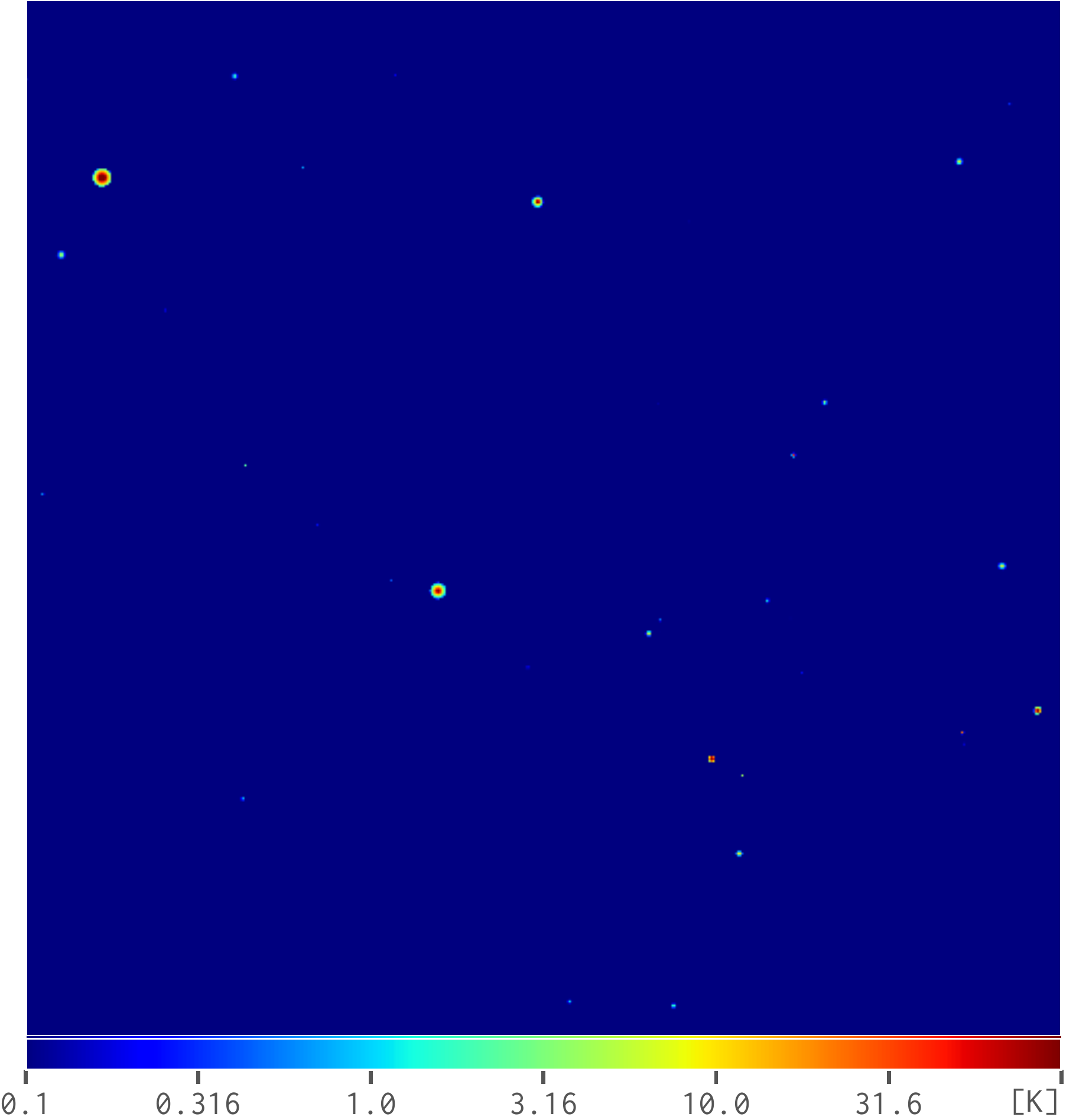}
  \caption{\label{fig:halos-example}%
    An example from the 100 simulation runs showing the simulated
    radio halos at \SI{158}{\MHz}.
    The sky region size is \SI[product-units=repeat]{10 x 10}{\degree},
    and the color bar is in units of \si{\kelvin}.
  }
\end{figure}

In order to characterize the uncertainty of the number density of bright
radio halos across the sky, we repeat the simulation of radio halos
for 100 times.
The medians and the corresponding 68\% uncertainties%
\footnote{%
  The 68\% uncertainty is derived from the 16$^{\text{th}}$
  and 84$^{\text{th}}$ percentiles because they are more robust than the
  mean and standard deviation for data with large dispersion.}
of the rms brightness temperature are
$\left(4.21_{-2.60}^{+11.2}\right) \times 10^3$ \si{\mK},
$\left(1.81_{-1.13}^{+5.28}\right) \times 10^3$ \si{\mK}, and
$\left(0.85_{-0.54}^{+2.74}\right) \times 10^3$ \si{\mK}
at \numlist{124; 158; 196} \si{\MHz}, respectively
(\autoref{tab:tb-rms}; see also \autoref{fig:halos-example} for an
example map of the simulated radio halos at \SI{158}{\MHz}).

\begin{deluxetable*}{cccc}
\tablecaption{\label{tab:tb-rms}%
  The rms Brightness Temperatures of All Components (unit: \si{\mK})
}
\tablehead{
  \colhead{Component} &
  \colhead{\SI{124}{\MHz}} &
  \colhead{\SI{158}{\MHz}} &
  \colhead{\SI{196}{\MHz}}
}
\startdata
Radio halos (100 simulations) &
  $\left(4.21_{-2.60}^{+11.2}\right) \times 10^3$ &
  $\left(1.81_{-1.13}^{+5.28}\right) \times 10^3$ &
  $\left(0.85_{-0.54}^{+2.74}\right) \times 10^3$ \\
Galactic synchrotron & \num{4.74e5} & \num{2.52e5} & \num{1.43e5} \\
Galactic free-free & \num{330} & \num{200} & \num{130} \\
Point sources & \num{29.7e7} & \num{5.90e7} & \num{1.39e7} \\
EoR signal & \num{15.1} & \num{11.3} & \num{3.77}
\enddata
\end{deluxetable*}

%========================================================================
\subsection{Other Foreground Components}
\label{sec:fg-other}
%========================================================================

Following our previous work \citep{wang2010}, we have also simulated
several other foreground components, including the Galactic synchrotron
and free-free emissions as well as the extragalactic point sources,
in order to carry out comparisons of power spectra between radio halos
and other foreground components as an effort to better characterize the
contribution of radio halos to the low-frequency radio sky.

The Galactic synchrotron map is simulated by extrapolating the
Haslam \SI{408}{\MHz} all-sky map as the template to lower frequencies
with a power-law spectrum.
We make use of the high-resolution version ($N_{\R{side}} = 2048$,
pixel size \SI{\sim 1.72}{\arcminute}) of the Haslam \SI{408}{\MHz}
map\footnote{The reprocessed Haslam \SI{408}{\MHz} map:
  \url{http://www.jb.man.ac.uk/research/cosmos/haslam_map/}},
which was reprocessed by \citet{remazeilles2015} using significantly
better instrument calibration and more accurate subtraction of
extragalactic sources.
We also use the all-sky synchrotron spectral index map made by
\citet{giardino2002} to account for the index variation with sky positions.
The Galactic free-free emission is deduced from the \Halpha{} survey
data \citep{finkbeiner2003}, which is corrected for dust absorption,
by employing the tight relation between the \Halpha{} and free-free
emissions due to their common origins
\citep[see][and references therein]{dickinson2003}.
Since the Galactic diffuse emissions vary remarkably across the sky,
we simulate them at position of
(R.A., decl.\@) = (\SI{0}{\degree}, \SI{-27}{\degree}), which locates at a
high galactic latitude ($b = \SI{-78.5}{\degree}$) and is expected to be
an appropriate choice for this study (see also \autoref{sec:obs-simu}).

The extragalactic point sources are simulated by taking into account the
following five types of sources: (1) star-forming and starburst galaxies,
(2) radio-quiet AGNs, (3) Fanaroff--Riley type I and type II AGNs,
(4) GHz-peaked spectrum AGNs, and (5) compact steep spectrum AGNs.
We simulate the former three types of sources by leveraging the simulation
results made by \citet{wilman2008}, and simulate the latter two types
by employing their corresponding luminosity functions and spectral models.
More details can be found in \citet{wang2010} and references therein.

\begin{figure*}
  \centering
  \includegraphics[width=\textwidth]{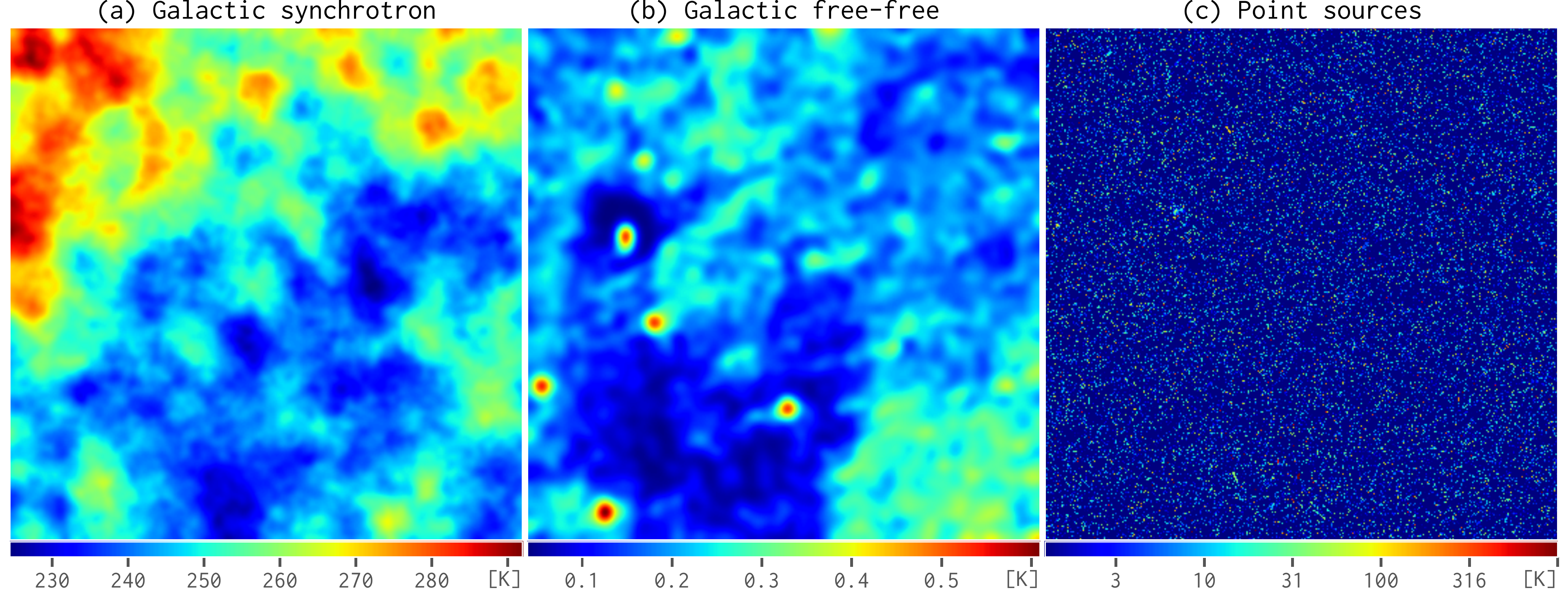}
  \caption{\label{fig:map-fg-other}%
    The sky maps of
    \textbf{(a)} the Galactic synchrotron emission,
    \textbf{(b)} the Galactic free-free emission, and
    \textbf{(c)} extragalactic point sources
    at \SI{158}{\MHz}.
    All the maps cover sky region of size
    \SI[product-units=repeat]{10 x 10}{\degree}
    and have color bars in units of \si{\kelvin}.
  }
\end{figure*}

The rms brightness temperatures of the Galactic synchrotron emission,
Galactic free-free emission, and extragalactic point sources are listed
in \autoref{tab:tb-rms}, and example maps simulated at \SI{158}{\MHz}
for these components are shown in \autoref{fig:map-fg-other}.

%========================================================================
\subsection{EoR Signal}
\label{sec:eor-signal}
%========================================================================

\begin{figure}
  \centering
  \includegraphics[width=\columnwidth]{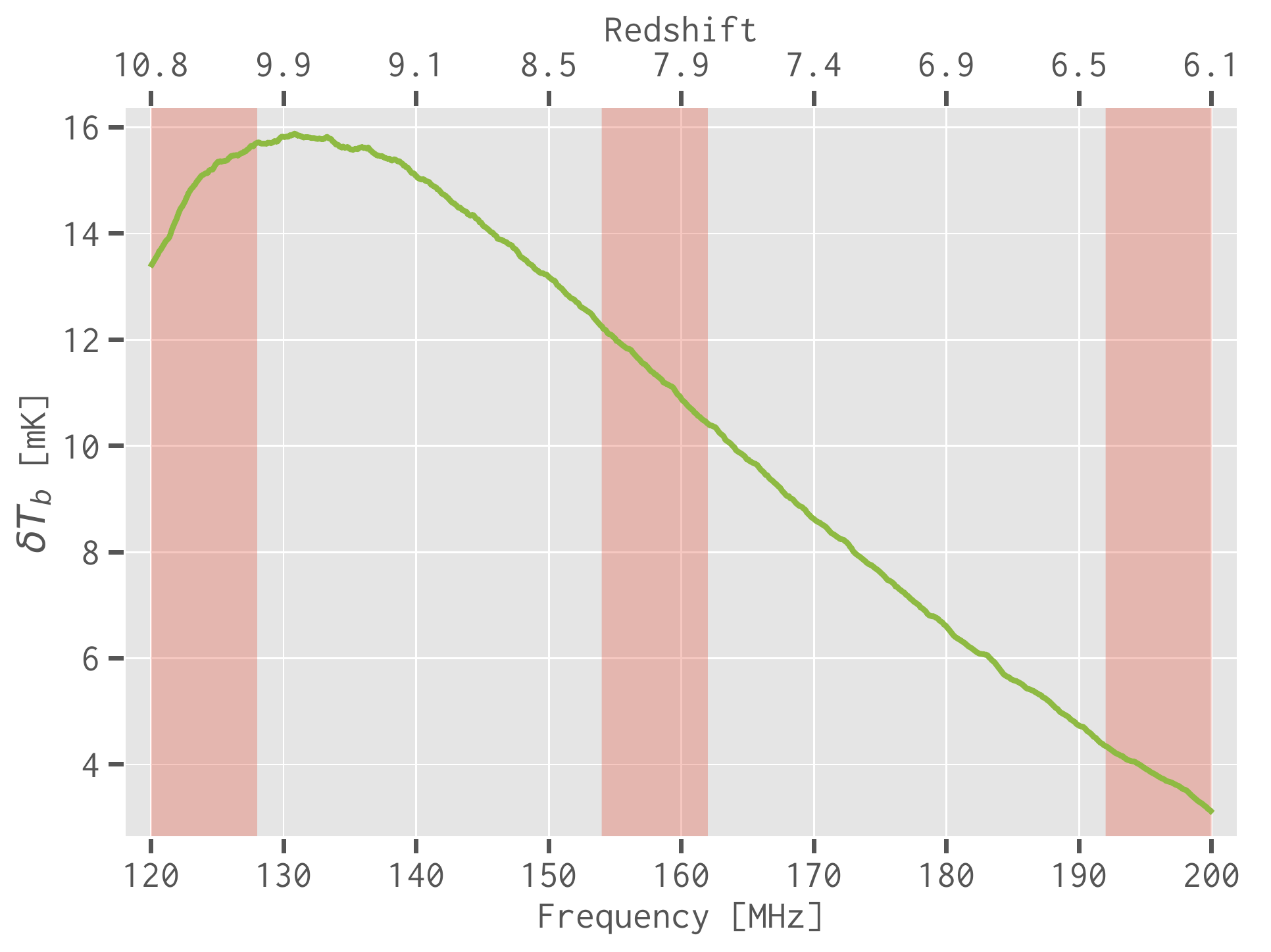}
  \caption{\label{fig:eor-tbrms}%
    The rms brightness temperatures of the EoR signal
    (solid green line) within \SIrange{120}{200}{\MHz}
    ($z = \numrange{6.1}{10.8}$).
    The red-shaded regions mark the three adopted frequency bands
    (\numrange{120}{128}, \numrange{154}{162}, and \numrange{192}{200}
    \si{\MHz}).
  }
\end{figure}

\begin{figure}
  \centering
  \includegraphics[width=\columnwidth]{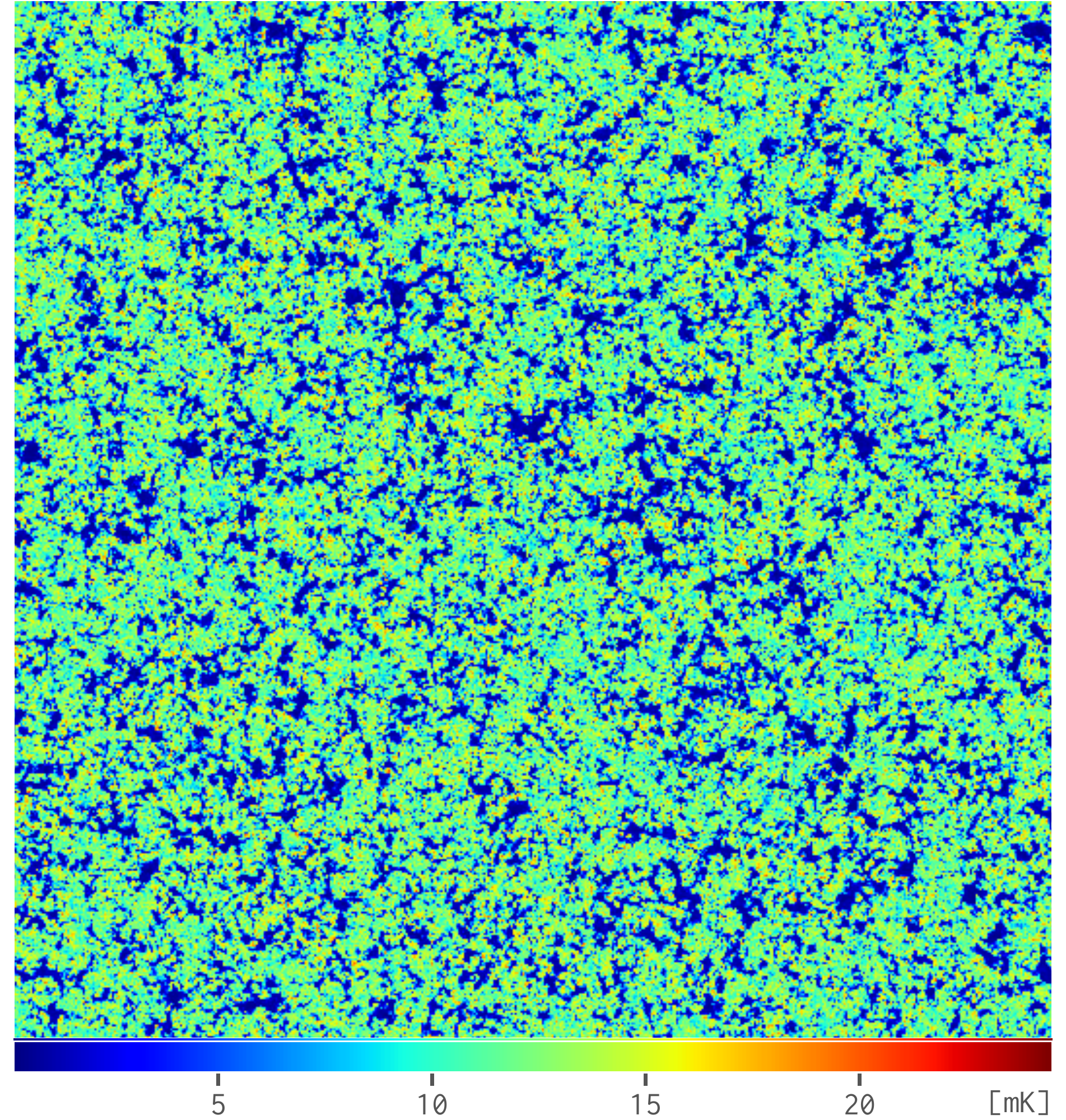}
  \caption{\label{fig:map-eor}%
    The sky map of the EoR signal at \SI{158}{\MHz}.
    The sky region size is \SI[product-units=repeat]{10 x 10}{\degree}
    and the color bar is in units of \si{\mK}.
  }
\end{figure}

The sky maps of the EoR signal are created using the 2016 data release
from the \emph{Evolution Of 21\,cm Structure} project\footnote{%
  Evolution Of 21\,cm Structure project:
  \url{http://homepage.sns.it/mesinger/EOS.html}},
which has made use of the \texttt{21cmFAST} to simulate the cosmic
reionization process from redshift 86.5 to 5.0 inside a large cube that is
1.6 comoving \si{\Gpc} (1024 cells) along each side \citep{mesinger2016}.
We extract the image slices at needed frequencies (i.e., redshifts) from
the light-cone cubes of the recommended \enquote{faint galaxies} case,
and then tile and rescale them to have the same sky coverage and
pixel size as our foreground maps.
\autoref{fig:eor-tbrms} shows the rms brightness temperatures of the
EoR signal among \SIrange{120}{200}{\MHz} ($z = \numrange{6.1}{10.8}$).
The corresponding rms brightness temperatures at the central
frequencies of the three adopted bands are given in \autoref{tab:tb-rms}
and the sky map of the EoR signal at \SI{158}{\MHz} is shown in
\autoref{fig:map-eor}.

%########################################################################
\section{Simulation of SKA Observations}
\label{sec:obs-simu}
%########################################################################

In order to properly evaluate the contamination of radio halos
on the EoR observations, it is essential to take account of the
practical instrumental effects of radio interferometers.
Therefore, we employ the latest SKA1-Low layout configuration%
\footnote{\raggedright%
  SKA1-Low Configuration Coordinates:
  \url{https://astronomers.skatelescope.org/wp-content/uploads/2016/09/SKA-TEL-SKO-0000422_02_SKA1_LowConfigurationCoordinates-1.pdf}
  (released on 2016 May 31)
}
to simulate the SKA observations of the above sky maps.
According to this layout configuration,
the SKA1-Low interferometer consists of 512 stations, with 224 of them
randomly distributed within the \enquote{core} of \SI{1000}{\meter} in
diameter, while the remaining stations are grouped into \enquote{clusters}
and placed on three spiral arms extending up to a radius of
\SI{\sim 35}{\kilo\meter}.
Each station has 256 antennas randomly distributed with a minimum separation
of $d_{\R{min}} = \SI{1.5}{\meter}$ inside a circular region of
\SI{35}{\meter} in diameter \citep[e.g.,][]{mort2017}.

The \SI{8}{\MHz} bandwidth of each frequency band is divided into 51
channels for a frequency resolution of \SI{160}{\kilo\hertz}.
For each component, we simulate the input sky maps at every frequency
channel, and then use the \texttt{OSKAR}\footnote{%
  OSKAR: \url{https://github.com/OxfordSKA/OSKAR} (v2.7.0)}
simulator \citep{mort2010} to perform observations for \SI{6}{\hour}.
The input sky maps are centered at sky position of
(R.A., decl.\@) = (\SI{0}{\degree}, \SI{-27}{\degree}),
which passes through the zenith of the SKA1-Low telescope and
is an ideal choice for the simulation of SKA observations.
The simulated visibility data are imaged through the
\texttt{WSClean}\footnote{%
  WSClean: \url{https://sourceforge.net/p/wsclean} (v2.6)}
imager \citep{offringa2014} using Briggs' weighting with a
robustness of zero \citep{briggs1995},
and the images created are cropped to keep only the central regions
because the marginal regions suffer from the problem of insufficient
CLEAN.
As the telescope's field of view (\fov) is inversely proportional to
the observing frequency, we choose to keep the central
\SI[product-units=repeat]{6 x 6}{\degree},
\SI[product-units=repeat]{5 x 5}{\degree}, and
\SI[product-units=repeat]{4 x 4}{\degree}
regions in the \numrange{120}{128}, \numrange{154}{162}, and
\numrange{192}{200} \si{\MHz} frequency bands, respectively.

The Galactic synchrotron and free-free emissions are combined for the
simulated observations because they have similar diffuse features.
Similar to the real-time peeling of the brightest point sources in
practical data analysis pipelines
\citep[e.g.,][]{mitchell2008,intema2009,mort2017},
we assume that extragalactic point sources with a \SI{158}{\MHz} flux
density $S_{158} > \SI{50}{\mJy}$ are removed
\citep[e.g.,][]{liu2009ps,pindor2011}.
Thus, the rms brightness temperatures of point sources are significantly
reduced to be about
\SIlist[%
  list-units=brackets,
  scientific-notation=fixed,
  fixed-exponent=4,
]{22.5e4; 9.81e4; 4.75e4}{\mK}
at \numlist{124; 158; 196} \si{\MHz}, respectively.
In addition, we create the foreground image cubes in each frequency band
using the CLEAN algorithm with joined-channel deconvolution in order
to ensure the spectral smoothness \citep{offringa2017}, which is crucial
to extract the faint EoR signal in the presence of overwhelming
foreground contamination.
For the EoR signal, we directly use the dirty images because the CLEAN
algorithm is not well applicable to such faint and diffuse emissions.
Hence we obtain the SKA \enquote{observed} image cubes of the EoR signal,
radio halos, the Galactic diffuse emission (with synchrotron and
free-free emissions combined), and the extragalactic point sources
(with the brightest ones removed) in the \numrange{120}{128},
\numrange{154}{162}, and \numrange{192}{200} \si{\MHz} frequency bands.

%########################################################################
\section{Power Spectra and EoR Window}
\label{sec:ps-eorw}
%########################################################################

In order to characterize the contamination of radio halos on the EoR
observations in terms of both foreground removal and avoidance methods,
we utilize the power spectra and EoR window to compare the powers of
radio halos with the EoR signal as well as other foreground components.
The redshifted 21~cm signal expected to be observed at different
frequencies represents a three-dimensional (3D) data cube, where the
two spatial dimensions describe the transverse distances across the sky
and the frequency dimension maps to the line-of-sight distance.
Within a limited redshift range (e.g., $\Delta z \sim 0.5$, corresponding
to a frequency bandwidth of \SI{\sim 8}{\MHz} at \SI{158}{\MHz})
during the EoR, the evolution effect of the universe is minor and the
\Hi{} distribution is believed to be isotropic.
The corresponding 3D power spectrum $P(k_x, k_y, k_z)$ of the EoR signal
should have spherical symmetry and can be averaged in spherical shells
of radii $k$, yielding the one-dimensional (1D) power spectrum $P(k)$,
which effectively increases the signal-to-noise ratio compared to
direct imaging observations \citep{morales2004,morales2006,datta2010}.
The dimensionless variant of the 1D power spectrum
$\Delta^2(k) = P(k) \,k^3 / (2\pi^2)$
is more commonly used in the literature.
To suppress the significant side lobes in the Fourier transform caused
by the sharp discontinuities at the ends of the finite frequency band,
we apply the Blackman--Nuttall window function to the frequency dimension
before calculating the power spectra \citep[e.g.,][]{trott2015,chapman2016}.

Since the two angular dimensions and the frequency dimension of the image
cubes of foreground continuum emissions are independent, which is
different from the image cube of the redshifted 21~cm signal, it is
appropriate to average the 3D power spectrum $P(k_x, k_y, k_z)$
over angular annuli of radii $\kperp \equiv \sqrt{k_x^2 + k_y^2}$
for each line-of-sight plane $\klos \equiv k_z$, which yields the
two-dimensional (2D) power spectrum $P(\kperp, \klos)$.
In the $(\kperp, \klos)$ plane, the spectral-smooth foreground emissions
are supposed to reside in the low-\klos{} region, although complicated
instrumental and observational effects (e.g., chromatic primary beams,
calibration errors) can throw some of the foreground contamination from
the purely angular (\kperp) modes into the line-of-sight (\klos)
dimension (i.e., mode mixing), which results in an expanded wedge-like
contamination region at the bottom right in the $(\kperp, \klos)$ plane
(i.e., foreground wedge; \citealt{datta2010,morales2012,liu2014}).
The region almost free of the foreground contamination, namely the
EoR window, is preserved at the top left in the $(\kperp, \klos)$ plane
and is described with
\begin{equation}
  \label{eq:eor-window}
  \klos \geq \frac{H(z) D_{\!M}(z)}{(1+z) c} \left[
    \kperp \sin\Theta + \frac{2\pi w f_{21}}{(1+z) D_{\!M}(z) B} \right]
\end{equation}
\citep{thyagarajan2013},
where
$B = \SI{8}{\MHz}$ is the frequency bandwidth of the image cube,
$f_{21} = \SI{1420.4}{\MHz}$ is the rest-frame frequency of the 21~cm line,
$z = f_{21}/f_c - 1$ is the signal redshift corresponding to the central
frequency ($f_c$) of the image cube,
$H(z)$ is the Hubble parameter at redshift $z$ [see \autoref{eq:hubble-z}],
$D_{\!M}(z)$ is the transverse comoving distance,
$w$ denotes the number of characteristic convolution widths
($\propto B^{-1}$) for the spillover region caused by the variations in
instrumental frequency response,
and $\Theta$ is the angular distance of foreground sources from the
field center.

%########################################################################
\section{Results}
\label{sec:results}
%########################################################################

We evaluate the contamination of radio halos on the EoR observations for
both foreground removal and avoidance methods.
First, we calculate the 1D power spectra and compare the powers of
radio halos with the EoR signal, which illustrates the impacts of radio
halos with the foreground removal methods.
Next, we calculate the 2D power spectra and carry out the comparison
between radio halos and the EoR signal inside the EoR window, from
which we evaluate the effects imposed by radio halos in adopting the
foreground avoidance methods to extract the EoR signal.

%========================================================================
\subsection{1D Power Spectra}
\label{sec:ps1d}
%========================================================================

\begin{figure*}
  \centering
  \includegraphics[width=\textwidth]{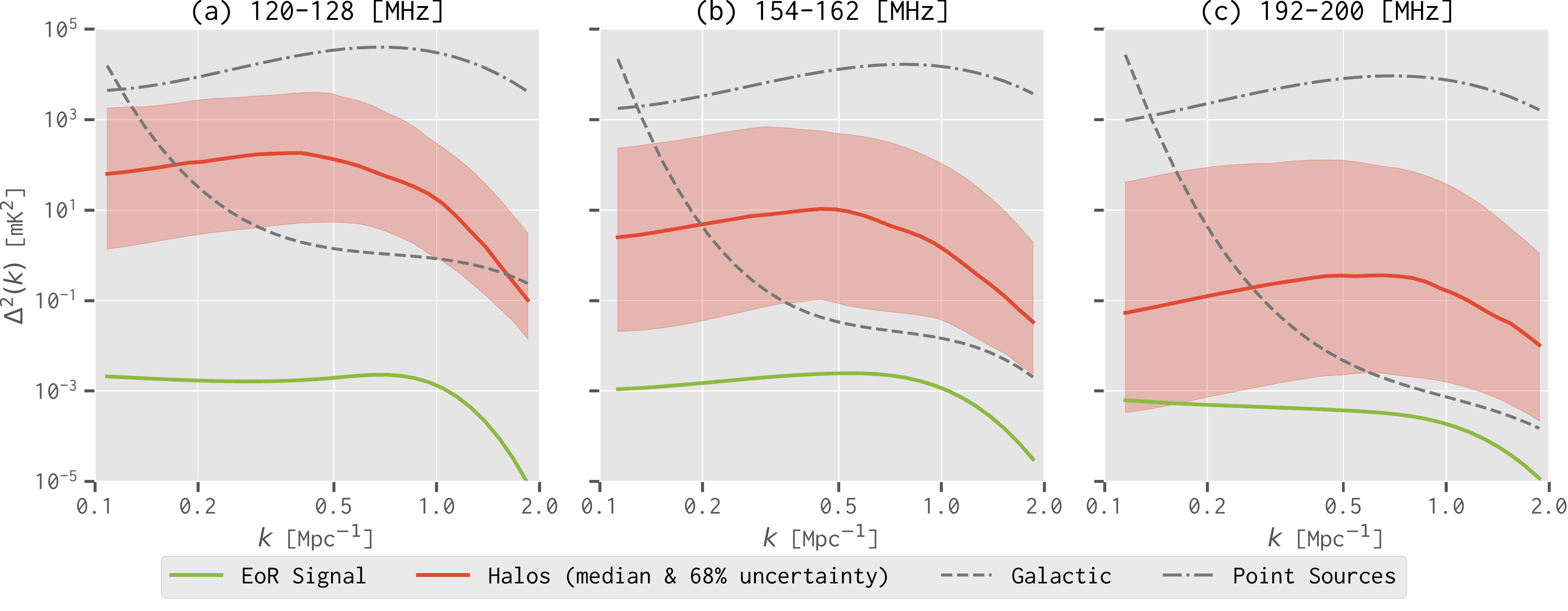}
  \caption{\label{fig:ps1d-3bands}%
    Comparisons of the 1D dimensionless power spectra $\Delta^2(k)$
    among the EoR signal (solid green line), radio halos (solid red line),
    Galactic diffuse emission (dashed gray line), and extragalactic point
    sources (dashed--dotted gray line) in the
    \textbf{(a)} \SIrange{120}{128}{\MHz},
    \textbf{(b)} \SIrange{154}{162}{\MHz}, and
    \textbf{(c)} \SIrange{192}{200}{\MHz} frequency bands.
    The solid red lines and shaded regions represent the median values
    and the corresponding 68\% uncertainties of the power
    spectra for radio halos estimated from the 100 simulation runs.
  }
\end{figure*}

We calculate the 1D dimensionless power spectra $\Delta^2(k)$ for each
image cube obtained in \autoref{sec:obs-simu}.
For radio halos, we make use of all 100 simulation runs
(\autoref{sec:cluster-halos}) to estimate the median power spectra and
the corresponding 68\% uncertainties.
The comparisons of the power spectra $\Delta^2(k)$ between radio halos
and the EoR signal in each frequency band are displayed
in \autoref{fig:ps1d-3bands}, where we also show the power spectra of
Galactic diffuse emission and extragalactic point sources for comparison.
The median power spectra (solid red lines) show that radio halos are
generally more luminous than the EoR signal by about \numlist{e4; e3; e2.5}
times on scales of $\SI{0.1}{\per\Mpc} < k < \SI{2}{\per\Mpc}$ in the
\numrange{120}{128}, \numrange{154}{162}, and \numrange{192}{200} \si{\MHz}
bands, respectively.
Given the large uncertainties in, e.g., brightness and number density,
of radio halos, the power spectra can vary by about \numrange{10}{100}
times with respect to the median values within the 68\%
uncertainties (red-shaded regions).
We also find that, although on large scales
($k \lesssim \SI{0.1}{\per\Mpc}$) the Galactic foreground is the
strongest contaminating source, its power deceases rapidly as the
scale becomes smaller and is weaker than the median power of radio halos
by a factor of about \numrange{10}{100} on scales of
$\SI{0.5}{\per\Mpc} \lesssim k \lesssim \SI{1}{\per\Mpc}$
in all the three bands.
These results evidently show that radio halos are severe foreground
contaminating sources.
Moreover, it can be a major challenge to accurately model and remove radio
halos, due to their diffuse and relatively complicated morphologies.

%========================================================================
\subsection{2D Power Spectra}
\label{sec:ps2d}
%========================================================================

\begin{figure}
  \centering
  \includegraphics[width=\columnwidth]{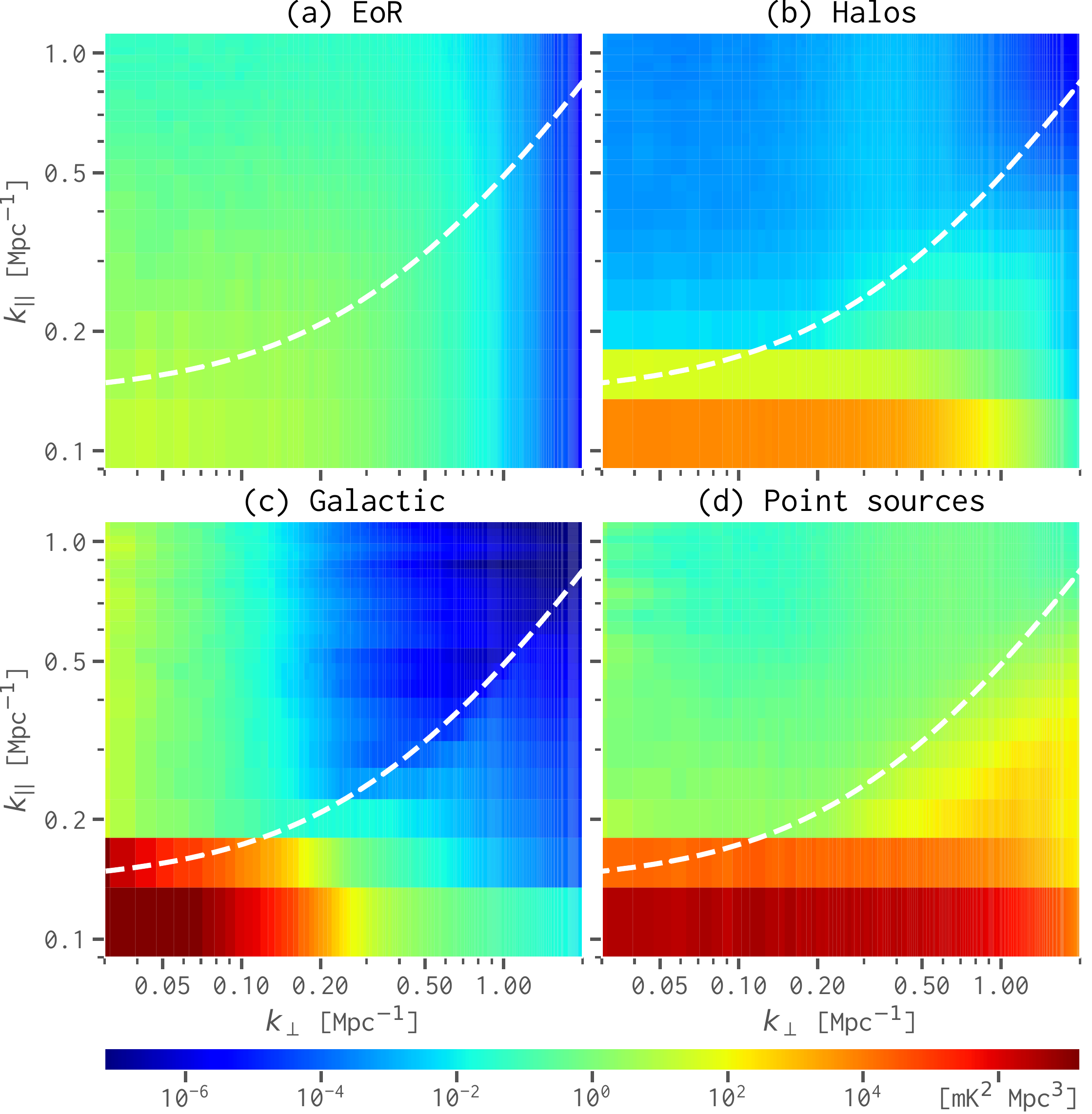}
  \caption{\label{fig:ps2d}%
    The \SIrange{154}{162}{\MHz} 2D power spectra $P(\kperp, \klos)$ of
    \textbf{(a)} the EoR signal,
    \textbf{(b)} radio halos (median of the 100 simulation runs),
    \textbf{(c)} Galactic diffuse emission,
    and
    \textbf{(d)} extragalactic point sources.
    All panels share the same logarithmic scale in units of
    [\si{\mK\squared\Mpc\cubed}].
    The dashed white lines mark the boundary between the EoR window
    (at the top left) and the contaminating wedge (at the bottom right).
  }
\end{figure}

\begin{figure*}
  \centering
  \includegraphics[width=\textwidth]{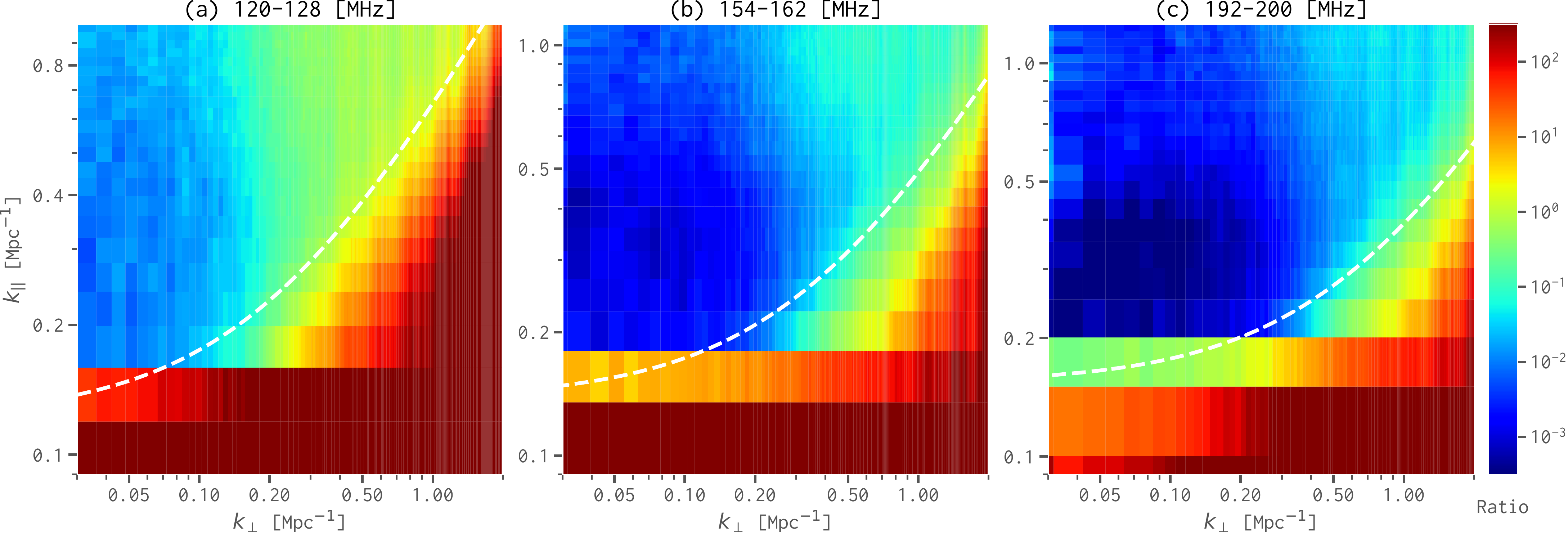}
  \caption{\label{fig:ps2d-ratio}%
    The 2D power spectrum ratios $R(\kperp, \klos)$ of radio halos to the
    EoR signal in the
    \textbf{(a)} \SIrange{120}{128}{\MHz},
    \textbf{(b)} \SIrange{154}{162}{\MHz}, and
    \textbf{(c)} \SIrange{192}{200}{\MHz} frequency bands.
    The median 2D power spectrum of 100 simulation runs for radio halos
    is used.
    All panels use the same color bar in logarithmic scale.
    The dashed white lines mark the EoR window boundaries.
  }
\end{figure*}

We take the \SIrange{154}{162}{\MHz} band as an example and show in
\autoref{fig:ps2d} the 2D power spectra $P(\kperp, \klos)$ of the EoR
signal, radio halos (the median power spectrum of the 100 simulation runs),
Galactic diffuse emission, and extragalactic point sources.
We find that, as shown in many previous works,
the EoR signal distributes its power across all \klos{} modes,
illustrating its rapid fluctuations along the line-of-sight dimension,
while the spectral-smooth foreground components dominate only in the
low-\klos{} regions ($\klos{} \lesssim \SI{0.2}{\per\Mpc}$).
With regard to the angular dimension, the power of radio halos appears in
the range of $\kperp \lesssim \SI{1}{\per\Mpc}$, showing a concentration
on the intermediate scales of $\kperp \sim \SI{0.5}{\per\Mpc}$.
Meanwhile, the powers of Galactic diffuse emission and extragalactic
point sources dominate on the large scales of
$\kperp \lesssim \SI{0.1}{\per\Mpc}$ and a broad angular scales of
$\kperp \gtrsim \SI{0.1}{\per\Mpc}$, respectively.
These results are also consistent with \autoref{fig:ps1d-3bands}(b).

In order to better evaluate the importance of radio halos as foreground
contaminating sources, we calculate the 2D power spectrum ratios
$R(\kperp, \klos)$ that are obtained by dividing the median 2D power
spectra of radio halos by those of the EoR signal in each frequency band.
We find that, as shown in \autoref{fig:ps2d-ratio}, the EoR measurements
will be significantly affected by radio halos on angular scales of
$\gtrsim \SI{0.1}{\per\Mpc}$, $\gtrsim \SI{0.3}{\per\Mpc}$, and
$\gtrsim \SI{0.5}{\per\Mpc}$ in the \numrange{120}{128},
\numrange{154}{162}, and \numrange{192}{200} \si{\MHz} bands, respectively.
It is also clearly shown that radio halos turn to cause stronger
contamination at lower frequencies (\SI{\sim 120}{\MHz}) than at higher
frequencies (\SI{\sim 200}{\MHz}).

\begin{figure}
  \centering
  \includegraphics[width=\columnwidth]{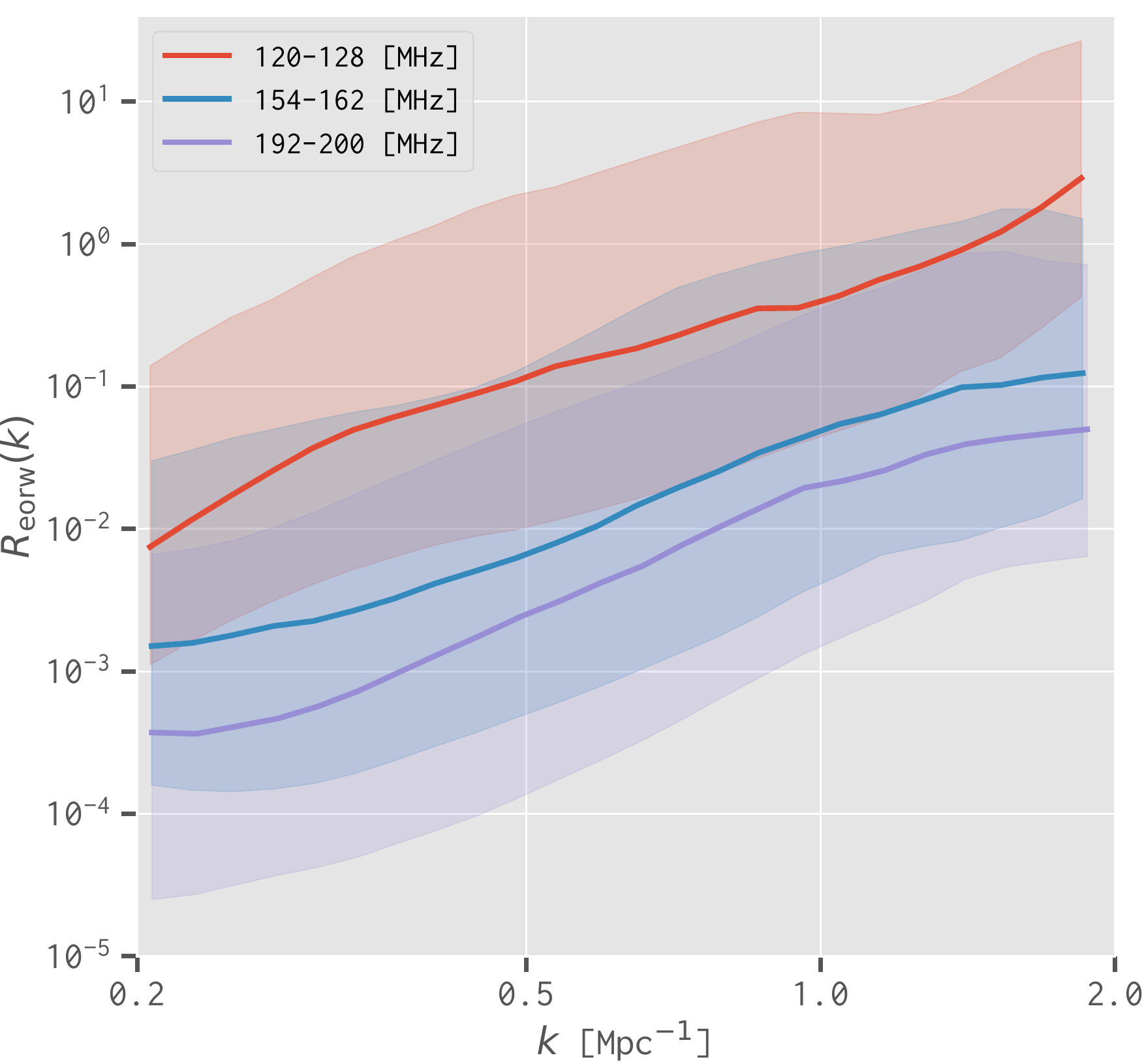}
  \caption{\label{fig:ps1d-ratio}%
    The 1D power ratios $R_{\R{eorw}}(k)$ inside the EoR window of
    radio halos to the EoR signal.
    The solid lines and shaded regions show the median values and
    corresponding 68\% uncertainties, respectively.
  }
\end{figure}

To further quantify the contamination caused by radio halos when
foreground avoidance methods are applied, we need to appropriately
define an EoR window in the $(\kperp, \klos)$ plane to avoid the
heavily contaminated areas and then compare the powers of radio halos
and the EoR signal derived inside the window.
We have tested multiple parameter configurations ($w$, $\Theta$) as
defined in \autoref{eq:eor-window}, and find that when $w = 3$ and
$\Theta$ being about 1.5 times the SKA1-Low's \fov{} (i.e.,
$\Theta = \SI{7.5}{\degree}$, \SI{6.0}{\degree}, and \SI{4.8}{\degree}
in \numrange{120}{128}, \numrange{154}{162}, and \numrange{192}{200}
\si{\MHz} bands, respectively) are used, a conservative EoR window
boundary can be defined to well avoid the contaminating wedge
(Figures~\ref{fig:ps2d} and \ref{fig:ps2d-ratio}).
However, a significant part (about 45\%, 46\%, and 60\% in
\numrange{120}{128}, \numrange{154}{162}, and \numrange{192}{200}
\si{\MHz} bands, respectively) of the power of
the EoR signal is lost in the excised wedges.
By averaging the modes only inside the defined EoR window, we calculate
the 1D power spectrum ratios $R_{\R{eorw}}(k)$ of radio halos to the EoR
signal and present the results in \autoref{fig:ps1d-ratio}.
We find that, compared to \autoref{fig:ps1d-3bands}, the 1D power ratios
inside the EoR window $R_{\R{eorw}}(k)$ are suppressed by about four orders
of magnitude, which demonstrates that the EoR window is a powerful tool
in detecting the EoR signal.
For example, $R_{\R{eorw}}(k)$ on scales of $k \sim \SI{1}{\per\Mpc}$ are
generally about 45\%, 6\%, and 2\% in the \numrange{120}{128},
\numrange{154}{162}, and \numrange{192}{200} \si{\MHz} bands, respectively.
However, the power of radio halos leaked into the EoR window can still be
significant, considering that $R_{\R{eorw}}(k)$ on scales of
$\SI{0.5}{\per\Mpc} \lesssim k \lesssim \SI{1}{\per\Mpc}$ can be up to
about \numrange{230}{800}\%, \numrange{18}{95}\%, and
\numrange{7}{40}\% in the three frequency bands within the
68\% uncertainties (shaded regions).

Based on the above results, we conclude that radio halos are severe
foreground contaminating sources to EoR observations.
Even inside the EoR window where most of the strong foreground
contamination is avoided, radio halos can still imprint
non-negligible contamination on the EoR measurements, especially at
lower frequencies (\SI{\sim 120}{\MHz}).
Careful treatments of radio halos as well as other foreground contaminating
sources would be indispensable for obtaining an EoR window that is not only
sufficiently clean but also as large as possible to preseve maximum
information of the EoR signal.

%########################################################################
\section{Discussion}
\label{sec:discussions}
%########################################################################

In practical observations with low-frequency radio interferometers, the
situations are much more complicated than our simulations.
For example, calibration uncertainties (e.g., insufficient sky modeling)
as well as other complicated instrumental and observational effects
(e.g., cable signal reflections, ionospheric distortions) can cause
frequency artifacts in the derived image cubes.
Foreground sources located in the side lobes of the station beam can
also significantly reduce the imaging dynamical range and quality.
In this section, we investigate how the EoR measurements are affected
in these two situations if the contamination of radio halos is not
properly removed.

%========================================================================
% figures of this section
\begin{figure*}
  \centering
  \includegraphics[width=0.9\textwidth]{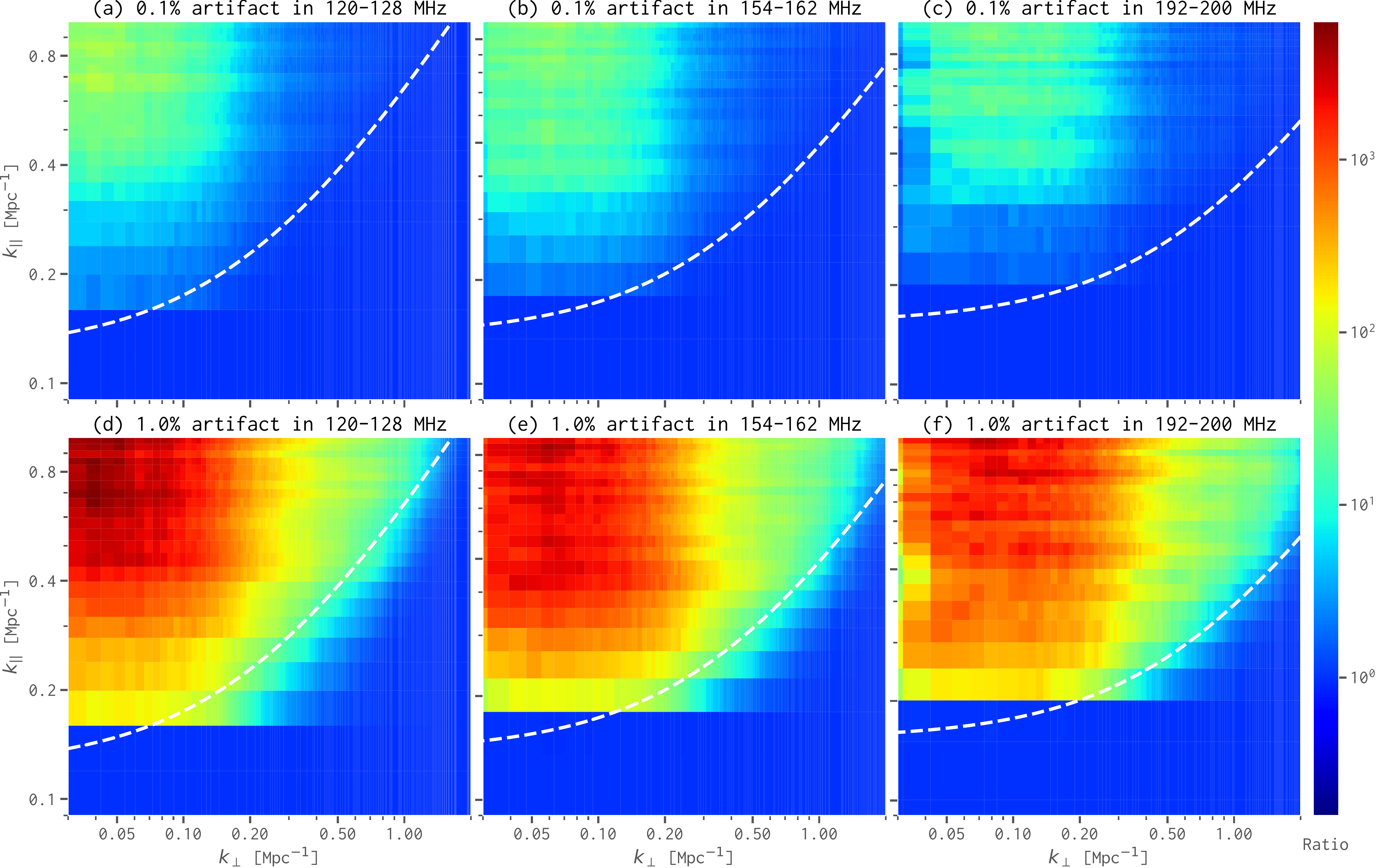}
  \caption{\label{fig:ps2d-ratio-crp}%
    The 2D power spectrum ratios $R_{\R{arti}}(\kperp, \klos)$ of radio
    halos that are obtained between the modified image cubes with
    frequency artifacts and the original ones.
    All the 100 simulation runs for radio halos are used to derive
    the median 2D power spectrum ratios that are presented here.
    The upper and lower rows show the cases of frequency artifacts
    being $A_{\R{arti}} = 0.1\%$ and $A_{\R{arti}} = 1\%$, respectively.
    The left, middle, and right columns show the power spectrum ratios
    in the \numrange{120}{128}, \numrange{154}{162}, and
    \numrange{192}{200} \si{\MHz} bands, respectively.
    The dashed white lines mark the EoR window boundaries.
    All panels share the same color bar in logarithmic scale.
  }
\end{figure*}

\begin{figure*}
  \centering
  \includegraphics[width=0.9\textwidth]{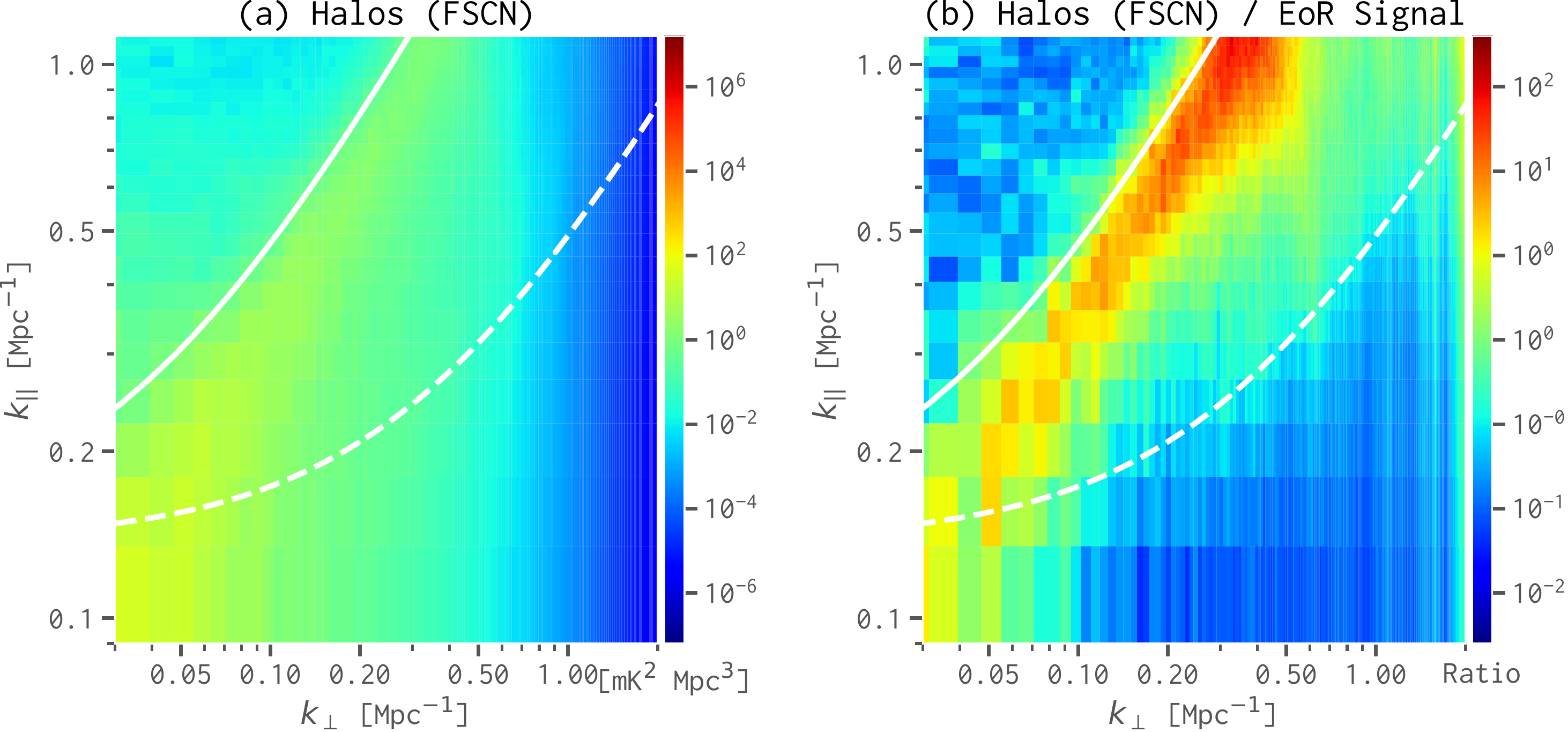}
  \caption{\label{fig:ps2d-fscn}%
    \textbf{(a)} The 2D power spectrum of the FSCN caused by radio halos
    in the far side lobes of the station beam.
    \textbf{(b)} The 2D power spectrum ratio of the FSCN to the EoR signal.
    The results are derived in the \SIrange{154}{162}{\MHz} band.
    The dashed and solid white lines mark the EoR window boundaries
    defined with $\Theta = \SI{6}{\degree}$ and \SI{90}{\degree},
    respectively.
  }
\end{figure*}

%========================================================================
\subsection{Impacts of Frequency Artifacts}
\label{sec:freq-artifacts}
%========================================================================

The smoothness along the frequency dimension is the most crucial feature
of various foreground components and is the key to extract the faint EoR
signal.
However, frequency artifacts may present in the obtained image cubes due
to calibration uncertainties and various instrumental and observational
effects, which break the spectral smoothness of the foreground emission
and hence damage the EoR measurements.

To evaluate the influence of the frequency artifacts on the power
spectra, we multiply each slice of the image cube by a random number
drawn from a Gaussian distribution with unity mean and then compare
the resulting power spectra \citep{chapman2016}.
Some simulation and observation studies have suggested that the residual
calibration errors in frequency channels may be about 0.1\%--1\%
\citep[e.g.,][]{barry2016,beardsley2016,ewallWice2017}.
We thus investigate two extreme cases here:
a frequency artifact of amplitude $A_{\R{arti}} = 0.1\%$ by
using $\sigma = 0.001$ for the Gaussian distribution
and a frequency artifact of $A_{\R{arti}} = 1\%$ with
$\sigma = 0.01$.

For each of the 100 simulation runs for radio halos, we calculate the
2D power spectrum ratios $R_{\R{arti}}(\kperp, \klos)$ of the modified
image cube with the frequency artifact to the original one
(\autoref{sec:obs-simu}),
and present the median 2D power spectrum ratios obtained in the
\numrange{120}{128}, \numrange{154}{162}, and \numrange{192}{200} \si{\MHz}
bands with either $A_{\R{arti}} = 0.1\%$ or $1\%$ in \autoref{fig:ps2d-ratio-crp}.
We find that, when the frequency artifact is added, the resulting 2D
power spectra are seriously damaged in all three frequency bands.
On scales of $\kperp \lesssim \SI{0.2}{\per\Mpc}$ and
$\klos \gtrsim \SI{0.3}{\per\Mpc}$,
adding frequency artifact of $A_{\R{arti}} = 0.1\%$
causes the power of radio halos to be about 17, 15, and 13 times stronger
in the \numrange{120}{128}, \numrange{154}{162}, and \numrange{192}{200}
\si{\MHz} bands, respectively, and the corresponding power increases are
about 1700, 1500, and 1300 times for frequency artifact of
$A_{\R{arti}} = 1\%$.
As a comparison, we add the same frequency artifacts
($A_{\R{arti}} = 0.1\%$ and $1\%$) to the image
cubes of the EoR signal, but find that the changes in the calculated
2D power spectra are negligible.
This is because the EoR signal already fluctuates remarkably along
the frequency dimension.
Consequently, even very minor ($\sim$\,0.1\%) instrumental or
calibration errors can make the contamination of radio halos
become much stronger, particularly inside the critical EoR window.
These results further support our conclusion made in \autoref{sec:ps2d}
that radio halos are important foreground sources and must be carefully
dealt with in EoR experiments.

%========================================================================
\subsection{Impacts of Far Side Lobes}
\label{sec:far-sidelobes}
%========================================================================

Phased arrays, which are widely used in low-frequency radio
interferometers (e.g., LOFAR, MWA, SKA1-Low), usually have complicated
beam profiles.
Sources far from the main lobe of the station beam can introduce
noise-like corruptions, known as the far side-lobe confusion noise
(FSCN; \citealt{smirnov2012}), to images through the multitude of
side lobes.
FSCN will not decrease once the $uv$ coverage of the observation no
longer improves, and can be the limiting factor in the noise
performance of interferometers \citep{mort2017}.

To investigate the impacts of FSCN contributed by the radio halos
located in the far side lobes of the station beam, we have generated
the corresponding sky model for the \texttt{OSKAR} simulator,
which evaluates the radio interferometer measurement equation
\citep{smirnov2011} and is able to perform full-sky simulations with
realistic beam profiles.
More details about the beam shapes and side-lobe properties of the
SKA1-Low can be found in \citet{mort2017}.
As an example, we simulate the radio halos in the \SIrange{154}{162}{\MHz}
band that cover the sky from the edge of the second side lobe
($\phi \sim \SI{10}{\degree}$ from the field center) to the horizon
($\phi = \SI{90}{\degree}$).
This emulates an ideal CLEAN procedure in practical data analysis that
removes all the radio halos in both the main lobe and the first side lobe
but leaves the ones in the far side lobes.
Using the \texttt{OSKAR} simulator and the \texttt{WSClean} imager as
described in \autoref{sec:obs-simu}, we obtain the dirty images of the
central \SI[product-units=repeat]{5 x 5}{\degree} region and then
calculate the 2D power spectrum.

In \autoref{fig:ps2d-fscn}, we present the 2D power spectrum of the FSCN
contributed by radio halos and the corresponding 2D power spectrum ratio
of the FSCN to the EoR signal.
We find that the FSCN contamination is very strong as its power can be
about 20 times the power of the EoR signal on scales of
$\kperp \sim \SI{0.3}{\per\Mpc}$ and $\klos \sim \SI{1.0}{\per\Mpc}$.
The wedge-shaped contamination region moves toward the top left in
the $(\kperp, \klos)$ plane and greatly reduces the EoR window.
In order to effectively avoid the FSCN contamination, we are forced to
employ a much more conservative EoR window boundary, such as the one
defined with $\Theta = \SI{90}{\degree}$ as marked in
\autoref{fig:ps2d-fscn} with solid white line, at the cost of losing
a remarkably larger portion of the EoR signal.
In consequence,
the serious FSCN contamination makes the selection of EoR sky region
a more challenging task since, in principle, neither bright radio halos
nor other strong sources are allowed in both the main and side lobes.
A highly accurate foreground model is hence crucial to mitigate the
impacts of FSCN.

%########################################################################
\section{Summary}
\label{sec:summary}
%########################################################################

Based on the Press--Schechter formalism and merger-induced turbulent
reacceleration model, we have simulated the emission maps of radio halos,
for which we have incorporated the SKA1-Low's instrumental effects by
utilizing its latest layout configuration.
By carrying out detailed comparisons of power spectra between radio halos
and the EoR signal as well as the Galactic diffuse emission and
extragalactic point sources in the \numrange{120}{128},
\numrange{154}{162}, and \numrange{192}{200} \si{\MHz} bands,
we have shown that radio halos are severe contaminating sources,
especially toward lower frequencies (\SI{\sim 120}{\MHz}).
Even inside the properly defined EoR windows, radio halos can still
be non-negligible contaminating sources to EoR observations.
Moreover, we have investigated the contamination resulted from
frequency artifacts and radio halos located inside the far side lobes,
both of which support our conclusion that radio halos are severe
foreground contaminating sources and need careful treatments in the
forthcoming deep EoR observations.

%%%%%%%%%%%%%%%%%%%%%%%%%%%%%%%%%%%%%%%%%%%%%%%%%%%%%%%%%%%%%%%%%%%%%%%%%
\acknowledgments

We gratefully acknowledge the reviewer for the constructive
comments that greatly help improve the manuscript.
We would like to thank
Fred Dulwich for providing the latest SKA1-Low layout configuration
as well as guidance on using the \texttt{OSKAR} simulator,
Andr\'e Offringa for help on interferometric imaging with
\texttt{WSClean},
Mathieu Remazeilles for providing the high-resolution Haslam
\SI{408}{\MHz} Galactic synchrotron map,
and Giovanna Giardino for the all-sky Galactic synchrotron spectral
index map.
We also acknowledge Emma Chapman, Uri Keshet, and Abhirup Datta for
their help.
Part of the work involving \texttt{OSKAR} and \texttt{WSClean} was
performed on the high-performance clusters
at Department of Astronomy, Shanghai Jiao Tong University and
at Shanghai Astronomical Observatory, Chinese Academy of Sciences.
This work is supported by
the Ministry of Science and Technology of China
(grant Nos.\@ 2018YFA0404601, 2017YFF0210903)
and the National Natural Science Foundation of China
(grant Nos.\@ 11433002, 11621303, 11835009, 61371147).

%% facilities: http://journals.aas.org/authors/aastex/facility.html
%
%% Use \facility{} or \facilities{} macros to list the keywords of facilities
%% used in the research for the paper.  Each keyword is check against the
%% master list during copy editing.  Individual instruments can be provided
%% in parentheses, after the keyword, but they are not verified.

\vspace{5mm}

%% The optional \software command to allow authors a place to specify which
%% programs were used during the creation of the manuscript.  Authors should
%% list each code and include either a citation or URL to the code inside ()s
%% when available.

\software{
  OSKAR \citep{mort2010},
  WSClean \citep{offringa2014},
  AstroPy \citep{astropy2013},
  HEALPix \citep{gorski2005},
  HMF \citep{murray2013},
  IPython (\url{https://ipython.org/}),
  Matplotlib (\url{https://matplotlib.org/}),
  NumPy (\url{http://www.numpy.org/}),
  SciPy (\url{https://scipy.org/}),
  Pandas (\url{https://pandas.pydata.org/}),
  Engauge Digitizer (\url{https://github.com/markummitchell/engauge-digitizer}).
}

\appendix

\renewcommand\thetable{\thesection\arabic{table}}
\setcounter{table}{0}

%########################################################################
\section{Supplemental Formulas}
\label{sec:formulas}
%########################################################################

In a flat \lcdm{} cosmology as adopted in this work, the critical linear
overdensity as a function of redshift $z$ is \citep{kitayama1996,randall2002}
\begin{equation}
  \label{eq:delta-crit}
  \delta_c(z) = \frac{D(z=0)}{D(z)} \frac{3}{20} (12\pi)^{2/3}
    \left[1 + 0.0123 \log_{10} \Omega_f(z) \right],
\end{equation}
where $\Omega_f(z)$ is the mass density ratio at redshift $z$ defined as
\begin{equation}
  \label{eq:omega-fz}
  \Omega_f(z) = \frac{\Omega_m(1+z)^3}{\Omega_m(1+z)^3 + \Omega_{\Lambda}},
\end{equation}
and $D(z)$ is the growth factor given by \citep[equation~(13.6)]{peebles1980}
\begin{equation}
  \label{eq:growth-factor}
  D(x) = \frac{(x^3 + 2)^{1/2}}{x^{3/2}}
    \int_0^x y^{3/2} (y^3 + 2)^{-3/2} \,\D{y},
\end{equation}
with $x_0 \equiv (2\Omega_{\Lambda}/\Omega_m)^{1/3}$ and $x = x_0 / (1+z)$.

The Hubble parameter at redshift $z$ is
\begin{equation}
  \label{eq:hubble-z}
  H(z) = H_0 E(z) = H_0 \sqrt{\Omega_m(1+z)^3 + \Omega_{\Lambda}} \,,
\end{equation}
where $E(z)$ is the redshift evolution factor \citep{hogg1999}.

The virial radius of a galaxy cluster at a redshift of $z$ is given by
\begin{equation}
  \label{eq:radius-virial}
  r_{\R{vir}} = \left[
    \frac{3 M_{\R{vir}}}{4\pi \,\Delta_{\R{vir}}(z) \rho_{\R{crit}}(z)}
  \right]^{1/3},
\end{equation}
where $M_{\R{vir}}$ is the virial mass of the cluster,
$\rho_{\R{crit}}(z) = 3 H^2(z) / (8\pi G)$ is the critical density,
$G$ being the gravitational constant, and $\Delta_{\R{vir}}(z)$ is the
virial overdensity given by \citep[e.g.,][]{bryan1998}
\begin{equation}
  \label{eq:delta-vir}
  \Delta_{\R{vir}}(z) = 18\pi^2 + 82x - 39x^2 ,
\end{equation}
where $x \equiv \Omega_f(z) - 1$.

%########################################################################
\section{Collection of Currently Observed Radio Halos}
\label{sec:halos-collection}
%########################################################################

\startlongtable
\begin{deluxetable*}{lcccr@{$\,\pm\,$}lr@{$\,\pm\,$}lll}
\tabletypesize{\scriptsize}
\tablecaption{\label{tab:halos-observed}%
  Currently Observed 71 Radio Halos and 9 Candidates (As of 2018 January)%
}
\tablehead{
  \colhead{Cluster} &
  \colhead{Redshift} &
  \colhead{kpc/\si{\arcsecond}} &
  \colhead{Size} &
  \multicolumn{2}{c}{$S_{\SI{1.4}{\GHz}}$} &
  \multicolumn{2}{c}{$P_{\SI{1.4}{\GHz}}$} &
  \colhead{Notes} &
  \colhead{References} \\
  \colhead{} &  % cluster
  \colhead{} &  % redshift
  \colhead{} &  % kpc/arcsec
  \colhead{(\si{\Mpc})} &  % size
  \multicolumn{2}{c}{(\si{\mJy})} &  % flux
  \multicolumn{2}{c}{(\SI{e24}{\watt\per\hertz})} &  % power
  \colhead{} &  % notes
  \colhead{} \\
  \colhead{(1)} &
  \colhead{(2)} &
  \colhead{(3)} &
  \colhead{(4)} &
  \multicolumn{2}{c}{(5)} &
  \multicolumn{2}{c}{(6)} &
  \colhead{(7)} &
  \colhead{(8)}
}

\startdata
% Name                 z        kpc/"  size    S1.4   Serr                    P1.4     Perr   Notes  Ref
1E 0657$-$56         & 0.2960 & 4.38 & 1.48 &  78.0 &  5.0                  & 21.33 &  1.49 &     & \citet{liang2000}  \\
Abell 141            & 0.2300 & 3.64 & 1.20 &   1.3 &  0.1\tablenotemark{a} &  0.25 &  0.02 &     & \citet{duchesne2017}  \\
Abell 209            & 0.2060 & 3.34 & 1.40 &  16.9 &  1.0                  &  2.04 &  0.12 & +cR & \citet{giovannini2009}  \\
Abell 399            & 0.0718 & 1.35 & 0.57 &  16.0 &  2.0                  &  0.20 &  0.03 &     & \citet{murgia2010}  \\
Abell 401            & 0.0737 & 1.38 & 0.49 &  17.0 &  1.0                  &  0.20 &  0.01 &     & \citet{bacchi2003}  \\
Abell 520            & 0.1990 & 3.25 & 0.99 &  34.4 &  1.5                  &  3.17 &  0.14 &     & \citet{govoni2001}  \\
Abell 521            & 0.2533 & 3.91 & 1.17 &   5.9 &  0.5                  &  1.12 &  0.09 & +R  & \citet{giovannini2009}  \\
Abell 523            & 0.1000 & 1.82 & 1.30 &  59.0 &  5.0                  &  1.47 &  0.12 &     & \citet{giovannini2011}  \\
Abell 545            & 0.1540 & 2.64 & 0.81 &  23.0 &  1.0                  &  1.25 &  0.05 &     & \citet{bacchi2003}  \\
Abell 665            & 0.1818 & 3.03 & 1.66 &  43.1 &  2.2                  &  3.28 &  0.17 &     & \citet{giovannini2000}  \\
Abell 697            & 0.2820 & 4.23 & 0.75 &   5.2 &  0.5                  &  2.20 &  0.21 &     & \citet{vanWeeren2011}  \\
Abell 746            & 0.2320 & 3.67 & 0.85 &  18.0 &  4.0                  &  3.80 &  0.84 & +R  & \citet{vanWeeren2011}  \\
Abell 754            & 0.0542 & 1.04 & 0.95 &  86.0 &  4.0                  &  0.56 &  0.03 & +R  & \citet{bacchi2003}  \\
Abell 773            & 0.2170 & 3.48 & 1.13 &  12.7 &  1.3                  &  1.39 &  0.14 &     & \citet{govoni2001}  \\
Abell 781            & 0.3004 & 4.42 & 1.60 &  20.5 &  5.0                  &  5.90 &  1.44 & +cR & \citet{govoni2011}  \\
Abell 800            & 0.2223 & 3.55 & 1.28 &  10.6 &  0.9                  &  1.52 &  0.13 &     & \citet{govoni2012}  \\
Abell 851            & 0.4069 & 5.40 & 1.08 &   3.7 &  0.3                  &  2.14 &  0.17 &     & \citet{giovannini2009}  \\
Abell 1132           & 0.1369 & 2.39 & 0.74 &   3.3 &  1.5                  &  0.16 &  0.07 &     & \citet{wilber2018}  \\
Abell 1213           & 0.0469 & 0.91 & 0.22 &  72.2 &  3.5                  &  0.36 &  0.02 &     & \citet{giovannini2009}  \\
Abell 1300           & 0.3100 & 4.52 & 0.92 &  20.0 &  2.0                  &  2.99 &  0.30 & +R  & \citet{reid1999}  \\
Abell 1351           & 0.3220 & 4.64 & 1.08 &  32.4 &  $\cdots$             & 11.37 &  $\cdots$ & & \citet{giacintucci2011b}  \\
Abell 1443           & 0.2700 & 4.10 & 1.10 &  11.0 &  1.1\tablenotemark{b} &  2.53 &  0.30 & cH  & \citet{bonafede2015}  \\
Abell 1451           & 0.1989 & 3.25 & 0.74 &   5.4 &  0.5                  &  0.62 &  0.07 & +cR & \citet{cuciti2018}  \\
Abell 1550           & 0.2540 & 3.92 & 1.41 &   7.7 &  1.6                  &  1.49 &  0.31 &     & \citet{govoni2012}  \\
Abell 1656           & 0.0232 & 0.46 & 0.58 & 530.0 & 50.0                  &  0.31 &  0.03 & +cR & \citet{kim1990}  \\
Abell 1682           & 0.2272 & 3.61 & 0.85 &   2.3 &  0.5\tablenotemark{c} &  0.41 &  0.08 & cH  & \citet{macario2013}  \\
Abell 1689           & 0.1832 & 3.05 & 0.73 &  10.0 &  2.9                  &  0.92 &  0.27 &     & \citet{vacca2011}  \\
Abell 1758A          & 0.2790 & 4.20 & 0.63 &   3.9 &  0.4                  &  0.93 &  0.10 & +R  & \citet{giovannini2009}  \\
Abell 1914           & 0.1712 & 2.88 & 1.16 &  64.0 &  3.0                  &  4.32 &  0.20 &     & \citet{bacchi2003}  \\
Abell 1995           & 0.3186 & 4.61 & 0.83 &   4.1 &  0.7                  &  1.35 &  0.23 &     & \citet{giovannini2009}  \\
Abell 2034           & 0.1130 & 2.03 & 0.60 &   7.3 &  2.0                  &  0.28 &  0.08 & +R  & \citet{vanWeeren2011}  \\
Abell 2061           & 0.0784 & 1.46 & 1.68 &  16.9 &  4.2                  &  0.25 &  0.06 & +R  & \citet{farnsworth2013}  \\
Abell 2065           & 0.0726 & 1.36 & 1.08 &  32.9 & 11.0                  &  0.41 &  0.14 &     & \citet{farnsworth2013}  \\
Abell 2069           & 0.1160 & 2.08 & 0.90 &   6.2 &  2.2\tablenotemark{d} &  0.25 &  0.05 &     & \citet{drabent2015}  \\
Abell 2142           & 0.0909 & 1.67 & 0.99 &  11.8 &  0.8                  &  1.12 &  0.08 &     & \citet{venturi2017}  \\
Abell 2163           & 0.2030 & 3.31 & 2.04 & 155.0 &  2.0                  & 14.93 &  0.20 & +R  & \citet{feretti2001}  \\
Abell 2218           & 0.1710 & 2.88 & 0.35 &   4.7 &  0.1                  &  0.32 &  0.01 &     & \citet{giovannini2000}  \\
Abell 2219           & 0.2256 & 3.59 & 1.54 &  81.0 &  4.0                  &  9.72 &  0.48 &     & \citet{bacchi2003}  \\
Abell 2254           & 0.1780 & 2.98 & 0.85 &  33.7 &  1.8                  &  2.43 &  0.13 &     & \citet{govoni2001}  \\
Abell 2255           & 0.0806 & 1.50 & 0.90 &  56.0 &  3.0                  &  0.87 &  0.05 & +R  & \citet{govoni2005}  \\
Abell 2256           & 0.0594 & 1.13 & 0.81 & 103.4 &  1.1                  &  0.82 &  0.01 & +R  & \citet{clarke2006}  \\
Abell 2294           & 0.1780 & 2.98 & 0.54 &   5.8 &  0.5                  &  0.51 &  0.04 &     & \citet{giovannini2009}  \\
Abell 2319           & 0.0524 & 1.01 & 0.93 & 153.0 &  8.0                  &  0.54 &  0.03 &     & \citet{feretti1997}  \\
Abell 2680           & 0.1901 & 3.14 & 0.57 &   1.8 &  0.6\tablenotemark{e} &  0.16 &  0.05 & cH  & \citet{duchesne2017}  \\
Abell 2693           & 0.1730 & 2.91 & 0.65 &   7.7 &  0.9\tablenotemark{f} &  0.61 &  0.07 & cH  & \citet{duchesne2017}  \\
Abell 2744           & 0.3080 & 4.50 & 1.62 &  57.1 &  2.9                  & 12.89 &  0.65 & +R  & \citet{govoni2001}  \\
Abell 2811           & 0.1080 & 1.95 & 0.48 &   3.4 &  0.7\tablenotemark{g} &  0.10 &  0.02 &     & \citet{duchesne2017}  \\
Abell 3411           & 0.1687 & 2.85 & 0.90 &   4.8 &  0.5                  &  0.46 &  0.05 & +R  & \citet{vanWeeren2013}  \\
Abell 3562           & 0.0480 & 0.93 & 0.44 &  20.0 &  2.0                  &  0.10 &  0.01 &     & \citet{venturi2003}  \\
Abell 3888           & 0.1510 & 2.60 & 0.99 &  27.6 &  3.1                  &  1.85 &  0.19 &     & \citet{shakouri2016}  \\
Abell S84            & 0.1080 & 1.95 & 0.49 &   2.1 &  0.3\tablenotemark{h} &  0.06 &  0.01 & cH  & \citet{duchesne2017}  \\
Abell S1121          & 0.3580 & 4.98 & 1.25 &   9.8 &  3.1\tablenotemark{h} &  4.54 &  1.44 &     & \citet{duchesne2017}  \\
ACT-CL J0102$-$4915  & 0.8700 & 7.73 & 2.17 &  10.7 &  1.1\tablenotemark{i} & 44.43 &  1.28 & +2R & \citet{lindner2014}  \\
ACT-CL J0256.5+0006  & 0.3430 & 4.84 & 0.79 &   2.1 &  0.5\tablenotemark{i} &  0.97 &  0.29 &     & \citet{knowles2016}  \\
CIZA J0107.7+5408    & 0.1066 & 1.93 & 1.10 &  55.0 &  5.0                  &  1.80 &  0.16 &     & \citet{vanWeeren2011}  \\
CIZA J0638.1+4747    & 0.1740 & 2.92 & 0.59 &   3.6 &  0.2                  &  0.30 &  0.02 &     & \citet{cuciti2018}  \\
CIZA J1938.3+5409    & 0.2600 & 3.99 & 0.72 &   1.6 &  0.2\tablenotemark{b} &  0.36 &  0.05 &     & \citet{bonafede2015}  \\
CIZA J2242.8+5301    & 0.1921 & 3.16 & 1.77 &  33.5 &  6.2\tablenotemark{j} &  3.40 &  0.97 & +2R & \citet{govoni2012}  \\
ClG 0016+16          & 0.5456 & 6.37 & 0.77 &   5.5 &  $\cdots$             &  4.42 &  $\cdots$ & & \citet{giovannini2000}  \\
ClG 0217+70          & 0.0655 & 1.24 & 0.73 &  58.6 &  0.9                  &  0.54 &  0.01 & +2R & \citet{brown2011}  \\
ClG 1446+26          & 0.3700 & 5.09 & 1.22 &   7.7 &  2.6                  &  3.57 &  1.21 & +R  & \citet{govoni2012}  \\
ClG 1821+64          & 0.2990 & 4.41 & 1.10 &  13.0 &  0.8\tablenotemark{k} &  3.70 &  0.10 &     & \citet{bonafede2014b}  \\
MACS J0416.1$-$2403  & 0.3960 & 5.31 & 0.64 &   1.7 &  0.8\tablenotemark{l} &  1.26 &  0.29 &     & \citet{pandeyPommier2015}  \\
MACS J0520.7$-$1328  & 0.3400 & 4.81 & 0.80 &   9.0 &  1.6                  &  3.38 &  0.60 & +cH & \citet{macario2014}  \\
MACS J0553.4$-$3342  & 0.4070 & 5.40 & 1.32 &   9.2 &  0.7\tablenotemark{b} &  6.73 &  0.61 &     & \citet{bonafede2012}  \\
MACS J0717.5+3745    & 0.5458 & 6.37 & 1.20 & 118.0 &  5.0                  & 50.00 & 10.00 & +R  & \citet{vanWeeren2009}  \\
MACS J0949.8+1708    & 0.3825 & 5.20 & 1.04 &   3.1 &  0.3\tablenotemark{b} &  1.63 &  0.15 &     & \citet{bonafede2015}  \\
MACS J1149.5+2223    & 0.5444 & 6.36 & 1.32 &   1.2 &  0.5                  &  1.95 &  0.93 & +cH, +2R & \citet{bonafede2012}  \\
MACS J1752.0+4440    & 0.3660 & 5.05 & 1.65 &  14.2 &  1.4\tablenotemark{m} &  9.50 &  0.91 & +2R & \citet{vanWeeren2012}  \\
MACS J2243.3$-$0935  & 0.4470 & 5.71 & 0.91 &   3.1 &  0.6\tablenotemark{n} &  3.11 &  0.58 & +cR & \citet{cantwell2016}  \\
PLCK G147.3$-$16.6   & 0.6500 & 6.92 & 1.80 &   2.5 &  0.4\tablenotemark{o} &  5.10 &  0.80 &     & \citet{vanWeeren2014}  \\
PLCK G171.9$-$40.7   & 0.2700 & 4.10 & 0.99 &  18.0 &  2.0                  &  4.76 &  0.10 &     & \citet{giacintucci2013}  \\
PLCK G285.0$-$23.7   & 0.3900 & 5.26 & 0.73 &   2.9 &  0.4\tablenotemark{p} &  1.67 &  0.21 &     & \citet{martinezAviles2016}  \\
PLCK G287.0+32.9     & 0.3900 & 5.26 & 1.30 &   8.8 &  0.9                  &  5.10 &  0.51 & +2R & \citet{bonafede2014a}  \\
PSZ1 G108.18$-$11.53 & 0.3347 & 4.77 & 0.84 &   6.8 &  0.2                  &  2.72 &  0.10 & +2R & \citet{deGasperin2015}  \\
RXC J1234.2+0947     & 0.2290 & 3.63 & 0.92 &   2.0 &  $\cdots$             &  0.30 &  $\cdots$ & cH  & \citet{govoni2012}  \\
RXC J1314.4$-$2515   & 0.2474 & 3.85 & 1.27 &  20.3 &  0.8                  &  1.45 &  0.06 & +2R & \citet{feretti2005}  \\
RXC J1514.9$-$1523   & 0.2226 & 3.55 & 1.38 &  10.0 &  2.0                  &  1.65 &  0.33 &     & \citet{giacintucci2011a}  \\
RXC J2003.5$-$2323   & 0.3171 & 4.59 & 1.38 &  35.0 &  2.0                  & 11.96 &  0.68 &     & \citet{giacintucci2009}  \\
RXC J2351.0$-$1954   & 0.2477 & 3.85 & 0.64 &   4.5 &  0.9\tablenotemark{q} &  0.89 &  0.18 & +cH & \citet{duchesne2017}  \\
\enddata

\tablecomments{%
  \textbf{(a)} Extrapolated from  \SI{168}{\MHz} with spectral index $\alpha=2.1$;
  \textbf{(b)} Extrapolated from  \SI{323}{\MHz} with spectral index $\alpha=1.3$;
  \textbf{(c)} Extrapolated from  \SI{153}{\MHz} with spectral index $\alpha=1.7$;
  \textbf{(d)} Extrapolated from  \SI{346}{\MHz} with spectral index $\alpha=1.0$;
  \textbf{(e)} Extrapolated from  \SI{168}{\MHz} with spectral index $\alpha=1.2$;
  \textbf{(f)} Extrapolated from  \SI{168}{\MHz} with spectral index $\alpha=0.88$;
  \textbf{(g)} Extrapolated from  \SI{168}{\MHz} with spectral index $\alpha=1.5$;
  \textbf{(h)} Extrapolated from  \SI{168}{\MHz} with spectral index $\alpha=1.3$;
  \textbf{(i)} Extrapolated from  \SI{610}{\MHz} with spectral index $\alpha=1.2$;
  \textbf{(j)} Extrapolated from  \SI{145}{\MHz} with spectral index $\alpha=1.03$;
  \textbf{(k)} Extrapolated from  \SI{325}{\MHz} with spectral index $\alpha=1.04$;
  \textbf{(l)} Extrapolated from  \SI{340}{\MHz} with spectral index $\alpha=1.5$;
  \textbf{(m)} Extrapolated from \SI{1714}{\MHz} with spectral index $\alpha=1.1$;
  \textbf{(n)} Extrapolated from  \SI{610}{\MHz} with spectral index $\alpha=1.4$;
  \textbf{(o)} Extrapolated from  \SI{610}{\MHz} with spectral index $\alpha=1.3$;
  \textbf{(p)} Extrapolated from \SI{1867}{\MHz} with spectral index $\alpha=1.3$;
  \textbf{(q)} Extrapolated from  \SI{168}{\MHz} with spectral index $\alpha=1.4$.
}

\textbf{Columns:}
  \textbf{(1)} galaxy cluster name;
  \textbf{(2)} redshift;
  \textbf{(3)} \si{\kpc} per \si{\arcsec} at the cluster's redshift
               (converted to our adopted cosmology);
  \textbf{(4)} largest linear size of the radio halo, in units of \si{\Mpc};
  \textbf{(5)} \SI{1.4}{\GHz} flux density;
  \textbf{(6)} \SI{1.4}{\GHz} radio power (converted to our adopted cosmology);
  \textbf{(7)} additional notes (cH, halo candidate; +R, with single relic;
    +cR, with single relic candidate; +2R, with double relics);
  \textbf{(8)} references to the quoted properties.
\end{deluxetable*}

%%%%%%%%%%%%%%%%%%%%%%%%%%%%%%%%%%%%%%%%%%%%%%%%%%%%%%%%%%%%%%%%%%%%%%%%%
\bibliography{main}

\begin{thebibliography}{}
\expandafter\ifx\csname natexlab\endcsname\relax\def\natexlab#1{#1}\fi
\providecommand{\url}[1]{\href{#1}{#1}}

\bibitem[{{Astropy Collaboration} {et~al.}(2013){Astropy Collaboration},
  {Robitaille}, {Tollerud}, {Greenfield}, {Droettboom}, {Bray}, {Aldcroft},
  {Davis}, {Ginsburg}, {Price-Whelan}, {Kerzendorf}, {Conley}, {Crighton},
  {Barbary}, {Muna}, {Ferguson}, {Grollier}, {Parikh}, {Nair}, {Unther},
  {Deil}, {Woillez}, {Conseil}, {Kramer}, {Turner}, {Singer}, {Fox}, {Weaver},
  {Zabalza}, {Edwards}, {Azalee Bostroem}, {Burke}, {Casey}, {Crawford},
  {Dencheva}, {Ely}, {Jenness}, {Labrie}, {Lim}, {Pierfederici}, {Pontzen},
  {Ptak}, {Refsdal}, {Servillat}, \& {Streicher}}]{astropy2013}
{Astropy Collaboration}, {Robitaille}, T.~P., {Tollerud}, E.~J., {et~al.} 2013,
  \href{http://dx.doi.org/10.1051/0004-6361/201322068}{\textcolor{magenta}{\aap}},
  \href{http://adsabs.harvard.edu/abs/2013A%26A...558A..33A}{558, A33}

\bibitem[{{Bacchi} {et~al.}(2003){Bacchi}, {Feretti}, {Giovannini}, \&
  {Govoni}}]{bacchi2003}
{Bacchi}, M., {Feretti}, L., {Giovannini}, G., \& {Govoni}, F. 2003,
  \href{http://dx.doi.org/10.1051/0004-6361:20030044}{\textcolor{magenta}{\aap}},
  \href{http://adsabs.harvard.edu/abs/2003A%26A...400..465B}{400, 465}

\bibitem[{{Barry} {et~al.}(2016){Barry}, {Hazelton}, {Sullivan}, {Morales}, \&
  {Pober}}]{barry2016}
{Barry}, N., {Hazelton}, B., {Sullivan}, I., {Morales}, M.~F., \& {Pober},
  J.~C. 2016,
  \href{http://dx.doi.org/10.1093/mnras/stw1380}{\textcolor{magenta}{\mnras}},
  \href{http://adsabs.harvard.edu/abs/2016MNRAS.461.3135B}{461, 3135}

\bibitem[{{Basu}(2012)}]{basu2012}
{Basu}, K. 2012,
  \href{http://dx.doi.org/10.1111/j.1745-3933.2012.01217.x}{\textcolor{magenta}{\mnras}},
  \href{http://adsabs.harvard.edu/abs/2012MNRAS.421L.112B}{421, L112}

\bibitem[{{Beardsley} {et~al.}(2016){Beardsley}, {Hazelton}, {Sullivan},
  {Carroll}, {Barry}, {Rahimi}, {Pindor}, {Trott}, {Line}, {Jacobs}, {Morales},
  {Pober}, {Bernardi}, {Bowman}, {Busch}, {Briggs}, {Cappallo}, {Corey}, {de
  Oliveira-Costa}, {Dillon}, {Emrich}, {Ewall-Wice}, {Feng}, {Gaensler},
  {Goeke}, {Greenhill}, {Hewitt}, {Hurley-Walker}, {Johnston-Hollitt},
  {Kaplan}, {Kasper}, {Kim}, {Kratzenberg}, {Lenc}, {Loeb}, {Lonsdale},
  {Lynch}, {McKinley}, {McWhirter}, {Mitchell}, {Morgan}, {Neben},
  {Thyagarajan}, {Oberoi}, {Offringa}, {Ord}, {Paul}, {Prabu}, {Procopio},
  {Riding}, {Rogers}, {Roshi}, {Udaya Shankar}, {Sethi}, {Srivani},
  {Subrahmanyan}, {Tegmark}, {Tingay}, {Waterson}, {Wayth}, {Webster},
  {Whitney}, {Williams}, {Williams}, {Wu}, \& {Wyithe}}]{beardsley2016}
{Beardsley}, A.~P., {Hazelton}, B.~J., {Sullivan}, I.~S., {et~al.} 2016,
  \href{http://dx.doi.org/10.3847/1538-4357/833/1/102}{\textcolor{magenta}{\apj}},
  \href{http://adsabs.harvard.edu/abs/2016ApJ...833..102B}{833, 102}

\bibitem[{{Beck} \& {Krause}(2005)}]{beck2005}
{Beck}, R., \& {Krause}, M. 2005,
  \href{http://dx.doi.org/10.1002/asna.200510366}{\textcolor{magenta}{Astronomische
  Nachrichten}}, \href{http://adsabs.harvard.edu/abs/2005AN....326..414B}{326,
  414}

\bibitem[{{Blasi} {et~al.}(2007){Blasi}, {Gabici}, \&
  {Brunetti}}]{blasi2007rev}
{Blasi}, P., {Gabici}, S., \& {Brunetti}, G. 2007,
  \href{http://dx.doi.org/10.1142/S0217751X0703529X}{\textcolor{magenta}{International
  Journal of Modern Physics A}},
  \href{http://adsabs.harvard.edu/abs/2007IJMPA..22..681B}{22, 681}

\bibitem[{{Bonafede} {et~al.}(2014{\natexlab{a}}){Bonafede}, {Intema},
  {Br{\"u}ggen}, {Girardi}, {Nonino}, {Kantharia}, {van Weeren}, \&
  {R{\"o}ttgering}}]{bonafede2014a}
{Bonafede}, A., {Intema}, H.~T., {Br{\"u}ggen}, M., {et~al.}
  2014{\natexlab{a}},
  \href{http://dx.doi.org/10.1088/0004-637X/785/1/1}{\textcolor{magenta}{\apj}},
  \href{http://adsabs.harvard.edu/abs/2014ApJ...785....1B}{785, 1}

\bibitem[{{Bonafede} {et~al.}(2012){Bonafede}, {Br{\"u}ggen}, {van Weeren},
  {Vazza}, {Giovannini}, {Ebeling}, {Edge}, {Hoeft}, \& {Klein}}]{bonafede2012}
{Bonafede}, A., {Br{\"u}ggen}, M., {van Weeren}, R., {et~al.} 2012,
  \href{http://dx.doi.org/10.1111/j.1365-2966.2012.21570.x}{\textcolor{magenta}{\mnras}},
  \href{http://adsabs.harvard.edu/abs/2012MNRAS.426...40B}{426, 40}

\bibitem[{{Bonafede} {et~al.}(2014{\natexlab{b}}){Bonafede}, {Intema},
  {Br{\"u}ggen}, {Russell}, {Ogrean}, {Basu}, {Sommer}, {van Weeren},
  {Cassano}, {Fabian}, \& {R{\"o}ttgering}}]{bonafede2014b}
{Bonafede}, A., {Intema}, H.~T., {Br{\"u}ggen}, M., {et~al.}
  2014{\natexlab{b}},
  \href{http://dx.doi.org/10.1093/mnrasl/slu110}{\textcolor{magenta}{\mnras}},
  \href{http://adsabs.harvard.edu/abs/2014MNRAS.444L..44B}{444, L44}

\bibitem[{{Bonafede} {et~al.}(2015){Bonafede}, {Intema}, {Br{\"u}ggen},
  {Vazza}, {Basu}, {Sommer}, {Ebeling}, {de Gasperin}, {R{\"o}ttgering}, {van
  Weeren}, \& {Cassano}}]{bonafede2015}
{Bonafede}, A., {Intema}, H., {Br{\"u}ggen}, M., {et~al.} 2015,
  \href{http://dx.doi.org/10.1093/mnras/stv2065}{\textcolor{magenta}{\mnras}},
  \href{http://adsabs.harvard.edu/abs/2015MNRAS.454.3391B}{454, 3391}

\bibitem[{{Bond} {et~al.}(1991){Bond}, {Cole}, {Efstathiou}, \&
  {Kaiser}}]{bond1991}
{Bond}, J.~R., {Cole}, S., {Efstathiou}, G., \& {Kaiser}, N. 1991,
  \href{http://dx.doi.org/10.1086/170520}{\textcolor{magenta}{\apj}},
  \href{http://adsabs.harvard.edu/abs/1991ApJ...379..440B}{379, 440}

\bibitem[{{Bowman} {et~al.}(2013){Bowman}, {Cairns}, {Kaplan}, {Murphy},
  {Oberoi}, {Staveley-Smith}, {Arcus}, {Barnes}, {Bernardi}, {Briggs}, {Brown},
  {Bunton}, {Burgasser}, {Cappallo}, {Chatterjee}, {Corey}, {Coster},
  {Deshpande}, {deSouza}, {Emrich}, {Erickson}, {Goeke}, {Gaensler},
  {Greenhill}, {Harvey-Smith}, {Hazelton}, {Herne}, {Hewitt},
  {Johnston-Hollitt}, {Kasper}, {Kincaid}, {Koenig}, {Kratzenberg}, {Lonsdale},
  {Lynch}, {Matthews}, {McWhirter}, {Mitchell}, {Morales}, {Morgan}, {Ord},
  {Pathikulangara}, {Prabu}, {Remillard}, {Robishaw}, {Rogers}, {Roshi},
  {Salah}, {Sault}, {Shankar}, {Srivani}, {Stevens}, {Subrahmanyan}, {Tingay},
  {Wayth}, {Waterson}, {Webster}, {Whitney}, {Williams}, {Williams}, \&
  {Wyithe}}]{bowman2013}
{Bowman}, J.~D., {Cairns}, I., {Kaplan}, D.~L., {et~al.} 2013,
  \href{http://dx.doi.org/10.1017/pas.2013.009}{\textcolor{magenta}{\pasa}},
  \href{http://adsabs.harvard.edu/abs/2013PASA...30...31B}{30, e031}

\bibitem[{{Briggs}(1995)}]{briggs1995}
{Briggs}, D.~S. 1995, PhD thesis, The New Mexico Institute of Mining and
  Technology

\bibitem[{{Brown} {et~al.}(2011){Brown}, {Duesterhoeft}, \&
  {Rudnick}}]{brown2011}
{Brown}, S., {Duesterhoeft}, J., \& {Rudnick}, L. 2011,
  \href{http://dx.doi.org/10.1088/2041-8205/727/1/L25}{\textcolor{magenta}{\apjl}},
  \href{http://adsabs.harvard.edu/abs/2011ApJ...727L..25B}{727, L25}

\bibitem[{{Brunetti} {et~al.}(2004){Brunetti}, {Blasi}, {Cassano}, \&
  {Gabici}}]{brunetti2004}
{Brunetti}, G., {Blasi}, P., {Cassano}, R., \& {Gabici}, S. 2004,
  \href{http://dx.doi.org/10.1111/j.1365-2966.2004.07727.x}{\textcolor{magenta}{\mnras}},
  \href{http://adsabs.harvard.edu/abs/2004MNRAS.350.1174B}{350, 1174}

\bibitem[{{Brunetti} {et~al.}(2009){Brunetti}, {Cassano}, {Dolag}, \&
  {Setti}}]{brunetti2009}
{Brunetti}, G., {Cassano}, R., {Dolag}, K., \& {Setti}, G. 2009,
  \href{http://dx.doi.org/10.1051/0004-6361/200912751}{\textcolor{magenta}{\aap}},
  \href{http://adsabs.harvard.edu/abs/2009A%26A...507..661B}{507, 661}

\bibitem[{{Brunetti} \& {Jones}(2014)}]{brunetti2014rev}
{Brunetti}, G., \& {Jones}, T.~W. 2014,
  \href{http://dx.doi.org/10.1142/S0218271814300079}{\textcolor{magenta}{International
  Journal of Modern Physics D}},
  \href{http://adsabs.harvard.edu/abs/2014IJMPD..2330007B}{23, 1430007}

\bibitem[{{Brunetti} \& {Lazarian}(2007)}]{brunetti2007}
{Brunetti}, G., \& {Lazarian}, A. 2007,
  \href{http://dx.doi.org/10.1111/j.1365-2966.2007.11771.x}{\textcolor{magenta}{\mnras}},
  \href{http://adsabs.harvard.edu/abs/2007MNRAS.378..245B}{378, 245}

\bibitem[{{Brunetti} \& {Lazarian}(2011)}]{brunetti2011}
---. 2011,
  \href{http://dx.doi.org/10.1111/j.1365-2966.2010.17457.x}{\textcolor{magenta}{\mnras}},
  \href{http://adsabs.harvard.edu/abs/2011MNRAS.410..127B}{410, 127}

\bibitem[{{Brunetti} {et~al.}(2001){Brunetti}, {Setti}, {Feretti}, \&
  {Giovannini}}]{brunetti2001}
{Brunetti}, G., {Setti}, G., {Feretti}, L., \& {Giovannini}, G. 2001,
  \href{http://dx.doi.org/10.1046/j.1365-8711.2001.03978.x}{\textcolor{magenta}{\mnras}},
  \href{http://adsabs.harvard.edu/abs/2001MNRAS.320..365B}{320, 365}

\bibitem[{{Bryan} \& {Norman}(1998)}]{bryan1998}
{Bryan}, G.~L., \& {Norman}, M.~L. 1998,
  \href{http://dx.doi.org/10.1086/305262}{\textcolor{magenta}{\apj}},
  \href{http://adsabs.harvard.edu/abs/1998ApJ...495...80B}{495, 80}

\bibitem[{{Cantwell} {et~al.}(2016){Cantwell}, {Scaife}, {Oozeer}, {Wen}, \&
  {Han}}]{cantwell2016}
{Cantwell}, T.~M., {Scaife}, A.~M.~M., {Oozeer}, N., {Wen}, Z.~L., \& {Han},
  J.~L. 2016,
  \href{http://dx.doi.org/10.1093/mnras/stw419}{\textcolor{magenta}{\mnras}},
  \href{http://adsabs.harvard.edu/abs/2016MNRAS.458.1803C}{458, 1803}

\bibitem[{{Cassano} \& {Brunetti}(2005)}]{cassano2005}
{Cassano}, R., \& {Brunetti}, G. 2005,
  \href{http://dx.doi.org/10.1111/j.1365-2966.2005.08747.x}{\textcolor{magenta}{\mnras}},
  \href{http://adsabs.harvard.edu/abs/2005MNRAS.357.1313C}{357, 1313}

\bibitem[{{Cassano} {et~al.}(2016){Cassano}, {Brunetti}, {Giocoli}, \&
  {Ettori}}]{cassano2016}
{Cassano}, R., {Brunetti}, G., {Giocoli}, C., \& {Ettori}, S. 2016,
  \href{http://dx.doi.org/10.1051/0004-6361/201628414}{\textcolor{magenta}{\aap}},
  \href{http://adsabs.harvard.edu/abs/2016A%26A...593A..81C}{593, A81}

\bibitem[{{Cassano} {et~al.}(2012){Cassano}, {Brunetti}, {Norris},
  {R{\"o}ttgering}, {Johnston-Hollitt}, \& {Trasatti}}]{cassano2012}
{Cassano}, R., {Brunetti}, G., {Norris}, R.~P., {et~al.} 2012,
  \href{http://dx.doi.org/10.1051/0004-6361/201220018}{\textcolor{magenta}{\aap}},
  \href{http://adsabs.harvard.edu/abs/2012A%26A...548A.100C}{548, A100}

\bibitem[{{Cassano} {et~al.}(2007){Cassano}, {Brunetti}, {Setti}, {Govoni}, \&
  {Dolag}}]{cassano2007}
{Cassano}, R., {Brunetti}, G., {Setti}, G., {Govoni}, F., \& {Dolag}, K. 2007,
  \href{http://dx.doi.org/10.1111/j.1365-2966.2007.11901.x}{\textcolor{magenta}{\mnras}},
  \href{http://adsabs.harvard.edu/abs/2007MNRAS.378.1565C}{378, 1565}

\bibitem[{{Cassano} {et~al.}(2013){Cassano}, {Ettori}, {Brunetti},
  {Giacintucci}, {Pratt}, {Venturi}, {Kale}, {Dolag}, \&
  {Markevitch}}]{cassano2013}
{Cassano}, R., {Ettori}, S., {Brunetti}, G., {et~al.} 2013,
  \href{http://dx.doi.org/10.1088/0004-637X/777/2/141}{\textcolor{magenta}{\apj}},
  \href{http://adsabs.harvard.edu/abs/2013ApJ...777..141C}{777, 141}

\bibitem[{{Cassano} {et~al.}(2015){Cassano}, {Bernardi}, {Brunetti},
  {Br{\"u}ggen}, {Clarke}, {Dallacasa}, {Dolag}, {Ettori}, {Giacintucci},
  {Giocoli}, {Gitti}, {Johnston-Hollitt}, {Kale}, {Markevich}, {Norris},
  {Pommier}, {Pratt}, {Rottgering}, \& {Venturi}}]{cassano2015}
{Cassano}, R., {Bernardi}, G., {Brunetti}, G., {et~al.} 2015,
  \href{http://adsabs.harvard.edu/abs/2015aska.confE..73C}{Advancing
  Astrophysics with the Square Kilometre Array (AASKA14), 73}

\bibitem[{{Cavaliere} \& {Fusco-Femiano}(1976)}]{cavaliere1976}
{Cavaliere}, A., \& {Fusco-Femiano}, R. 1976, \aap,
  \href{http://cdsads.u-strasbg.fr/abs/1976A%26A....49..137C}{49, 137}

\bibitem[{{Cavaliere} {et~al.}(1998){Cavaliere}, {Menci}, \&
  {Tozzi}}]{cavaliere1998}
{Cavaliere}, A., {Menci}, N., \& {Tozzi}, P. 1998,
  \href{http://dx.doi.org/10.1086/305839}{\textcolor{magenta}{\apj}},
  \href{http://adsabs.harvard.edu/abs/1998ApJ...501..493C}{501, 493}

\bibitem[{{Chang} \& {Cooper}(1970)}]{chang1970}
{Chang}, J.~S., \& {Cooper}, G. 1970,
  \href{http://dx.doi.org/10.1016/0021-9991(70)90001-X}{\textcolor{magenta}{Journal
  of Computational Physics}},
  \href{http://adsabs.harvard.edu/abs/1970JCoPh...6....1C}{6, 1}

\bibitem[{{Chapman} {et~al.}(2016){Chapman}, {Zaroubi}, {Abdalla}, {Dulwich},
  {Jeli{\'c}}, \& {Mort}}]{chapman2016}
{Chapman}, E., {Zaroubi}, S., {Abdalla}, F.~B., {et~al.} 2016,
  \href{http://dx.doi.org/10.1093/mnras/stw161}{\textcolor{magenta}{\mnras}},
  \href{http://adsabs.harvard.edu/abs/2016MNRAS.458.2928C}{458, 2928}

\bibitem[{{Clarke} \& {Ensslin}(2006)}]{clarke2006}
{Clarke}, T.~E., \& {Ensslin}, T.~A. 2006,
  \href{http://dx.doi.org/10.1086/504076}{\textcolor{magenta}{\aj}},
  \href{http://adsabs.harvard.edu/abs/2006AJ....131.2900C}{131, 2900}

\bibitem[{{Cuciti} {et~al.}(2018){Cuciti}, {Brunetti}, {van Weeren},
  {Bonafede}, {Dallacasa}, {Cassano}, {Venturi}, \& {Kale}}]{cuciti2018}
{Cuciti}, V., {Brunetti}, G., {van Weeren}, R., {et~al.} 2018,
  \href{http://dx.doi.org/10.1051/0004-6361/201731174}{\textcolor{magenta}{\aap}},
  \href{http://adsabs.harvard.edu/abs/2018A%26A...609A..61C}{609, A61}

\bibitem[{{Datta} {et~al.}(2010){Datta}, {Bowman}, \& {Carilli}}]{datta2010}
{Datta}, A., {Bowman}, J.~D., \& {Carilli}, C.~L. 2010,
  \href{http://dx.doi.org/10.1088/0004-637X/724/1/526}{\textcolor{magenta}{\apj}},
  \href{http://adsabs.harvard.edu/abs/2010ApJ...724..526D}{724, 526}

\bibitem[{{Dayal} \& {Ferrara}(2018)}]{dayal2018}
{Dayal}, P., \& {Ferrara}, A. 2018,
  \href{http://dx.doi.org/10.1016/j.physrep.2018.10.002}{\textcolor{magenta}{\physrep}},
  \href{http://adsabs.harvard.edu/abs/2018PhR...780....1D}{780, 1}

\bibitem[{{de Gasperin} {et~al.}(2015){de Gasperin}, {Intema}, {van Weeren},
  {Dawson}, {Golovich}, {Wittman}, {Bonafede}, \&
  {Br{\"u}ggen}}]{deGasperin2015}
{de Gasperin}, F., {Intema}, H.~T., {van Weeren}, R.~J., {et~al.} 2015,
  \href{http://dx.doi.org/10.1093/mnras/stv1873}{\textcolor{magenta}{\mnras}},
  \href{http://adsabs.harvard.edu/abs/2015MNRAS.453.3483D}{453, 3483}

\bibitem[{{DeBoer} {et~al.}(2017){DeBoer}, {Parsons}, {Aguirre}, {Alexander},
  {Ali}, {Beardsley}, {Bernardi}, {Bowman}, {Bradley}, {Carilli}, {Cheng}, {de
  Lera Acedo}, {Dillon}, {Ewall-Wice}, {Fadana}, {Fagnoni}, {Fritz},
  {Furlanetto}, {Glendenning}, {Greig}, {Grobbelaar}, {Hazelton}, {Hewitt},
  {Hickish}, {Jacobs}, {Julius}, {Kariseb}, {Kohn}, {Lekalake}, {Liu}, {Loots},
  {MacMahon}, {Malan}, {Malgas}, {Maree}, {Martinot}, {Mathison}, {Matsetela},
  {Mesinger}, {Morales}, {Neben}, {Patra}, {Pieterse}, {Pober}, {Razavi-Ghods},
  {Ringuette}, {Robnett}, {Rosie}, {Sell}, {Smith}, {Syce}, {Tegmark},
  {Thyagarajan}, {Williams}, \& {Zheng}}]{deboer2017}
{DeBoer}, D.~R., {Parsons}, A.~R., {Aguirre}, J.~E., {et~al.} 2017,
  \href{http://dx.doi.org/10.1088/1538-3873/129/974/045001}{\textcolor{magenta}{\pasp}},
  \href{http://adsabs.harvard.edu/abs/2017PASP..129d5001D}{129, 045001}

\bibitem[{{Di Matteo} {et~al.}(2004){Di Matteo}, {Ciardi}, \&
  {Miniati}}]{diMatteo2004}
{Di Matteo}, T., {Ciardi}, B., \& {Miniati}, F. 2004,
  \href{http://dx.doi.org/10.1111/j.1365-2966.2004.08443.x}{\textcolor{magenta}{\mnras}},
  \href{http://adsabs.harvard.edu/abs/2004MNRAS.355.1053D}{355, 1053}

\bibitem[{{Dickinson} {et~al.}(2003){Dickinson}, {Davies}, \&
  {Davis}}]{dickinson2003}
{Dickinson}, C., {Davies}, R.~D., \& {Davis}, R.~J. 2003,
  \href{http://dx.doi.org/10.1046/j.1365-8711.2003.06439.x}{\textcolor{magenta}{\mnras}},
  \href{http://adsabs.harvard.edu/abs/2003MNRAS.341..369D}{341, 369}

\bibitem[{{Dolag} {et~al.}(2002){Dolag}, {Bartelmann}, \& {Lesch}}]{dolag2002}
{Dolag}, K., {Bartelmann}, M., \& {Lesch}, H. 2002,
  \href{http://dx.doi.org/10.1051/0004-6361:20020241}{\textcolor{magenta}{\aap}},
  \href{http://adsabs.harvard.edu/abs/2002A%26A...387..383D}{387, 383}

\bibitem[{{Donnert} \& {Brunetti}(2014)}]{donnert2014}
{Donnert}, J., \& {Brunetti}, G. 2014,
  \href{http://dx.doi.org/10.1093/mnras/stu1417}{\textcolor{magenta}{\mnras}},
  \href{http://adsabs.harvard.edu/abs/2014MNRAS.443.3564D}{443, 3564}

\bibitem[{{Drabent} {et~al.}(2015){Drabent}, {Hoeft}, {Pizzo}, {Bonafede}, {van
  Weeren}, \& {Klein}}]{drabent2015}
{Drabent}, A., {Hoeft}, M., {Pizzo}, R.~F., {et~al.} 2015,
  \href{http://dx.doi.org/10.1051/0004-6361/201424828}{\textcolor{magenta}{\aap}},
  \href{http://adsabs.harvard.edu/abs/2015A%26A...575A...8D}{575, A8}

\bibitem[{{Duchesne} {et~al.}(2017){Duchesne}, {Johnston-Hollitt}, {Offringa},
  {Pratt}, {Zheng}, \& {Dehghan}}]{duchesne2017}
{Duchesne}, S.~W., {Johnston-Hollitt}, M., {Offringa}, A.~R., {et~al.} 2017,
  \href{http://adsabs.harvard.edu/abs/2017arXiv170703517D}{ArXiv e-prints,
  arXiv:1707.03517}

\bibitem[{{Duffy} {et~al.}(2008){Duffy}, {Schaye}, {Kay}, \& {Dalla
  Vecchia}}]{duffy2008}
{Duffy}, A.~R., {Schaye}, J., {Kay}, S.~T., \& {Dalla Vecchia}, C. 2008,
  \href{http://dx.doi.org/10.1111/j.1745-3933.2008.00537.x}{\textcolor{magenta}{\mnras}},
  \href{http://adsabs.harvard.edu/abs/2008MNRAS.390L..64D}{390, L64}

\bibitem[{{Eilek} \& {Hughes}(1991)}]{eilek1991}
{Eilek}, J.~A., \& {Hughes}, P.~A. 1991, Particle acceleration and magnetic
  field evolution, ed. P.~A. {Hughes} (Cambridge University Press), 428

\bibitem[{{Ettori} {et~al.}(2013){Ettori}, {Donnarumma}, {Pointecouteau},
  {Reiprich}, {Giodini}, {Lovisari}, \& {Schmidt}}]{ettori2013}
{Ettori}, S., {Donnarumma}, A., {Pointecouteau}, E., {et~al.} 2013,
  \href{http://dx.doi.org/10.1007/s11214-013-9976-7}{\textcolor{magenta}{\ssr}},
  \href{http://adsabs.harvard.edu/abs/2013SSRv..177..119E}{177, 119}

\bibitem[{{Ewall-Wice} {et~al.}(2017){Ewall-Wice}, {Dillon}, {Liu}, \&
  {Hewitt}}]{ewallWice2017}
{Ewall-Wice}, A., {Dillon}, J.~S., {Liu}, A., \& {Hewitt}, J. 2017,
  \href{http://dx.doi.org/10.1093/mnras/stx1221}{\textcolor{magenta}{\mnras}},
  \href{http://adsabs.harvard.edu/abs/2017MNRAS.470.1849E}{470, 1849}

\bibitem[{{Fan} {et~al.}(2006){Fan}, {Carilli}, \& {Keating}}]{fan2006rev}
{Fan}, X., {Carilli}, C.~L., \& {Keating}, B. 2006,
  \href{http://dx.doi.org/10.1146/annurev.astro.44.051905.092514}{\textcolor{magenta}{\araa}},
  \href{http://adsabs.harvard.edu/abs/2006ARA%26A..44..415F}{44, 415}

\bibitem[{{Farnsworth} {et~al.}(2013){Farnsworth}, {Rudnick}, {Brown}, \&
  {Brunetti}}]{farnsworth2013}
{Farnsworth}, D., {Rudnick}, L., {Brown}, S., \& {Brunetti}, G. 2013,
  \href{http://dx.doi.org/10.1088/0004-637X/779/2/189}{\textcolor{magenta}{\apj}},
  \href{http://adsabs.harvard.edu/abs/2013ApJ...779..189F}{779, 189}

\bibitem[{{Feretti} {et~al.}(2001){Feretti}, {Fusco-Femiano}, {Giovannini}, \&
  {Govoni}}]{feretti2001}
{Feretti}, L., {Fusco-Femiano}, R., {Giovannini}, G., \& {Govoni}, F. 2001,
  \href{http://dx.doi.org/10.1051/0004-6361:20010581}{\textcolor{magenta}{\aap}},
  \href{http://adsabs.harvard.edu/abs/2001A%26A...373..106F}{373, 106}

\bibitem[{{Feretti} {et~al.}(1997){Feretti}, {Giovannini}, \&
  {B{\"o}hringer}}]{feretti1997}
{Feretti}, L., {Giovannini}, G., \& {B{\"o}hringer}, H. 1997,
  \href{http://dx.doi.org/10.1016/S1384-1076(97)00034-1}{\textcolor{magenta}{\na}},
  \href{http://adsabs.harvard.edu/abs/1997NewA....2..501F}{2, 501}

\bibitem[{{Feretti} {et~al.}(2012){Feretti}, {Giovannini}, {Govoni}, \&
  {Murgia}}]{feretti2012rev}
{Feretti}, L., {Giovannini}, G., {Govoni}, F., \& {Murgia}, M. 2012,
  \href{http://dx.doi.org/10.1007/s00159-012-0054-z}{\textcolor{magenta}{\aapr}},
  \href{http://adsabs.harvard.edu/abs/2012A%26ARv..20...54F}{20, 54}

\bibitem[{{Feretti} {et~al.}(2005){Feretti}, {Schuecker}, {B{\"o}hringer},
  {Govoni}, \& {Giovannini}}]{feretti2005}
{Feretti}, L., {Schuecker}, P., {B{\"o}hringer}, H., {Govoni}, F., \&
  {Giovannini}, G. 2005,
  \href{http://dx.doi.org/10.1051/0004-6361:20052808}{\textcolor{magenta}{\aap}},
  \href{http://adsabs.harvard.edu/abs/2005A%26A...444..157F}{444, 157}

\bibitem[{{Finkbeiner}(2003)}]{finkbeiner2003}
{Finkbeiner}, D.~P. 2003,
  \href{http://dx.doi.org/10.1086/374411}{\textcolor{magenta}{\apjs}},
  \href{http://adsabs.harvard.edu/abs/2003ApJS..146..407F}{146, 407}

\bibitem[{{Fujita} {et~al.}(2003){Fujita}, {Takizawa}, \&
  {Sarazin}}]{fujita2003}
{Fujita}, Y., {Takizawa}, M., \& {Sarazin}, C.~L. 2003,
  \href{http://dx.doi.org/10.1086/345599}{\textcolor{magenta}{\apj}},
  \href{http://adsabs.harvard.edu/abs/2003ApJ...584..190F}{584, 190}

\bibitem[{{Furlanetto}(2016)}]{furlanetto2016rev}
{Furlanetto}, S.~R. 2016,
  \href{http://dx.doi.org/10.1007/978-3-319-21957-8_9}{\textcolor{magenta}{Understanding
  the Epoch of Cosmic Reionization: Challenges and Progress}},
  \href{http://adsabs.harvard.edu/abs/2016ASSL..423..247F}{423, 247}

\bibitem[{{Furlanetto} {et~al.}(2006){Furlanetto}, {Oh}, \&
  {Briggs}}]{furlanetto2006rev}
{Furlanetto}, S.~R., {Oh}, S.~P., \& {Briggs}, F.~H. 2006,
  \href{http://dx.doi.org/10.1016/j.physrep.2006.08.002}{\textcolor{magenta}{\physrep}},
  \href{http://adsabs.harvard.edu/abs/2006PhR...433..181F}{433, 181}

\bibitem[{{Giacintucci} {et~al.}(2011){Giacintucci}, {Dallacasa}, {Venturi},
  {Brunetti}, {Cassano}, {Markevitch}, \& {Athreya}}]{giacintucci2011a}
{Giacintucci}, S., {Dallacasa}, D., {Venturi}, T., {et~al.} 2011,
  \href{http://dx.doi.org/10.1051/0004-6361/201117820}{\textcolor{magenta}{\aap}},
  \href{http://adsabs.harvard.edu/abs/2011A%26A...534A..57G}{534, A57}

\bibitem[{{Giacintucci} {et~al.}(2013){Giacintucci}, {Kale}, {Wik}, {Venturi},
  \& {Markevitch}}]{giacintucci2013}
{Giacintucci}, S., {Kale}, R., {Wik}, D.~R., {Venturi}, T., \& {Markevitch}, M.
  2013,
  \href{http://dx.doi.org/10.1088/0004-637X/766/1/18}{\textcolor{magenta}{\apj}},
  \href{http://adsabs.harvard.edu/abs/2013ApJ...766...18G}{766, 18}

\bibitem[{{Giacintucci} {et~al.}(2009{\natexlab{a}}){Giacintucci}, {Venturi},
  {Brunetti}, {Dallacasa}, {Mazzotta}, {Cassano}, {Bardelli}, \&
  {Zucca}}]{giacintucci2009}
{Giacintucci}, S., {Venturi}, T., {Brunetti}, G., {et~al.} 2009{\natexlab{a}},
  \href{http://dx.doi.org/10.1051/0004-6361/200912301}{\textcolor{magenta}{\aap}},
  \href{http://adsabs.harvard.edu/abs/2009A%26A...505...45G}{505, 45}

\bibitem[{{Giacintucci} {et~al.}(2009{\natexlab{b}}){Giacintucci}, {Venturi},
  {Cassano}, {Dallacasa}, \& {Brunetti}}]{giacintucci2011b}
{Giacintucci}, S., {Venturi}, T., {Cassano}, R., {Dallacasa}, D., \&
  {Brunetti}, G. 2009{\natexlab{b}},
  \href{http://dx.doi.org/10.1088/0004-637X/704/1/L54}{\textcolor{magenta}{\apjl}},
  \href{http://adsabs.harvard.edu/abs/2009ApJ...704L..54G}{704, L54}

\bibitem[{{Giardino} {et~al.}(2002){Giardino}, {Banday}, {G{\'o}rski},
  {Bennett}, {Jonas}, \& {Tauber}}]{giardino2002}
{Giardino}, G., {Banday}, A.~J., {G{\'o}rski}, K.~M., {et~al.} 2002,
  \href{http://dx.doi.org/10.1051/0004-6361:20020285}{\textcolor{magenta}{\aap}},
  \href{http://adsabs.harvard.edu/abs/2002A%26A...387...82G}{387, 82}

\bibitem[{{Giovannini} {et~al.}(2009){Giovannini}, {Bonafede}, {Feretti},
  {Govoni}, {Murgia}, {Ferrari}, \& {Monti}}]{giovannini2009}
{Giovannini}, G., {Bonafede}, A., {Feretti}, L., {et~al.} 2009,
  \href{http://dx.doi.org/10.1051/0004-6361/200912667}{\textcolor{magenta}{\aap}},
  \href{http://adsabs.harvard.edu/abs/2009A%26A...507.1257G}{507, 1257}

\bibitem[{{Giovannini} \& {Feretti}(2000)}]{giovannini2000}
{Giovannini}, G., \& {Feretti}, L. 2000,
  \href{http://dx.doi.org/10.1016/S1384-1076(00)00034-8}{\textcolor{magenta}{\na}},
  \href{http://adsabs.harvard.edu/abs/2000NewA....5..335G}{5, 335}

\bibitem[{{Giovannini} {et~al.}(2011){Giovannini}, {Feretti}, {Girardi},
  {Govoni}, {Murgia}, {Vacca}, \& {Bagchi}}]{giovannini2011}
{Giovannini}, G., {Feretti}, L., {Girardi}, M., {et~al.} 2011,
  \href{http://dx.doi.org/10.1051/0004-6361/201116930}{\textcolor{magenta}{\aap}},
  \href{http://adsabs.harvard.edu/abs/2011A%26A...530L...5G}{530, L5}

\bibitem[{{Gleser} {et~al.}(2008){Gleser}, {Nusser}, \& {Benson}}]{gleser2008}
{Gleser}, L., {Nusser}, A., \& {Benson}, A.~J. 2008,
  \href{http://dx.doi.org/10.1111/j.1365-2966.2008.13897.x}{\textcolor{magenta}{\mnras}},
  \href{http://adsabs.harvard.edu/abs/2008MNRAS.391..383G}{391, 383}

\bibitem[{{G{\'o}rski} {et~al.}(2005){G{\'o}rski}, {Hivon}, {Banday},
  {Wandelt}, {Hansen}, {Reinecke}, \& {Bartelmann}}]{gorski2005}
{G{\'o}rski}, K.~M., {Hivon}, E., {Banday}, A.~J., {et~al.} 2005,
  \href{http://dx.doi.org/10.1086/427976}{\textcolor{magenta}{\apj}},
  \href{http://adsabs.harvard.edu/abs/2005ApJ...622..759G}{622, 759}

\bibitem[{{Govoni} \& {Feretti}(2004)}]{govoni2004}
{Govoni}, F., \& {Feretti}, L. 2004,
  \href{http://dx.doi.org/10.1142/S0218271804005080}{\textcolor{magenta}{International
  Journal of Modern Physics D}},
  \href{http://adsabs.harvard.edu/abs/2004IJMPD..13.1549G}{13, 1549}

\bibitem[{{Govoni} {et~al.}(2001){Govoni}, {Feretti}, {Giovannini},
  {B{\"o}hringer}, {Reiprich}, \& {Murgia}}]{govoni2001}
{Govoni}, F., {Feretti}, L., {Giovannini}, G., {et~al.} 2001,
  \href{http://dx.doi.org/10.1051/0004-6361:20011016}{\textcolor{magenta}{\aap}},
  \href{http://adsabs.harvard.edu/abs/2001A%26A...376..803G}{376, 803}

\bibitem[{{Govoni} {et~al.}(2012){Govoni}, {Ferrari}, {Feretti}, {Vacca},
  {Murgia}, {Giovannini}, {Perley}, \& {Benoist}}]{govoni2012}
{Govoni}, F., {Ferrari}, C., {Feretti}, L., {et~al.} 2012,
  \href{http://dx.doi.org/10.1051/0004-6361/201219151}{\textcolor{magenta}{\aap}},
  \href{http://adsabs.harvard.edu/abs/2012A%26A...545A..74G}{545, A74}

\bibitem[{{Govoni} {et~al.}(2005){Govoni}, {Murgia}, {Feretti}, {Giovannini},
  {Dallacasa}, \& {Taylor}}]{govoni2005}
{Govoni}, F., {Murgia}, M., {Feretti}, L., {et~al.} 2005,
  \href{http://dx.doi.org/10.1051/0004-6361:200400113}{\textcolor{magenta}{\aap}},
  \href{http://adsabs.harvard.edu/abs/2005A%26A...430L...5G}{430, L5}

\bibitem[{{Govoni} {et~al.}(2011){Govoni}, {Murgia}, {Giovannini}, {Vacca}, \&
  {Bonafede}}]{govoni2011}
{Govoni}, F., {Murgia}, M., {Giovannini}, G., {Vacca}, V., \& {Bonafede}, A.
  2011,
  \href{http://dx.doi.org/10.1051/0004-6361/201016042}{\textcolor{magenta}{\aap}},
  \href{http://adsabs.harvard.edu/abs/2011A%26A...529A..69G}{529, A69}

\bibitem[{{Gunn} \& {Gott}(1972)}]{gunn1972}
{Gunn}, J.~E., \& {Gott}, III, J.~R. 1972,
  \href{http://dx.doi.org/10.1086/151605}{\textcolor{magenta}{\apj}},
  \href{http://adsabs.harvard.edu/abs/1972ApJ...176....1G}{176, 1}

\bibitem[{{Hogg}(1999)}]{hogg1999}
{Hogg}, D.~W. 1999,
  \href{http://adsabs.harvard.edu/abs/1999astro.ph..5116H}{ArXiv Astrophysics
  e-prints, astro-ph/9905116}

\bibitem[{{Intema} {et~al.}(2009){Intema}, {van der Tol}, {Cotton}, {Cohen},
  {van Bemmel}, \& {R{\"o}ttgering}}]{intema2009}
{Intema}, H.~T., {van der Tol}, S., {Cotton}, W.~D., {et~al.} 2009,
  \href{http://dx.doi.org/10.1051/0004-6361/200811094}{\textcolor{magenta}{\aap}},
  \href{http://adsabs.harvard.edu/abs/2009A%26A...501.1185I}{501, 1185}

\bibitem[{{Jeli{\'c}} {et~al.}(2008){Jeli{\'c}}, {Zaroubi}, {Labropoulos},
  {Thomas}, {Bernardi}, {Brentjens}, {de Bruyn}, {Ciardi}, {Harker},
  {Koopmans}, {Pandey}, {Schaye}, \& {Yatawatta}}]{jelic2008}
{Jeli{\'c}}, V., {Zaroubi}, S., {Labropoulos}, P., {et~al.} 2008,
  \href{http://dx.doi.org/10.1111/j.1365-2966.2008.13634.x}{\textcolor{magenta}{\mnras}},
  \href{http://adsabs.harvard.edu/abs/2008MNRAS.389.1319J}{389, 1319}

\bibitem[{{Jones} \& {Forman}(1984)}]{jones1984}
{Jones}, C., \& {Forman}, W. 1984,
  \href{http://dx.doi.org/10.1086/161591}{\textcolor{magenta}{\apj}},
  \href{http://adsabs.harvard.edu/abs/1984ApJ...276...38J}{276, 38}

\bibitem[{{Keshet} {et~al.}(2004){Keshet}, {Waxman}, \& {Loeb}}]{keshet2004rev}
{Keshet}, U., {Waxman}, E., \& {Loeb}, A. 2004,
  \href{http://dx.doi.org/10.1016/j.newar.2004.09.032}{\textcolor{magenta}{\nar}},
  \href{http://adsabs.harvard.edu/abs/2004NewAR..48.1119K}{48, 1119}

\bibitem[{{Kim} {et~al.}(1990){Kim}, {Kronberg}, {Dewdney}, \&
  {Landecker}}]{kim1990}
{Kim}, K.-T., {Kronberg}, P.~P., {Dewdney}, P.~E., \& {Landecker}, T.~L. 1990,
  \href{http://dx.doi.org/10.1086/168737}{\textcolor{magenta}{\apj}},
  \href{http://adsabs.harvard.edu/abs/1990ApJ...355...29K}{355, 29}

\bibitem[{{Kitayama} \& {Suto}(1996)}]{kitayama1996}
{Kitayama}, T., \& {Suto}, Y. 1996,
  \href{http://dx.doi.org/10.1086/177797}{\textcolor{magenta}{\apj}},
  \href{http://adsabs.harvard.edu/abs/1996ApJ...469..480K}{469, 480}

\bibitem[{{Knowles} {et~al.}(2016){Knowles}, {Intema}, {Baker}, {Bharadwaj},
  {Bond}, {Cress}, {Gupta}, {Hajian}, {Hilton}, {Hincks}, {Hlozek}, {Hughes},
  {Lindner}, {Marriage}, {Menanteau}, {Moodley}, {Niemack}, {Reese}, {Sievers},
  {Sif{\'o}n}, {Srianand}, \& {Wollack}}]{knowles2016}
{Knowles}, K., {Intema}, H.~T., {Baker}, A.~J., {et~al.} 2016,
  \href{http://dx.doi.org/10.1093/mnras/stw795}{\textcolor{magenta}{\mnras}},
  \href{http://adsabs.harvard.edu/abs/2016MNRAS.459.4240K}{459, 4240}

\bibitem[{{Koopmans} {et~al.}(2015){Koopmans}, {Pritchard}, {Mellema},
  {Aguirre}, {Ahn}, {Barkana}, {van Bemmel}, {Bernardi}, {Bonaldi}, {Briggs},
  {de Bruyn}, {Chang}, {Chapman}, {Chen}, {Ciardi}, {Dayal}, {Ferrara},
  {Fialkov}, {Fiore}, {Ichiki}, {Illiev}, {Inoue}, {Jelic}, {Jones}, {Lazio},
  {Maio}, {Majumdar}, {Mack}, {Mesinger}, {Morales}, {Parsons}, {Pen},
  {Santos}, {Schneider}, {Semelin}, {de Souza}, {Subrahmanyan}, {Takeuchi},
  {Vedantham}, {Wagg}, {Webster}, {Wyithe}, {Datta}, \&
  {Trott}}]{koopmans2015rev}
{Koopmans}, L., {Pritchard}, J., {Mellema}, G., {et~al.} 2015,
  \href{http://adsabs.harvard.edu/abs/2015aska.confE...1K}{Advancing
  Astrophysics with the Square Kilometre Array (AASKA14), 1}

\bibitem[{{Lacey} \& {Cole}(1993)}]{lacey1993}
{Lacey}, C., \& {Cole}, S. 1993,
  \href{http://dx.doi.org/10.1093/mnras/262.3.627}{\textcolor{magenta}{\mnras}},
  \href{http://adsabs.harvard.edu/abs/1993MNRAS.262..627L}{262, 627}

\bibitem[{{Large} {et~al.}(1959){Large}, {Mathewson}, \& {Haslam}}]{large1959}
{Large}, M.~I., {Mathewson}, D.~S., \& {Haslam}, C.~G.~T. 1959,
  \href{http://dx.doi.org/10.1038/1831663a0}{\textcolor{magenta}{\nat}},
  \href{http://adsabs.harvard.edu/abs/1959Natur.183.1663L}{183, 1663}

\bibitem[{{Liang} {et~al.}(2000){Liang}, {Hunstead}, {Birkinshaw}, \&
  {Andreani}}]{liang2000}
{Liang}, H., {Hunstead}, R.~W., {Birkinshaw}, M., \& {Andreani}, P. 2000,
  \href{http://dx.doi.org/10.1086/317223}{\textcolor{magenta}{\apj}},
  \href{http://adsabs.harvard.edu/abs/2000ApJ...544..686L}{544, 686}

\bibitem[{{Lindner} {et~al.}(2014){Lindner}, {Baker}, {Hughes}, {Battaglia},
  {Gupta}, {Knowles}, {Marriage}, {Menanteau}, {Moodley}, {Reese}, \&
  {Srianand}}]{lindner2014}
{Lindner}, R.~R., {Baker}, A.~J., {Hughes}, J.~P., {et~al.} 2014,
  \href{http://dx.doi.org/10.1088/0004-637X/786/1/49}{\textcolor{magenta}{\apj}},
  \href{http://adsabs.harvard.edu/abs/2014ApJ...786...49L}{786, 49}

\bibitem[{{Liu} {et~al.}(2014){Liu}, {Parsons}, \& {Trott}}]{liu2014}
{Liu}, A., {Parsons}, A.~R., \& {Trott}, C.~M. 2014,
  \href{http://dx.doi.org/10.1103/PhysRevD.90.023018}{\textcolor{magenta}{\prd}},
  \href{http://adsabs.harvard.edu/abs/2014PhRvD..90b3018L}{90, 023018}

\bibitem[{{Liu} \& {Tegmark}(2012)}]{liu2012}
{Liu}, A., \& {Tegmark}, M. 2012,
  \href{http://dx.doi.org/10.1111/j.1365-2966.2011.19989.x}{\textcolor{magenta}{\mnras}},
  \href{http://adsabs.harvard.edu/abs/2012MNRAS.419.3491L}{419, 3491}

\bibitem[{{Liu} {et~al.}(2009){Liu}, {Tegmark}, \& {Zaldarriaga}}]{liu2009ps}
{Liu}, A., {Tegmark}, M., \& {Zaldarriaga}, M. 2009,
  \href{http://dx.doi.org/10.1111/j.1365-2966.2009.14426.x}{\textcolor{magenta}{\mnras}},
  \href{http://adsabs.harvard.edu/abs/2009MNRAS.394.1575L}{394, 1575}

\bibitem[{{Macario} {et~al.}(2013){Macario}, {Venturi}, {Intema}, {Dallacasa},
  {Brunetti}, {Cassano}, {Giacintucci}, {Ferrari}, {Ishwara-Chandra}, \&
  {Athreya}}]{macario2013}
{Macario}, G., {Venturi}, T., {Intema}, H.~T., {et~al.} 2013,
  \href{http://dx.doi.org/10.1051/0004-6361/201220667}{\textcolor{magenta}{\aap}},
  \href{http://adsabs.harvard.edu/abs/2013A%26A...551A.141M}{551, A141}

\bibitem[{{Macario} {et~al.}(2014){Macario}, {Intema}, {Ferrari}, {Bourdin},
  {Giacintucci}, {Venturi}, {Mazzotta}, {Bartalucci}, {Johnston-Hollitt},
  {Cassano}, {Dallacasa}, {Pratt}, {Kale}, \& {Brown}}]{macario2014}
{Macario}, G., {Intema}, H.~T., {Ferrari}, C., {et~al.} 2014,
  \href{http://dx.doi.org/10.1051/0004-6361/201323275}{\textcolor{magenta}{\aap}},
  \href{http://adsabs.harvard.edu/abs/2014A%26A...565A..13M}{565, A13}

\bibitem[{{Martinez Aviles} {et~al.}(2016){Martinez Aviles}, {Ferrari},
  {Johnston-Hollitt}, {Pratley}, {Macario}, {Venturi}, {Brunetti}, {Cassano},
  {Dallacasa}, {Intema}, {Giacintucci}, {Hurier}, {Aghanim}, {Douspis}, \&
  {Langer}}]{martinezAviles2016}
{Martinez Aviles}, G., {Ferrari}, C., {Johnston-Hollitt}, M., {et~al.} 2016,
  \href{http://dx.doi.org/10.1051/0004-6361/201628788}{\textcolor{magenta}{\aap}},
  \href{http://adsabs.harvard.edu/abs/2016A%26A...595A.116M}{595, A116}

\bibitem[{{Mellema} {et~al.}(2013){Mellema}, {Koopmans}, {Abdalla}, {Bernardi},
  {Ciardi}, {Daiboo}, {de Bruyn}, {Datta}, {Falcke}, {Ferrara}, {Iliev},
  {Iocco}, {Jeli{\'c}}, {Jensen}, {Joseph}, {Labroupoulos}, {Meiksin},
  {Mesinger}, {Offringa}, {Pandey}, {Pritchard}, {Santos}, {Schwarz},
  {Semelin}, {Vedantham}, {Yatawatta}, \& {Zaroubi}}]{mellema2013rev}
{Mellema}, G., {Koopmans}, L.~V.~E., {Abdalla}, F.~A., {et~al.} 2013,
  \href{http://dx.doi.org/10.1007/s10686-013-9334-5}{\textcolor{magenta}{Experimental
  Astronomy}}, \href{http://adsabs.harvard.edu/abs/2013ExA....36..235M}{36,
  235}

\bibitem[{{Mesinger} {et~al.}(2016){Mesinger}, {Greig}, \&
  {Sobacchi}}]{mesinger2016}
{Mesinger}, A., {Greig}, B., \& {Sobacchi}, E. 2016,
  \href{http://dx.doi.org/10.1093/mnras/stw831}{\textcolor{magenta}{\mnras}},
  \href{http://adsabs.harvard.edu/abs/2016MNRAS.459.2342M}{459, 2342}

\bibitem[{{Miniati}(2015)}]{miniati2015}
{Miniati}, F. 2015,
  \href{http://dx.doi.org/10.1088/0004-637X/800/1/60}{\textcolor{magenta}{\apj}},
  \href{http://adsabs.harvard.edu/abs/2015ApJ...800...60M}{800, 60}

\bibitem[{{Miniati} \& {Beresnyak}(2015)}]{miniati2015ss}
{Miniati}, F., \& {Beresnyak}, A. 2015,
  \href{http://dx.doi.org/10.1038/nature14552}{\textcolor{magenta}{\nat}},
  \href{http://adsabs.harvard.edu/abs/2015Natur.523...59M}{523, 59}

\bibitem[{{Mitchell} {et~al.}(2008){Mitchell}, {Greenhill}, {Wayth}, {Sault},
  {Lonsdale}, {Cappallo}, {Morales}, \& {Ord}}]{mitchell2008}
{Mitchell}, D.~A., {Greenhill}, L.~J., {Wayth}, R.~B., {et~al.} 2008,
  \href{http://dx.doi.org/10.1109/JSTSP.2008.2005327}{\textcolor{magenta}{IEEE
  Journal of Selected Topics in Signal Processing}},
  \href{http://adsabs.harvard.edu/abs/2008ISTSP...2..707M}{2, 707}

\bibitem[{{Morales} {et~al.}(2006){Morales}, {Bowman}, {Cappallo}, {Hewitt}, \&
  {Lonsdale}}]{morales2006}
{Morales}, M.~F., {Bowman}, J.~D., {Cappallo}, R., {Hewitt}, J.~N., \&
  {Lonsdale}, C.~J. 2006,
  \href{http://dx.doi.org/10.1016/j.newar.2005.11.033}{\textcolor{magenta}{\nar}},
  \href{http://adsabs.harvard.edu/abs/2006NewAR..50..173M}{50, 173}

\bibitem[{{Morales} {et~al.}(2012){Morales}, {Hazelton}, {Sullivan}, \&
  {Beardsley}}]{morales2012}
{Morales}, M.~F., {Hazelton}, B., {Sullivan}, I., \& {Beardsley}, A. 2012,
  \href{http://dx.doi.org/10.1088/0004-637X/752/2/137}{\textcolor{magenta}{\apj}},
  \href{http://adsabs.harvard.edu/abs/2012ApJ...752..137M}{752, 137}

\bibitem[{{Morales} \& {Hewitt}(2004)}]{morales2004}
{Morales}, M.~F., \& {Hewitt}, J. 2004,
  \href{http://dx.doi.org/10.1086/424437}{\textcolor{magenta}{\apj}},
  \href{http://adsabs.harvard.edu/abs/2004ApJ...615....7M}{615, 7}

\bibitem[{{Morales} \& {Wyithe}(2010)}]{morales2010rev}
{Morales}, M.~F., \& {Wyithe}, J.~S.~B. 2010,
  \href{http://dx.doi.org/10.1146/annurev-astro-081309-130936}{\textcolor{magenta}{\araa}},
  \href{http://adsabs.harvard.edu/abs/2010ARA%26A..48..127M}{48, 127}

\bibitem[{{Mort} {et~al.}(2017){Mort}, {Dulwich}, {Razavi-Ghods}, {de Lera
  Acedo}, \& {Grainge}}]{mort2017}
{Mort}, B., {Dulwich}, F., {Razavi-Ghods}, N., {de Lera Acedo}, E., \&
  {Grainge}, K. 2017,
  \href{http://dx.doi.org/10.1093/mnras/stw2814}{\textcolor{magenta}{\mnras}},
  \href{http://adsabs.harvard.edu/abs/2017MNRAS.465.3680M}{465, 3680}

\bibitem[{{Mort} {et~al.}(2010){Mort}, {Dulwich}, {Salvini}, {Adami}, \&
  {Jones}}]{mort2010}
{Mort}, B.~J., {Dulwich}, F., {Salvini}, S., {Adami}, K.~Z., \& {Jones}, M.~E.
  2010, in 2010 IEEE International Symposium on Phased Array Systems and
  Technology, 690--694

\bibitem[{{Murgia} {et~al.}(2010){Murgia}, {Govoni}, {Feretti}, \&
  {Giovannini}}]{murgia2010}
{Murgia}, M., {Govoni}, F., {Feretti}, L., \& {Giovannini}, G. 2010,
  \href{http://dx.doi.org/10.1051/0004-6361/200913414}{\textcolor{magenta}{\aap}},
  \href{http://adsabs.harvard.edu/abs/2010A%26A...509A..86M}{509, A86}

\bibitem[{{Murgia} {et~al.}(2009){Murgia}, {Govoni}, {Markevitch}, {Feretti},
  {Giovannini}, {Taylor}, \& {Carretti}}]{murgia2009}
{Murgia}, M., {Govoni}, F., {Markevitch}, M., {et~al.} 2009,
  \href{http://dx.doi.org/10.1051/0004-6361/200911659}{\textcolor{magenta}{\aap}},
  \href{http://adsabs.harvard.edu/abs/2009A%26A...499..679M}{499, 679}

\bibitem[{{Murray} {et~al.}(2013){Murray}, {Power}, \& {Robotham}}]{murray2013}
{Murray}, S.~G., {Power}, C., \& {Robotham}, A.~S.~G. 2013,
  \href{http://dx.doi.org/10.1016/j.ascom.2013.11.001}{\textcolor{magenta}{Astronomy
  and Computing}},
  \href{http://adsabs.harvard.edu/abs/2013A%26C.....3...23M}{3, 23}

\bibitem[{{Murray} {et~al.}(2017){Murray}, {Trott}, \& {Jordan}}]{murray2017}
{Murray}, S.~G., {Trott}, C.~M., \& {Jordan}, C.~H. 2017,
  \href{http://dx.doi.org/10.3847/1538-4357/aa7d0a}{\textcolor{magenta}{\apj}},
  \href{http://adsabs.harvard.edu/abs/2017ApJ...845....7M}{845, 7}

\bibitem[{{Navarro} {et~al.}(1997){Navarro}, {Frenk}, \& {White}}]{navarro1997}
{Navarro}, J.~F., {Frenk}, C.~S., \& {White}, S.~D.~M. 1997,
  \href{http://dx.doi.org/10.1086/304888}{\textcolor{magenta}{\apj}},
  \href{http://adsabs.harvard.edu/abs/1997ApJ...490..493N}{490, 493}

\bibitem[{{Offringa} \& {Smirnov}(2017)}]{offringa2017}
{Offringa}, A.~R., \& {Smirnov}, O. 2017,
  \href{http://dx.doi.org/10.1093/mnras/stx1547}{\textcolor{magenta}{\mnras}},
  \href{http://adsabs.harvard.edu/abs/2017MNRAS.471..301O}{471, 301}

\bibitem[{{Offringa} {et~al.}(2014){Offringa}, {McKinley}, {Hurley-Walker},
  {Briggs}, {Wayth}, {Kaplan}, {Bell}, {Feng}, {Neben}, {Hughes}, {Rhee},
  {Murphy}, {Bhat}, {Bernardi}, {Bowman}, {Cappallo}, {Corey}, {Deshpande},
  {Emrich}, {Ewall-Wice}, {Gaensler}, {Goeke}, {Greenhill}, {Hazelton},
  {Hindson}, {Johnston-Hollitt}, {Jacobs}, {Kasper}, {Kratzenberg}, {Lenc},
  {Lonsdale}, {Lynch}, {McWhirter}, {Mitchell}, {Morales}, {Morgan},
  {Kudryavtseva}, {Oberoi}, {Ord}, {Pindor}, {Procopio}, {Prabu}, {Riding},
  {Roshi}, {Shankar}, {Srivani}, {Subrahmanyan}, {Tingay}, {Waterson},
  {Webster}, {Whitney}, {Williams}, \& {Williams}}]{offringa2014}
{Offringa}, A.~R., {McKinley}, B., {Hurley-Walker}, N., {et~al.} 2014,
  \href{http://dx.doi.org/10.1093/mnras/stu1368}{\textcolor{magenta}{\mnras}},
  \href{http://adsabs.harvard.edu/abs/2014MNRAS.444..606O}{444, 606}

\bibitem[{{Pandey-Pommier} {et~al.}(2015){Pandey-Pommier}, {van Weeren},
  {Ogrean}, {Combes}, {Johnston-Hollitt}, {Richard}, {Bagchi}, {Guiderdoni},
  {Jacob}, {Dwarakanath}, {Narasimha}, {Edge}, {Ebeling}, {Clarke}, \&
  {Mroczkowski}}]{pandeyPommier2015}
{Pandey-Pommier}, M., {van Weeren}, R.~J., {Ogrean}, G.~A., {et~al.} 2015, in
  SF2A-2015: Proceedings of the Annual meeting of the French Society of
  Astronomy and Astrophysics, ed. F.~{Martins}, S.~{Boissier}, V.~{Buat},
  L.~{Cambr{\'e}sy}, \& P.~{Petit}, 247--252

\bibitem[{{Park} \& {Petrosian}(1996)}]{park1996}
{Park}, B.~T., \& {Petrosian}, V. 1996,
  \href{http://dx.doi.org/10.1086/192278}{\textcolor{magenta}{\apjs}},
  \href{http://adsabs.harvard.edu/abs/1996ApJS..103..255P}{103, 255}

\bibitem[{{Parsons} {et~al.}(2010){Parsons}, {Backer}, {Foster}, {Wright},
  {Bradley}, {Gugliucci}, {Parashare}, {Benoit}, {Aguirre}, {Jacobs},
  {Carilli}, {Herne}, {Lynch}, {Manley}, \& {Werthimer}}]{parsons2010}
{Parsons}, A.~R., {Backer}, D.~C., {Foster}, G.~S., {et~al.} 2010,
  \href{http://dx.doi.org/10.1088/0004-6256/139/4/1468}{\textcolor{magenta}{\aj}},
  \href{http://adsabs.harvard.edu/abs/2010AJ....139.1468P}{139, 1468}

\bibitem[{{Peebles}(1980)}]{peebles1980}
{Peebles}, P.~J.~E. 1980, The Large-scale Structure of the Universe (Princeton
  University Press)

\bibitem[{{Petrosian}(2001)}]{petrosian2001}
{Petrosian}, V. 2001,
  \href{http://dx.doi.org/10.1086/321557}{\textcolor{magenta}{\apj}},
  \href{http://adsabs.harvard.edu/abs/2001ApJ...557..560P}{557, 560}

\bibitem[{{Pindor} {et~al.}(2011){Pindor}, {Wyithe}, {Mitchell}, {Ord},
  {Wayth}, \& {Greenhill}}]{pindor2011}
{Pindor}, B., {Wyithe}, J.~S.~B., {Mitchell}, D.~A., {et~al.} 2011,
  \href{http://dx.doi.org/10.1071/AS10023}{\textcolor{magenta}{\pasa}},
  \href{http://adsabs.harvard.edu/abs/2011PASA...28...46P}{28, 46}

\bibitem[{{Pinzke} {et~al.}(2017){Pinzke}, {Oh}, \& {Pfrommer}}]{pinzke2017}
{Pinzke}, A., {Oh}, S.~P., \& {Pfrommer}, C. 2017,
  \href{http://dx.doi.org/10.1093/mnras/stw3024}{\textcolor{magenta}{\mnras}},
  \href{http://adsabs.harvard.edu/abs/2017MNRAS.465.4800P}{465, 4800}

\bibitem[{{Press} \& {Schechter}(1974)}]{press1974}
{Press}, W.~H., \& {Schechter}, P. 1974,
  \href{http://dx.doi.org/10.1086/152650}{\textcolor{magenta}{\apj}},
  \href{http://adsabs.harvard.edu/abs/1974ApJ...187..425P}{187, 425}

\bibitem[{{Randall} {et~al.}(2002){Randall}, {Sarazin}, \&
  {Ricker}}]{randall2002}
{Randall}, S.~W., {Sarazin}, C.~L., \& {Ricker}, P.~M. 2002,
  \href{http://dx.doi.org/10.1086/342239}{\textcolor{magenta}{\apj}},
  \href{http://adsabs.harvard.edu/abs/2002ApJ...577..579R}{577, 579}

\bibitem[{{Reid} {et~al.}(1999){Reid}, {Hunstead}, {Lemonon}, \&
  {Pierre}}]{reid1999}
{Reid}, A.~D., {Hunstead}, R.~W., {Lemonon}, L., \& {Pierre}, M.~M. 1999,
  \href{http://dx.doi.org/10.1046/j.1365-8711.1999.02177.x}{\textcolor{magenta}{\mnras}},
  \href{http://adsabs.harvard.edu/abs/1999MNRAS.302..571R}{302, 571}

\bibitem[{{Remazeilles} {et~al.}(2015){Remazeilles}, {Dickinson}, {Banday},
  {Bigot-Sazy}, \& {Ghosh}}]{remazeilles2015}
{Remazeilles}, M., {Dickinson}, C., {Banday}, A.~J., {Bigot-Sazy}, M.-A., \&
  {Ghosh}, T. 2015,
  \href{http://dx.doi.org/10.1093/mnras/stv1274}{\textcolor{magenta}{\mnras}},
  \href{http://adsabs.harvard.edu/abs/2015MNRAS.451.4311R}{451, 4311}

\bibitem[{{Rybicki} \& {Lightman}(1979)}]{rybicki1979}
{Rybicki}, G.~B., \& {Lightman}, A.~P. 1979, Radiative Processes in
  Astrophysics, A Wiley-Interscience publication (Wiley)

\bibitem[{{Ryu} {et~al.}(2008){Ryu}, {Kang}, {Cho}, \& {Das}}]{ryu2008}
{Ryu}, D., {Kang}, H., {Cho}, J., \& {Das}, S. 2008,
  \href{http://dx.doi.org/10.1126/science.1154923}{\textcolor{magenta}{Science}},
  \href{http://adsabs.harvard.edu/abs/2008Sci...320..909R}{320, 909}

\bibitem[{{Sanderson} \& {Ponman}(2003)}]{sanderson2003}
{Sanderson}, A.~J.~R., \& {Ponman}, T.~J. 2003,
  \href{http://dx.doi.org/10.1046/j.1365-2966.2003.07040.x}{\textcolor{magenta}{\mnras}},
  \href{http://adsabs.harvard.edu/abs/2003MNRAS.345.1241S}{345, 1241}

\bibitem[{{Sarazin}(1999)}]{sarazin1999}
{Sarazin}, C.~L. 1999,
  \href{http://dx.doi.org/10.1086/307501}{\textcolor{magenta}{\apj}},
  \href{http://adsabs.harvard.edu/abs/1999ApJ...520..529S}{520, 529}

\bibitem[{{Sarazin}(2002)}]{sarazin2002}
{Sarazin}, C.~L. 2002, in Astrophysics and Space Science Library, Vol. 272,
  Merging Processes in Galaxy Clusters, ed. L.~{Feretti}, I.~M. {Gioia}, \&
  G.~{Giovannini}, 1--38

\bibitem[{{Schlickeiser}(2002)}]{schlickeiser2002}
{Schlickeiser}, R. 2002, Cosmic Ray Astrophysics (Springer)

\bibitem[{{Shakouri} {et~al.}(2016){Shakouri}, {Johnston-Hollitt}, \&
  {Pratt}}]{shakouri2016}
{Shakouri}, S., {Johnston-Hollitt}, M., \& {Pratt}, G.~W. 2016,
  \href{http://dx.doi.org/10.1093/mnras/stw812}{\textcolor{magenta}{\mnras}},
  \href{http://adsabs.harvard.edu/abs/2016MNRAS.459.2525S}{459, 2525}

\bibitem[{{Shaver} {et~al.}(1999){Shaver}, {Windhorst}, {Madau}, \& {de
  Bruyn}}]{shaver1999}
{Shaver}, P.~A., {Windhorst}, R.~A., {Madau}, P., \& {de Bruyn}, A.~G. 1999,
  \aap, \href{http://adsabs.harvard.edu/abs/1999A%26A...345..380S}{345, 380}

\bibitem[{{Smirnov}(2011)}]{smirnov2011}
{Smirnov}, O.~M. 2011,
  \href{http://dx.doi.org/10.1051/0004-6361/201016082}{\textcolor{magenta}{\aap}},
  \href{http://adsabs.harvard.edu/abs/2011A%26A...527A.106S}{527, A106}

\bibitem[{{Smirnov} {et~al.}(2012){Smirnov}, {Frank}, {Theron}, \&
  {Wood}}]{smirnov2012}
{Smirnov}, O.~M., {Frank}, B.~S., {Theron}, I.~P., \& {Wood}, I.~H. 2012, in
  2012 International Conference on Electromagnetics in Advanced Applications,
  586--590

\bibitem[{{Spinelli} {et~al.}(2018){Spinelli}, {Bernardi}, \&
  {Santos}}]{spinelli2018}
{Spinelli}, M., {Bernardi}, G., \& {Santos}, M.~G. 2018,
  \href{http://dx.doi.org/10.1093/mnras/sty1457}{\textcolor{magenta}{\mnras}},
  \href{https://ui.adsabs.harvard.edu/abs/2018MNRAS.479..275S}{479, 275}

\bibitem[{{Thyagarajan} {et~al.}(2013){Thyagarajan}, {Udaya Shankar},
  {Subrahmanyan}, {Arcus}, {Bernardi}, {Bowman}, {Briggs}, {Bunton},
  {Cappallo}, {Corey}, {deSouza}, {Emrich}, {Gaensler}, {Goeke}, {Greenhill},
  {Hazelton}, {Herne}, {Hewitt}, {Johnston-Hollitt}, {Kaplan}, {Kasper},
  {Kincaid}, {Koenig}, {Kratzenberg}, {Lonsdale}, {Lynch}, {McWhirter},
  {Mitchell}, {Morales}, {Morgan}, {Oberoi}, {Ord}, {Pathikulangara},
  {Remillard}, {Rogers}, {Anish Roshi}, {Salah}, {Sault}, {Srivani}, {Stevens},
  {Thiagaraj}, {Tingay}, {Wayth}, {Waterson}, {Webster}, {Whitney}, {Williams},
  {Williams}, \& {Wyithe}}]{thyagarajan2013}
{Thyagarajan}, N., {Udaya Shankar}, N., {Subrahmanyan}, R., {et~al.} 2013,
  \href{http://dx.doi.org/10.1088/0004-637X/776/1/6}{\textcolor{magenta}{\apj}},
  \href{http://adsabs.harvard.edu/abs/2013ApJ...776....6T}{776, 6}

\bibitem[{{Tingay} {et~al.}(2013){Tingay}, {Goeke}, {Bowman}, {Emrich}, {Ord},
  {Mitchell}, {Morales}, {Booler}, {Crosse}, {Wayth}, {Lonsdale}, {Tremblay},
  {Pallot}, {Colegate}, {Wicenec}, {Kudryavtseva}, {Arcus}, {Barnes},
  {Bernardi}, {Briggs}, {Burns}, {Bunton}, {Cappallo}, {Corey}, {Deshpande},
  {Desouza}, {Gaensler}, {Greenhill}, {Hall}, {Hazelton}, {Herne}, {Hewitt},
  {Johnston-Hollitt}, {Kaplan}, {Kasper}, {Kincaid}, {Koenig}, {Kratzenberg},
  {Lynch}, {Mckinley}, {Mcwhirter}, {Morgan}, {Oberoi}, {Pathikulangara},
  {Prabu}, {Remillard}, {Rogers}, {Roshi}, {Salah}, {Sault}, {Udaya-Shankar},
  {Schlagenhaufer}, {Srivani}, {Stevens}, {Subrahmanyan}, {Waterson},
  {Webster}, {Whitney}, {Williams}, {Williams}, \& {Wyithe}}]{tingay2013}
{Tingay}, S.~J., {Goeke}, R., {Bowman}, J.~D., {et~al.} 2013,
  \href{http://dx.doi.org/10.1017/pasa.2012.007}{\textcolor{magenta}{\pasa}},
  \href{http://adsabs.harvard.edu/abs/2013PASA...30....7T}{30, e007}

\bibitem[{{Tormen} {et~al.}(2004){Tormen}, {Moscardini}, \&
  {Yoshida}}]{tormen2004}
{Tormen}, G., {Moscardini}, L., \& {Yoshida}, N. 2004,
  \href{http://dx.doi.org/10.1111/j.1365-2966.2004.07736.x}{\textcolor{magenta}{\mnras}},
  \href{http://adsabs.harvard.edu/abs/2004MNRAS.350.1397T}{350, 1397}

\bibitem[{{Trott} \& {Tingay}(2015)}]{trott2015}
{Trott}, C.~M., \& {Tingay}, S.~J. 2015,
  \href{http://dx.doi.org/10.1088/0004-637X/814/1/27}{\textcolor{magenta}{\apj}},
  \href{http://adsabs.harvard.edu/abs/2015ApJ...814...27T}{814, 27}

\bibitem[{{Vacca} {et~al.}(2011){Vacca}, {Govoni}, {Murgia}, {Giovannini},
  {Feretti}, {Tugnoli}, {Verheijen}, \& {Taylor}}]{vacca2011}
{Vacca}, V., {Govoni}, F., {Murgia}, M., {et~al.} 2011,
  \href{http://dx.doi.org/10.1051/0004-6361/201117607}{\textcolor{magenta}{\aap}},
  \href{http://adsabs.harvard.edu/abs/2011A%26A...535A..82V}{535, A82}

\bibitem[{{van Haarlem} {et~al.}(2013){van Haarlem}, {Wise}, {Gunst}, {Heald},
  {McKean}, {Hessels}, {de Bruyn}, {Nijboer}, {Swinbank}, {Fallows},
  {Brentjens}, {Nelles}, {Beck}, {Falcke}, {Fender}, {H{\"o}randel},
  {Koopmans}, {Mann}, {Miley}, {R{\"o}ttgering}, {Stappers}, {Wijers},
  {Zaroubi}, {van den Akker}, {Alexov}, {Anderson}, {Anderson}, {van Ardenne},
  {Arts}, {Asgekar}, {Avruch}, {Batejat}, {B{\"a}hren}, {Bell}, {Bell}, {van
  Bemmel}, {Bennema}, {Bentum}, {Bernardi}, {Best}, {B{\^i}rzan}, {Bonafede},
  {Boonstra}, {Braun}, {Bregman}, {Breitling}, {van de Brink}, {Broderick},
  {Broekema}, {Brouw}, {Br{\"u}ggen}, {Butcher}, {van Cappellen}, {Ciardi},
  {Coenen}, {Conway}, {Coolen}, {Corstanje}, {Damstra}, {Davies}, {Deller},
  {Dettmar}, {van Diepen}, {Dijkstra}, {Donker}, {Doorduin}, {Dromer}, {Drost},
  {van Duin}, {Eisl{\"o}ffel}, {van Enst}, {Ferrari}, {Frieswijk}, {Gankema},
  {Garrett}, {de Gasperin}, {Gerbers}, {de Geus}, {Grie{\ss}meier}, {Grit},
  {Gruppen}, {Hamaker}, {Hassall}, {Hoeft}, {Holties}, {Horneffer}, {van der
  Horst}, {van Houwelingen}, {Huijgen}, {Iacobelli}, {Intema}, {Jackson},
  {Jelic}, {de Jong}, {Juette}, {Kant}, {Karastergiou}, {Koers}, {Kollen},
  {Kondratiev}, {Kooistra}, {Koopman}, {Koster}, {Kuniyoshi}, {Kramer},
  {Kuper}, {Lambropoulos}, {Law}, {van Leeuwen}, {Lemaitre}, {Loose}, {Maat},
  {Macario}, {Markoff}, {Masters}, {McFadden}, {McKay-Bukowski}, {Meijering},
  {Meulman}, {Mevius}, {Middelberg}, {Millenaar}, {Miller-Jones}, {Mohan},
  {Mol}, {Morawietz}, {Morganti}, {Mulcahy}, {Mulder}, {Munk}, {Nieuwenhuis},
  {van Nieuwpoort}, {Noordam}, {Norden}, {Noutsos}, {Offringa}, {Olofsson},
  {Omar}, {Orr{\'u}}, {Overeem}, {Paas}, {Pandey-Pommier}, {Pandey}, {Pizzo},
  {Polatidis}, {Rafferty}, {Rawlings}, {Reich}, {de Reijer}, {Reitsma},
  {Renting}, {Riemers}, {Rol}, {Romein}, {Roosjen}, {Ruiter}, {Scaife}, {van
  der Schaaf}, {Scheers}, {Schellart}, {Schoenmakers}, {Schoonderbeek},
  {Serylak}, {Shulevski}, {Sluman}, {Smirnov}, {Sobey}, {Spreeuw}, {Steinmetz},
  {Sterks}, {Stiepel}, {Stuurwold}, {Tagger}, {Tang}, {Tasse}, {Thomas},
  {Thoudam}, {Toribio}, {van der Tol}, {Usov}, {van Veelen}, {van der Veen},
  {ter Veen}, {Verbiest}, {Vermeulen}, {Vermaas}, {Vocks}, {Vogt}, {de Vos},
  {van der Wal}, {van Weeren}, {Weggemans}, {Weltevrede}, {White}, {Wijnholds},
  {Wilhelmsson}, {Wucknitz}, {Yatawatta}, {Zarka}, {Zensus}, \& {van
  Zwieten}}]{vanHaarlem2013}
{van Haarlem}, M.~P., {Wise}, M.~W., {Gunst}, A.~W., {et~al.} 2013,
  \href{http://dx.doi.org/10.1051/0004-6361/201220873}{\textcolor{magenta}{\aap}},
  \href{http://adsabs.harvard.edu/abs/2013A%26A...556A...2V}{556, A2}

\bibitem[{{van Weeren} {et~al.}(2012){van Weeren}, {Bonafede}, {Ebeling},
  {Edge}, {Br{\"u}ggen}, {Giovannini}, {Hoeft}, \&
  {R{\"o}ttgering}}]{vanWeeren2012}
{van Weeren}, R.~J., {Bonafede}, A., {Ebeling}, H., {et~al.} 2012,
  \href{http://dx.doi.org/10.1111/j.1745-3933.2012.01301.x}{\textcolor{magenta}{\mnras}},
  \href{http://adsabs.harvard.edu/abs/2012MNRAS.425L..36V}{425, L36}

\bibitem[{{van Weeren} {et~al.}(2011){van Weeren}, {Br{\"u}ggen},
  {R{\"o}ttgering}, {Hoeft}, {Nuza}, \& {Intema}}]{vanWeeren2011}
{van Weeren}, R.~J., {Br{\"u}ggen}, M., {R{\"o}ttgering}, H.~J.~A., {et~al.}
  2011,
  \href{http://dx.doi.org/10.1051/0004-6361/201117149}{\textcolor{magenta}{\aap}},
  \href{http://adsabs.harvard.edu/abs/2011A%26A...533A..35V}{533, A35}

\bibitem[{{van Weeren} {et~al.}(2009){van Weeren}, {R{\"o}ttgering},
  {Br{\"u}ggen}, \& {Cohen}}]{vanWeeren2009}
{van Weeren}, R.~J., {R{\"o}ttgering}, H.~J.~A., {Br{\"u}ggen}, M., \& {Cohen},
  A. 2009,
  \href{http://dx.doi.org/10.1051/0004-6361/200912528}{\textcolor{magenta}{\aap}},
  \href{http://adsabs.harvard.edu/abs/2009A%26A...505..991V}{505, 991}

\bibitem[{{van Weeren} {et~al.}(2013){van Weeren}, {Fogarty}, {Jones},
  {Forman}, {Clarke}, {Br{\"u}ggen}, {Kraft}, {Lal}, {Murray}, \&
  {R{\"o}ttgering}}]{vanWeeren2013}
{van Weeren}, R.~J., {Fogarty}, K., {Jones}, C., {et~al.} 2013,
  \href{http://dx.doi.org/10.1088/0004-637X/769/2/101}{\textcolor{magenta}{\apj}},
  \href{http://adsabs.harvard.edu/abs/2013ApJ...769..101V}{769, 101}

\bibitem[{{van Weeren} {et~al.}(2014){van Weeren}, {Intema}, {Lal}, {Bonafede},
  {Jones}, {Forman}, {R{\"o}ttgering}, {Br{\"u}ggen}, {Stroe}, {Hoeft}, {Nuza},
  \& {de Gasperin}}]{vanWeeren2014}
{van Weeren}, R.~J., {Intema}, H.~T., {Lal}, D.~V., {et~al.} 2014,
  \href{http://dx.doi.org/10.1088/2041-8205/781/2/L32}{\textcolor{magenta}{\apjl}},
  \href{http://adsabs.harvard.edu/abs/2014ApJ...781L..32V}{781, L32}

\bibitem[{{Vazza} {et~al.}(2011){Vazza}, {Brunetti}, {Gheller}, {Brunino}, \&
  {Br{\"u}ggen}}]{vazza2011}
{Vazza}, F., {Brunetti}, G., {Gheller}, C., {Brunino}, R., \& {Br{\"u}ggen}, M.
  2011,
  \href{http://dx.doi.org/10.1051/0004-6361/201016015}{\textcolor{magenta}{\aap}},
  \href{http://adsabs.harvard.edu/abs/2011A%26A...529A..17V}{529, A17}

\bibitem[{{Vazza} {et~al.}(2015){Vazza}, {Ferrari}, {Bonafede}, {Br{\"u}ggen},
  {Gheller}, {Braun}, \& {Brown}}]{vazza2015}
{Vazza}, F., {Ferrari}, C., {Bonafede}, A., {et~al.} 2015,
  \href{http://adsabs.harvard.edu/abs/2015aska.confE..97V}{Advancing
  Astrophysics with the Square Kilometre Array (AASKA14), 97}

\bibitem[{{Vazza} {et~al.}(2012){Vazza}, {Roediger}, \&
  {Br{\"u}ggen}}]{vazza2012}
{Vazza}, F., {Roediger}, E., \& {Br{\"u}ggen}, M. 2012,
  \href{http://dx.doi.org/10.1051/0004-6361/201118688}{\textcolor{magenta}{\aap}},
  \href{http://adsabs.harvard.edu/abs/2012A%26A...544A.103V}{544, A103}

\bibitem[{{Venturi} {et~al.}(2003){Venturi}, {Bardelli}, {Dallacasa},
  {Brunetti}, {Giacintucci}, {Hunstead}, \& {Morganti}}]{venturi2003}
{Venturi}, T., {Bardelli}, S., {Dallacasa}, D., {et~al.} 2003,
  \href{http://dx.doi.org/10.1051/0004-6361:20030345}{\textcolor{magenta}{\aap}},
  \href{http://adsabs.harvard.edu/abs/2003A%26A...402..913V}{402, 913}

\bibitem[{{Venturi} {et~al.}(2017){Venturi}, {Rossetti}, {Brunetti},
  {Farnsworth}, {Gastaldello}, {Giacintucci}, {Lal}, {Rudnick}, {Shimwell},
  {Eckert}, {Molendi}, \& {Owers}}]{venturi2017}
{Venturi}, T., {Rossetti}, M., {Brunetti}, G., {et~al.} 2017,
  \href{http://dx.doi.org/10.1051/0004-6361/201630014}{\textcolor{magenta}{\aap}},
  \href{http://adsabs.harvard.edu/abs/2017A%26A...603A.125V}{603, A125}

\bibitem[{{Wang} {et~al.}(2010){Wang}, {Xu}, {Gu}, {An}, {Cui}, {Li}, {Zhang},
  {Zheng}, \& {Wu}}]{wang2010}
{Wang}, J., {Xu}, H., {Gu}, J., {et~al.} 2010,
  \href{http://dx.doi.org/10.1088/0004-637X/723/1/620}{\textcolor{magenta}{\apj}},
  \href{http://adsabs.harvard.edu/abs/2010ApJ...723..620W}{723, 620}

\bibitem[{{Wang} {et~al.}(2013){Wang}, {Xu}, {An}, {Gu}, {Guo}, {Li}, {Wang},
  {Liu}, {Martineau-Huynh}, \& {Wu}}]{wang2013}
{Wang}, J., {Xu}, H., {An}, T., {et~al.} 2013,
  \href{http://dx.doi.org/10.1088/0004-637X/763/2/90}{\textcolor{magenta}{\apj}},
  \href{http://adsabs.harvard.edu/abs/2013ApJ...763...90W}{763, 90}

\bibitem[{{Wilber} {et~al.}(2018){Wilber}, {Br{\"u}ggen}, {Bonafede}, {Savini},
  {Shimwell}, {van Weeren}, {Rafferty}, {Mechev}, {Intema}, {Andrade-Santos},
  {Clarke}, {Mahony}, {Morganti}, {Prandoni}, {Brunetti}, {R{\"o}ttgering},
  {Mandal}, {de Gasperin}, \& {Hoeft}}]{wilber2018}
{Wilber}, A., {Br{\"u}ggen}, M., {Bonafede}, A., {et~al.} 2018,
  \href{http://dx.doi.org/10.1093/mnras/stx2568}{\textcolor{magenta}{\mnras}},
  \href{http://adsabs.harvard.edu/abs/2018MNRAS.473.3536W}{473, 3536}

\bibitem[{{Wilman} {et~al.}(2008){Wilman}, {Miller}, {Jarvis}, {Mauch},
  {Levrier}, {Abdalla}, {Rawlings}, {Kl{\"o}ckner}, {Obreschkow}, {Olteanu}, \&
  {Young}}]{wilman2008}
{Wilman}, R.~J., {Miller}, L., {Jarvis}, M.~J., {et~al.} 2008,
  \href{http://dx.doi.org/10.1111/j.1365-2966.2008.13486.x}{\textcolor{magenta}{\mnras}},
  \href{http://adsabs.harvard.edu/abs/2008MNRAS.388.1335W}{388, 1335}

\bibitem[{{Wyithe} \& {Loeb}(2004)}]{wyithe2004}
{Wyithe}, J.~S.~B., \& {Loeb}, A. 2004,
  \href{http://dx.doi.org/10.1038/nature03033}{\textcolor{magenta}{\nat}},
  \href{http://adsabs.harvard.edu/abs/2004Natur.432..194W}{432, 194}

\bibitem[{{Zaroubi}(2013)}]{zaroubi2013rev}
{Zaroubi}, S. 2013, in Astrophysics and Space Science Library, Vol. 396, The
  First Galaxies, ed. T.~{Wiklind}, B.~{Mobasher}, \& V.~{Bromm}, 45

\bibitem[{{Zheng} {et~al.}(2016){Zheng}, {Wu}, {Johnston-Hollitt}, {Gu}, \&
  {Xu}}]{zheng2016}
{Zheng}, Q., {Wu}, X.-P., {Johnston-Hollitt}, M., {Gu}, J.-h., \& {Xu}, H.
  2016,
  \href{http://dx.doi.org/10.3847/0004-637X/832/2/190}{\textcolor{magenta}{\apj}},
  \href{http://adsabs.harvard.edu/abs/2016ApJ...832..190Z}{832, 190}

\end{thebibliography}


\end{document}